\documentclass[a4paper,fleqn,usenatbib]{mnras}
\usepackage{mathptmx}
\usepackage[T1]{fontenc}
\usepackage{ae,aecompl}

\usepackage[pdftex]{graphicx}
\usepackage{amsmath}	
\usepackage{amssymb}	
\usepackage{subfig}

\def\Meraxes{\textsc{Meraxes}}
\def\DRAGONS{DRAGONS}
\def\Tiamat{{\em Tiamat}}
\def\MediTiamat{{\em Medi Tiamat}}
\def\TinyTiamat{{\em Tiny Tiamat}}
\def\tocf{\textsc{21cmFAST}}
\def\fiducial{{\em fiducial}}
\def\fiducialLine{\includegraphics[scale=0.7]{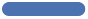}}
\def\nofb{{\em no feedback}}
\def\nofbLine{\includegraphics[scale=0.7]{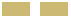}}
\def\recalibnofb{{\em recalibrated no feedback}}
\def\recalibnofbLine{\includegraphics[scale=0.7]{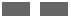}}
\def\nosnfb{{\em no supernova feedback}}
\def\nosnfbLine{\includegraphics[scale=0.7]{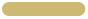}}
\def\recalibnosnfb{{\em recalibrated no supernova feedback}}
\def\recalibnosnfbLine{\includegraphics[scale=0.7]{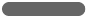}}
\def\noreionfb{{\em no reionization feedback}}
\def\noreionfbLine{\includegraphics[scale=0.7]{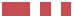}}
\def\homogeneous{{\em homogeneous}}
\def\homogeneousLine{\includegraphics[scale=0.7]{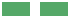}}
\def\doublefesc{{\em double $f_{\rm esc}$}}
\def\doublefescLine{\includegraphics[scale=0.7]{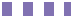}}
\def\halffesc{{\em half $f_{\rm esc}$}}
\def\halffescLine{\includegraphics[scale=0.7]{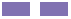}}
\def\zvaryingfesc{{\em $z$-varying $f_{\rm esc}$}}
\def\zvaryingfescLine{\includegraphics[scale=0.7]{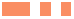}}

\title[DRAGONS -- Galaxy formation and the EoR]{
  Dark-ages~reionization~and~galaxy~formation~simulation - III. Modelling galaxy
  formation and the epoch of reionization}
\author[Mutch et al.]{Simon J. Mutch$^{1}$\thanks{E-mail: smutch@unimelb.edu.au},
  Paul M. Geil$^{1}$,
  Gregory B. Poole$^{1}$,
  Paul W. Angel$^{1}$,\newauthor
  Alan R. Duffy$^{2,1}$,
  Andrei Mesinger$^{3}$,
  J. Stuart B. Wyithe$^{1}$\\
  $^{1}$School of Physics, The University of Melbourne, Parkville, VIC 3010,
  Australia\\
  $^{2}$Centre for Astrophysics and Supercomputing, Swinburne University of
  Technology, PO Box 218, Hawthorn, VIC 3122, Australia\\
  $^{3}$Scuola Normale Superiore, Piazza dei Cavalieri 7, I-56126 Pisa, Italy
}

\begin{document}


\pagerange{\pageref{firstpage}--\pageref{lastpage}} \pubyear{2014}

\maketitle

\label{firstpage}

\begin{abstract}
%
We introduce \Meraxes, a new, purpose-built semi-analytic galaxy formation
model designed for studying galaxy growth during reionization. \Meraxes\ is the
first model of its type to include a temporally and spatially coupled treatment
of reionization and is built upon a custom (100\,Mpc)$^3$ \textit{N}-body
simulation with high temporal and mass resolution, allowing us to resolve the
galaxy and star formation physics relevant to early galaxy formation.  Our
fiducial model with supernova feedback reproduces the observed optical depth to
electron scattering and evolution of the galaxy stellar mass function between
$z$=5 and 7, predicting that a broad range of halo masses contribute to
reionization.  Using a constant escape fraction and global recombination rate,
our model is unable to simultaneously match the observed ionizing emissivity at
$z{\la}6$.  However, the use of an evolving escape fraction of 0.05--0.1 at
$z{\sim}6$, increasing towards higher redshift, is able to satisfy these three
constraints.  We also demonstrate that photoionization suppression of low mass
galaxy formation during reionization has only a small effect on the ionization
history of the inter-galactic medium.  This lack of `self-regulation' arises
due to the already efficient quenching of star formation by supernova feedback.
It is only in models with gas supply-limited star formation that reionization
feedback is effective at regulating galaxy growth.  We similarly find that
reionization has only a small effect on the stellar mass function, with no
observationally detectable imprint at $M_{\rm *}{>}10^{7.5}\,{\rm M_{{\sun}}}$.
However, patchy reionization has significant effects on individual galaxy
masses, with variations of factors of 2--3 at $z$=5 that correlate with
environment.
%
%

\end{abstract}



\section{Introduction}
\label{sec:introduction}

There are several key observational areas in which substantial progress
will be made in the study of the first galaxies during the coming
decade. Of particular importance will be forthcoming programmes searching
for galaxies beyond the current redshift frontier using the {\em Hubble
Space Telescope} and, in the future, the {\em James Webb Space Telescope}
\citep[e.g.][]{Bouwens2011,McLure2013,Schenker2013}.  However, even next
generation surveys will not extend to the faint luminosities of the faintest
galaxies thought to drive the reionization of inter-galactic neutral hydrogen
in the early Universe \citep{Robertson2013, Duffy2014}. Thus, alongside new
probes provided by high redshift gamma-ray bursts \citep[e.g.][]{Trenti2015}
and metal pollution of the inter-galactic medium \citep[IGM; e.g.][]{Diaz2014},
an important new observational window for study of the first galaxies will be
provided by experiments to measure the redshifted $21\, {\rm cm}$ radio signal
\citep{Furlanetto2006,Morales2010}. These observations will both provide the
first direct probe of the neutral hydrogen content in the high redshift Universe
and, through modelling, provide a route to study the early dwarf galaxies
thought to exist during reionization alongside their more massive counterparts
whose star formation can be directly detected.

Within this context, the development of theoretical models that include a
self-consistent treatment of the physics of galaxy formation and intergalactic
hydrogen will play a key role. Traditional approaches to the study of galaxies
and their effects on the IGM utilize either numerical simulation or analytic
modelling. The latter allows investigation of average behaviours on large
scales but the calculations are inherently linear, meaning that complex
feedback processes cannot be addressed
\citep[e.g.][]{Furlanetto2004,Wyithe2004,Wyithe2013}. Numerical simulations, on
the other hand, include non-linear effects but at the expense of computational
cost. To achieve a volume sufficiently large to study ionized structure, a
popular and effective approach to simulating reionization is to begin with a
collisionless \textit{N}-body simulation \citep[e.g.][]{Ciardi2003,
Sokasian2003, Iliev2007, Iliev2008, Trac2007, Zahn2007, Shin2008, Trac2008} and
use a simple prescription to relate halo mass to ionizing luminosity. A
radiative transfer method (for example ray-tracing algorithms) can then be used
to calculate the ionization structure on large scales. In recent years, new
hybrid, or semi-numerical models \citep{Mesinger2007,Geil2008,Kim2013} have
been developed that combine \textit{N}-body simulations with analytical methods
to enable the calculation of reionization structure in very large volumes with
high efficiency.  These methods have elucidated the primary features of the
ionization structure during reionization, but do not capture the physics of
galaxy formation.

Therefore, to better understand the physics of galaxy formation, many
authors have performed hydrodynamic simulations of galaxy formation
\citep[e.g.]{Finlator2011, Salvaterra2011, Jaacks2012} which are able to
directly model the growth of stellar mass in high-redshift galaxies when
coupled with sub-grid models for processes including metal enrichment and
feedback. These simulations are able to broadly reproduce the luminosity
function of galaxies at high redshift, however, computational expense limits
their ability to self-consistently model reionization in volumes large enough
to statistically describe the spatial evolution of this process.  Instead,
a common approach is to impose a simple parametrized model to approximate
the average ionizing background as a function of redshift, independent of
the properties of the ionizing source population \citep[e.g.][]{Feng2015} or
their spatial distribution \citep[e.g.][]{Genel2014, Schaye2015}.  Recently,
hydrodynamical simulations of galaxy formation with coupled radiative transfer
have been used to compute the effects of reionization on galaxy formation
self-consistently for the first time \citep[e.g.][]{So2014, Norman2015, Ocvirk2015,
  Pawlik2015}. However, the extreme computational expense of these
simulations limit their size to relatively small volumes and/or few variations
on galaxy formation physics or reionization scenarios that can be explored.
In addition, the modelling of sub-grid physical processes remains uncertain,
requiring systematic studies of the available parameter space in order to draw
robust conclusions.  Such studies represent an extreme computational challenge
which has yet to be overcome.

Another approach to the realistic modelling of high redshift galaxies has been
through the use of semi-analytic galaxy formation models \citep{Benson2006,
Lacey2011, Raicevic2011, Zhou2013}. While large volumes are available to
such models, until now they have not been fully coupled to an accurate
description of reionization.  This is in part due to the structure of
most existing semi-analytic models, which utilize so-called `vertical'
halo merger trees \citep[e.g.][]{Springel2005, Bower2006, Harker2006,
De-Lucia2007} in which galaxies belonging to each tree branch are evolved
independently from the rest of the simulation volume.  Since galaxies
drive the process of reionization, which in turn affects their subsequent
evolution, galaxies spatially separated by tens of Mpc cannot be considered and
evolved independently as has traditionally been the case \citep{Wyithe2004}.
Self-consistently studying reionization instead requires a semi-analytic
model designed to run on `horizontal' merger trees where all haloes at each
snapshot of the parent \textit{N}-body simulation are processed simultaneously.
Additionally, the reduced dynamical time of dark matter haloes at high redshift
requires snapshots with a much higher cadence than is needed to model galaxy
formation at lower redshifts.

This is the third paper in a series describing the Dark-ages Reionization And
Galaxy Observables from Numerical Simulations (\DRAGONS)
project\footnote{http://dragons.ph.unimelb.edu.au/}, which integrates detailed
semi-analytic models constructed specifically to study galaxy formation at high
redshift, with semi-numerical models of the galaxy--reionization process
interaction. In this work, we introduce \Meraxes, the new semi-analytic model
of galaxy formation developed for \DRAGONS, integrating the \tocf\
semi-numerical model for ionization structure described in
\citet{Mesinger2007}. \Meraxes\ is implemented within the large-volume,
high-resolution, and high-cadence \Tiamat\ \textit{N}-body simulation described
in \citet[][hereafter Paper I]{Poole2016} and \citet[][Paper II]{Angel2016}.
In subsequent papers we will use \Meraxes\ to carry out a range of studies
including the investigation of the high redshift galaxy luminosity function
\citep[][Paper IV]{Liu2015}, and the ionization structure of the IGM
\citep[][Paper V]{Geil2015}. Complimentary, high resolution hydrodynamic
simulations are described in \citet{Duffy2014}.

The outline of the paper is as follows. In Section \ref{sec:meraxes} we provide
a full description of \Meraxes\ including an overview of the \Tiamat\
simulation and associated merger trees which act as input for our semi-analytic
model, the physical prescriptions employed, and the methodology of our
reionization coupling and the integration of \tocf.  In
Section~\ref{sec:model_calibration} we then go on to describe the calibration
of the model's free parameters against the evolution of the high-redshift
galaxy stellar mass function.  In Section~\ref{sec:results} we investigate a
range of different extreme reionization and galaxy physics modifications in
order to elucidate the roles of reionization suppression and galactic feedback
processes in the build up of stellar mass and the evolution of the global
neutral hydrogen fraction.  We also highlight the important consequences of
utilizing a patchy, self-consistent reionization model compared to more
commonly employed, parametrized descriptions of reionization.  Finally, in
Section \ref{sec:conclusions} we summarize our study and conclusions.
Throughout this work, we employ a standard, spatially flat $\Lambda$ cold dark
matter cosmology with the most up-to-date cosmological parameters as determined
by \citet{Planck-Collaboration2015}: ($h$, $\Omega_{\rm{m}}$,
  $\Omega_{\rm{b}}$, $\Omega_\Lambda$, $\sigma_8$, $n_{\rm{s}}$)=(0.678, 0.308,
  0.0484, 0.692, 0.815,
0.968).


\section{Meraxes}
\label{sec:meraxes}

Modern semi-analytic galaxy formation models are capable of providing
statistically accurate representations of the global properties of galaxies
across a broad range of redshifts \citep[e.g.][]{Baugh2006, Mutch2013,
Henriques2015}, and are therefore able to describe the distribution and
evolution of the ionizing photons which drive the process of cosmic
reionization.  These photons generate regions of ionized hydrogen (H\,\textsc{ii})
with characteristic sizes of tens of ${\rm Mpc}$ during reionization
\citep{Wyithe2004}.  Thus, in order to take advantage of this information and to
self-consistently model the effect of these photons on the growth of galaxies,
one must consider the contributions of galaxies separated by similar scales.

Traditionally, semi-analytic models have therefore used parametrized
descriptions to include the average effect of reionization and the
associated photo-suppression of baryonic infall on the growth of galaxies
\citep{Benson2006, Croton2006, Somerville2008}.  These parametrizations are
typically calibrated using radiative transfer simulations and are provided as
a function of redshift and halo mass alone \citep[e.g.][]{Gnedin2000}.  Whilst
it is computationally efficient to include reionization in this manner, there
are a number of important drawbacks.  First, the progression of reionization
is not self-consistently modified by the growth of the galaxies which are
driving it.  Therefore it is impossible to investigate how different galaxy
physics affect the ionization state of the IGM or to quantify
the potential back-reaction on galaxy evolution.  Secondly, these simple
reionization prescriptions miss the potentially important effects of spatially
dependent self-regulation \citep{Iliev2007, Sobacchi2013b}, whereby massive
galaxies located at the peaks in the density distribution can reionize their
surroundings, delaying or preventing the onset of star formation in nearby lower
mass haloes.

Our new semi-analytic galaxy formation model, \Meraxes, has been written from 
the ground up to facilitate these modelling requirements.  Its key features 
include the `horizontal' processing of merger trees constructed from a purpose 
run \textit{N}-body simulation (\Tiamat; see Section~\ref{sub:n_body_simulation} below) 
and the incorporation of the semi-numerical reionization algorithm, \tocf, as a 
core component.  When combined, these features allow \Meraxes\ to efficiently 
couple the growth of galaxies to the process of reionization, both temporally 
and spatially.  It can therefore be used to investigate the potentially complex 
effects of various reionization models on the properties of high-$z$ galaxies, 
as well as to test for observational discriminants of different galaxy physics 
in the distribution and evolution of inter-galactic neutral hydrogen.

In order to develop confidence in our newly developed framework, as well
as provide a solid foundation for future additions and improvements, our
initial implementation of the baryonic physics processes in \Meraxes\ is
heavily based on the well-studied \textsc{L-Galaxies} semi-analytic model
\citep{Kauffmann1996, De-Lucia2007, Guo2013, Henriques2015}, in particular
the version described in \citet{Croton2006} and extended in \citet{Guo2011}.
However, as well as our improved treatment of reionization, the excellent
temporal resolution provided to us by the \Tiamat\ merger trees has also
necessitated the development of a number of important updates to the treatment
of supernova feedback and stellar mass recycling.

In the following sub-sections we describe \Meraxes\ in full, including its
input data set in the form of halo merger trees extracted from the \Tiamat\
\textit{N}-body simulation, the details of the implemented galaxy physics
prescriptions, and our methodology for integrating \tocf\ to self-consistently
model reionization.

\subsection{Input -- the \Tiamat\ \textit{N}-body simulations}
\label{sub:n_body_simulation}

The \Tiamat\ collisionless N-body simulation has been designed for the \DRAGONS\
study of high-redshift galaxy formation and the epoch of reionization (EoR).  It
contains $2160^3$ dark matter particles within a $100\,{\rm Mpc}$ (comoving)
periodic box and was run using a modified version of the {\small GADGET-2}
N-body code and the latest {\em Planck} 2015 \citep{Planck-Collaboration2015}
cosmology.  The volume of \Tiamat\ allows for the investigation of the
statistical signatures of reionization and its $21\,{\rm cm}$ observational
signal, whilst the resulting particle mass of $3.89\times 10^6\,{\rm M_{{\sun}}}$
provides the necessary resolution to identify the low-mass sources thought to be
driving this process.  Furthermore, \Tiamat\ provides high temporal resolution
in the form of 100 output snapshots evenly spaced in cosmic time between
$z{=}35$ and 5, resulting in a cadence of $11.1\,{\rm Myr}$ per snapshot.  This
level of temporal resolution is a unique feature of \Tiamat\ which allows
our semi-analytic model to accurately simulate the stochastic nature of star
formation in a regime where the dynamical time of a typical galactic disc is
shorter than the lifetime of the least massive Type II supernova progenitor
($\sim 40\,{\rm Myr}$).

In addition to the main \Tiamat\ volume, a suite of smaller, higher
mass resolution \textit{N}-body simulations have been run as part of the
DRAGONS programme (Paper-I).  For this work we make particular use of the
\TinyTiamat\ and \MediTiamat\ volumes in order to quantify the effect of
resolution on our results (see Section~\ref{sub:results-relative_eff} and
Appendix~\ref{sec:appendix-resolution}). \TinyTiamat\ is the highest resolution
simulation of the \DRAGONS\ suite, with a particle mass of $10^5\,{\rm
M_{{\sun}}}$ in a small box of side length $14.8\,{\rm Mpc}$, whilst \MediTiamat\
bridges the resolution gap with the main simulation by providing a particle
mass of $1.16\times 10^{6}\,{\rm M_{{\sun}}}$ in a $33.3\,{\rm Mpc}$ box.  Both
simulations maintain the same snapshot cadence as the main \Tiamat\ volume and
are described in detail in Paper-I.

Halo identification in all simulations used in this work was carried out using
the \textsc{Subfind} \citep{Springel2001} real-space halo finder down to a
minimum mass of 32 particles (corresponding to $3.71{\times} 10^8$,
$1.25{\times} 10^8$, $3.2{\times} 10^6\, {\rm M_{{\sun}}}$ for \Tiamat,
\MediTiamat, and \TinyTiamat\ respectively).  The resulting halo catalogues
comprise of friends-of-friends (FoF) groups of gravitationally bound
particles which themselves are made up of a single mass dominant `central'
subhalo along with zero or more sub-dominant `satellite' subhaloes.  For
further details, interested readers are referred to Papers I and II.

\subsubsection{Merger trees}
\label{ssub:input_merger_trees}

The formation history of subhaloes, in the form of hierarchical merger trees,
acts as the raw input to \Meraxes\ and is used to define the positions and
growth of galaxies.  Many traditional semi-analytic models, such as the
\textsc{L-Galaxies} \citep[e.g.][]{De-Lucia2007, Guo2011, Henriques2015} and
\textsc{GALFORM} \citep[e.g.][]{Bower2006, Lagos2012, Kim2013b} variants,
process such trees in a depth-first (or `vertical') order, whereby small
collections of directly interacting dark matter haloes are processed one after
the other from high to low redshift and independently of each other.  Whilst
computationally efficient in terms of minimizing the memory overhead required
to process the simulation, the inherent assumption is that haloes (and by
extension galaxies) which do not directly interact do not affect each other's
evolution.  This assumption breaks down when considering the process of
reionization during which ionizing photons from galaxies tens of Mpc away can
heat the IGM, raising the local Jeans mass and altering the accretion rate of
baryons \citep{Dijkstra2004}.  \Meraxes\ instead processes trees breadth-first
(or `horizontally').  In this method all of the haloes in the entire volume
are loaded into memory and the associated galaxies evolved for each snapshot
sequentially.  This allows \Meraxes\ to more efficiently model reionization
than previous comparable works \citep[e.g.][]{Kim2013}.  More detailed
information, including the precise details of our merger tree construction
technique, can be found in Poole et al. (in preparation).

\subsection{Baryonic infall}
\label{sub:baryonic_infall}

We begin by making the standard assumption that as FoF groups grow, any freshly
accreted mass, always carries with it the universal baryon fraction, $f_{\rm b}
{=} \Omega_{\rm b} / \Omega_{\rm m}$, in the form of pristine primordial gas.
However, the fraction of these infalling baryons which will remain bound to the
FoF group and participate in galaxy formation may be reduced by a number of
factors.  In particular, ionizing ultraviolet background (UVB) radiation from
both local and external sources can heat the IGM, increasing the local Jeans
mass and leading to a non-negligible reduction in the amount of baryons
successfully captured by low mass systems \citep{Dijkstra2004}.  We parametrize
this reduction in terms of a baryon fraction modifier, $f_{\rm mod}$, which
represents the attenuation of the total baryon mass that could have ever been
successfully captured by an FoF group in its lifetime:
\begin{equation}
  m_{\rm infall} = f_{\rm mod}f_{\rm b}M_{\rm vir} - \sum_{i=0}^{N_{\rm gal}-1} 
  m^i_* + m^i_{\rm cold} + m^i_{\rm hot} + m^i_{\rm ejected}\ ,
  \label{eqn:fmod_defn}
\end{equation}
where $m_{\rm infall}$ is the infalling baryonic mass, $0 {\le} f_{\rm mod}
{\le} 1$, and $N_{\rm gal}$ is the number of galaxies in the FoF group.  The
baryonic reservoirs $m_*$, $m^i_{\rm cold}$, $m^i_{\rm hot}$, and $m^i_{\rm
ejected}$ are described in the following sections along with the physical
prescriptions which govern their evolution.  If the mass of the FoF group or
the value of $f_{\rm mod}$ decreases then it is possible for $m_{\rm infall}$
to become negative.  In this case baryons are stripped from the system as
described in Section~\ref{sub:halo_infall_and_gas_stripping}.  An accurate
spatially and temporally dependent calculation of the value of $f_{\rm mod}$
is a key feature of \Meraxes\ and a subject which we return to in detail in
Section~\ref{sub:reionization}.

Any baryons which are successfully captured are assumed to be shocked to the
virial temperature of the host FoF group and added to a diffuse hydrostatic hot
reservoir where they mix with any already present hot gas.

\subsection{Cooling}
\label{sub:cooling}

At each time step in the simulation some fraction of the hydrostatic hot
reservoir may cool and condense down into the central regions of the group where
it can then participate in galaxy formation.  In order to calculate the rate
at which this occurs we follow the commonly employed methodology outlined in
\citet{White1991}.  In this model, the cooling time of a quasi-static isothermal
hot halo is given by the ratio of the specific thermal energy to cooling rate
per unit volume:
\begin{equation}
  t_{\rm cool}(r) = \frac{1.5 \bar{\mu}m_{\rm p}kT}{\rho_{\rm
      hot}(r)\Lambda(T, Z)}\ ,
  \label{eqn:tcool}
\end{equation}
where $\bar \mu m_{\rm p}$ is the mean particle mass ($9.868{\times} 10^{-25}\,
{\rm g}$ for a fully ionized gas), $k$ is the Boltzmann constant, $\Lambda$
is the cooling function \citep{Sutherland1993}, $T$ is the temperature of
the gas and $\rho_{\rm hot}(r)$ is its density profile.  As mentioned above,
we assume that the hot gas is shocked to the virial temperature of the FoF
group, therefore we set $T{=}T_{\rm vir}{=}35.9(V_{\rm vir}/{\rm km\,s^{-1}})^2
{\rm K}$.  For simplicity, we also assume that the hot gas follows a singular
isothermal sphere density profile:
\begin{equation}
  \rho_{\rm hot}(r) = \frac{m_{\rm hot}}{4 {\rm \pi} R_{\rm vir} r^2}\ .
\end{equation}

With knowledge of the cooling time, we can define an appropriate cooling
radius, $r_{\rm cool}$, within which there is enough time for the material
to lose pressure support and condense to the system centre.  Following
\citet{Croton2006}, we take this to be the radius at which $t_{\rm cool}$ is
equal to the dynamical time of the host FoF group, $t_{\rm dyn}^{\rm FoF} {=} R_{\rm
vir}/V_{\rm vir}$.  As discussed by \citet{White1991}, this model for cooling
naturally leads to three distinct regimes.
\begin{enumerate}
  \item When $r_{\rm cool}{\ge}R_{\rm vir}$, any infalling gas will cool so
    rapidly that there will be no time for a stable shock to form and thus for
    the gas to reach hydrostatic equilibrium.  In this case we assume that the
    infalling material flows directly into the central regions of the halo over
    a dynamical (free-fall) time, $\dot m_{\rm cool} {=} m_{\rm hot}/t_{\rm 
      dyn}$.
  \item When $r_{\rm cool}{<}R_{\rm vir}$ the cooling time will be sufficiently
    long that a quasi-static hot atmosphere will form.  The cooling rate from
    this atmosphere can then be calculated from a simple continuity equation for
    the mass flux across the evolving cooling radius:
    \begin{eqnarray}
      \dot m_{\rm cool} & = & 4{\rm \pi} \rho_{\rm hot}(r_{\rm cool})r^2_{\rm cool} \dot
      r_{\rm cool} \nonumber \\
      & = & m_{\rm hot} \frac{r_{\rm cool}}{R_{\rm 
          vir}}\frac{1}{t_{\rm dyn}^{\rm FoF}}
    \end{eqnarray}
  \item When $T_{\rm vir}{\le}10^4\,{\rm K}$, we set $r_{\rm cool}{=}0$ and no 
    cooling occurs.  In the standard model of galaxy formation, haloes with this
    temperature represent the lowest mass scale for galaxy formation.  Below
    this, the primary mechanism for gas cooling is via molecular hydrogen which
    is easily photo-dissociated by trace amounts of star formation, making
    it an inefficient pathway for Pop II star formation.  Above this temperature,
    atomic line cooling provides an efficient mechanism to dissipate energy and
    remove pressure support \citep{Barkana2001}.  The mass resolution of our
    input \textit{N}-body simulation, \Tiamat, was chosen such that the minimum halo
    mass at $z{=}5$ is close to the atomic cooling mass threshold of $T_{\rm
    vir}{=}10^4\,{\rm K}$.  Although earlier Pop III and Pop II star formation
    is possible in smaller haloes, the level of contribution of these objects to 
    reionization remains unclear, being heavily dependent on the masses of the 
    first supernov\ae\ which could potentially delay future star formation by
    tens to hundreds of Myr \citep[e.g.][]{Chen2014,Jeon2014}. We therefore do not
    include these objects in our current model.

\end{enumerate}

All material which successfully cools into the central regions of the FoF group
is assumed to be deposited directly into the cold gas reservoir of the galaxy
hosted by the central halo.  This assumption is commonly employed by a number of
other semi-analytic models \citep[e.g.][]{Bower2006, De-Lucia2007, Somerville2008,
Lu2011} which utilize halo catalogues created by the \textsc{Subfind} halo
finder.  It is also warranted in the vast majority of systems where the central
halo dominates the mass of the FoF group and ensures a physically meaningful
match between the galaxy formation physics of \Meraxes\ and the substructure
hierarchy produced by the halo finder employed for this programme (see Paper I
where this issue is raised).

\subsection{Star formation}
\label{sub:star_formation}

As discussed in the previous section, gas which cools from the FoF group hot
reservoir is assumed to be deposited into the galaxy hosted by the central halo 
of the group.  Here we assume that it settles into a rotationally supported cold 
gas disc with an exponential surface density profile.  Under the simplifying
assumption of full conservation of specific angular momentum, the scale radius
of the disc can be approximated from the spin of the host dark matter halo,
$\lambda$, to be $r_{\rm s} = R_{\rm vir}(\lambda/\sqrt{2})$, where we use the 
definition of $\lambda$ provided by \citet{Bullock2001}.

Based on the well-established observational work of \citet{Kennicutt1998}, the
star formation rate of local spiral galaxies can be related to the surface
density of cold gas above a given threshold.  The value of this threshold
can be understood in terms of the gravitational instability required to form
massive star-forming clouds \citep{Kennicutt1989}.  Assuming a constant gas
velocity dispersion and a flat rotation curve with a circular velocity equal to
that of the host dark matter halo, $V_{\rm vir}$, \cite{Kauffmann1996}
demonstrated that this stability criterion can be expressed as
\begin{equation}
  \Sigma_{\rm crit}(r) = \Sigma_{\rm norm} \left( \frac{V_{\rm vir}}{\rm km\,s^{-1}}
  \right) \left( \frac{r}{\rm kpc} \right)^{-1}
  {\rm M_{{\sun}}\,pc^{-2}}\ .
  \label{eqn:sigma_crit}
\end{equation}
\citet{Kauffmann1996} originally assumed a thin isothermal disc with a
gas velocity dispersion appropriate for low-redshift spiral galaxies of
$6\,{\rm km\,s^{-1}}$, resulting in $\Sigma_{\rm norm} = 0.59$.  However,
both observations and simulations of high-redshift star forming galaxies
indicate that they typically possess highly turbulent, clumpy discs
\citep[e.g.][]{Wisnioski2011,Glazebrook2013,Bournaud2014}.  This suggests the
need for a modified $\Sigma_{\rm norm}$ value.  Given the uncertainty in what
value this should take, we choose to leave it as a free parameter in our model.
We note that \citet{Henriques2015} also advocate allowing freedom in the choice
of $\Sigma_{\rm norm}$; however, they instead motivate this by the observation
that star formation is more closely linked to molecular, rather than total,
gas density \citep[e.g.][]{Leroy2008}.  This suggests that the surface density
of total gas required for star formation could plausibly be lower than the
\citet{Kauffmann1996} value.

Equation~(\ref{eqn:sigma_crit}) can be converted to a total critical mass, $m_{\rm 
crit}$, by
integrating out to the disc radius, $r_{\rm disc}$:
\begin{equation}
  m_{\rm crit} = 2{\rm \pi}\,\Sigma_{\rm norm} \left(\frac{V_{\rm vir}}{\rm km\,
      s^{-1}}\right) \left(\frac{r_{\rm disc}}{\rm kpc}\right) 10^6 M_{{\sun}}\ .
\end{equation}
Following \citet{Croton2006}, we assume $r_{\rm disc}{=}3r_{\rm s}$.  The factor
of 3 was chosen by \citet{Croton2006} based on the properties of the Milky Way
and therefore may not be representative of the high-redshift galaxies which we
consider in this work.  However, instead of considering this to be another
free parameter of the model we note that $m_{\rm crit} {\propto} \Sigma_{\rm
norm}r_{\rm disc}$, meaning any such freedom can already be considered to be
included in $\Sigma_{\rm norm}$.  In future work, we will compare our predicted 
disc sizes to high-$z$ observations and investigate the success of these simple 
scaling relations in detail.

If the total amount of cold gas in the disc is greater than the critical mass,
the star formation rate is assumed to be given by
\begin{equation}
  \dot m_* = \alpha_{\rm SF}\frac{(m_{\rm cold} - m_{\rm crit})}{t_{\rm 
      dyn}^{\rm disc}}\ ,
  \label{eqn:new_stars}
\end{equation}
where $t_{\rm dyn}^{\rm disc} = r_{\rm disc}/V_{\rm vir}$ is the dynamical time of the disc
and $\alpha_{\rm SF}$ is a free parameter describing the efficiency of star
formation in the form of the formation time-scale in units of the dynamical time.

In summary, the star formation prescription we employ in \Meraxes\ is almost 
identical to that of \citet{Croton2006} with the addition of $\Sigma_{\rm norm}$ 
as an extra free parameter \citep[as previously advocated by][]{Henriques2015}.

\subsection{Supernova feedback}
\label{sub:supernova_feedback}

The radiative and mechanical energy liberated by supernov\ae\ can have a profound
impact on galaxy evolution, potentially heating significant amounts of gas and
even ejecting it from a galaxy or host dark matter halo entirely.  This is 
especially so at high redshift where haloes are on average less massive than in 
the local Universe, and possess correspondingly shallower potential wells.  
Supernov\ae\ also enrich the inter stellar medium (ISM), altering the chemical 
composition of future stellar generations and changing the efficiency with which 
gas can cool (c.f. Equation~\ref{eqn:tcool}).

\subsubsection{Delayed supernova feedback}
\label{ssub:delayed_supernova_feedback}

Many semi-analytic models make the simplifying assumption that all supernova
feedback energy is released instantaneously, during the same snapshot in which
the relevant stars formed.  This approximation is motivated by the reasonable
further assumption that the majority of supernova feedback energy is released
by massive stars ($m_*{>}8\,{\rm M_{{\sun}}}$) which have short \citep[$\la
40\,{\rm Myr}$;][]{Portinari1998} lifetimes, ending in violent SN-II.  In the
cases where the time span between each simulation snapshot is large \citep[e.g.
$\approx$ 250 Myr in the case of the Millennium Simulation;][]{Springel2005},
the approximation of the instantaneous deposition of all supernova energy into
the ISM is valid.  However, motivated by the short
dynamical time of systems at high redshift, the separation between snapshots in
our input simulation is approximately 11.1 Myr.  It therefore takes at least
three snapshots after a single coeval star formation episode for all stars more
massive than $8\,{\rm M_{\sun}}$ to have gone supernova.  In order to
accommodate this matching of time-scales in \Meraxes, we have implemented a
simple delayed supernova feedback scheme which we outline in this section.

We begin with the standard assumption that all supernova feedback energy is
released by SN-II which are the end result of the evolution of stars with
initial masses greater than $8\,{\rm M_{\sun}}$. The basic methodology of our
delayed feedback scheme is then to calculate the total amount of energy which
should be injected into the ISM by a single star formation episode, and to
release this energy gradually over time in proportion to the fraction of SN-II
which will have occurred.

We assume a standard \cite{Salpeter1955} initial mass function (IMF) with upper and lower mass limits
of $0.1\,{\rm M_{\sun}}$ and $120\,{\rm M_{\sun}}$ respectively:
\begin{equation}
\phi(m) = \phi_{\rm norm} m^{-2.35}\ ,
\end{equation}
where, by definition
\begin{equation}
  \int_{0.1 {\rm M_{{\sun}}}}^{120 {\rm M_{{\sun}}}}m\phi(m)\,{\rm d}m = 1\ ,
\end{equation}
and thus $\phi_{\rm norm}$ = 0.1706.  With this choice of IMF, the number
fraction of stars that will end their lives as type II supernov\ae\ ($\eta_{\rm
SNII}$) is given by:
\begin{equation}
\eta_{\rm SNII} = \int_{8 {\rm M_{{\sun}}}}^{120 {\rm M_{{\sun}}}}
\phi(m)\,{\rm d}m = 7.432\times 10^{-3}\, \rm{M_{{\sun}}^{-1}}.
\label{eqn:total_SNII}
\end{equation}

If we further assume that each supernova produced injects a constant $E_{\rm
nova} = 10^{51}\,\rm{erg}$ of energy into the ISM then, for a burst of mass
$\Delta m_*$, the total amount of energy deposited into the ISM ($\Delta E_{\rm
total}$) is
\begin{equation}
  \Delta E_{\rm total} = \epsilon_{\rm energy} \Delta m_* \eta_{\rm SNII}
  E_{\rm nova}\ ,
  \label{eqn:SN_energy}
\end{equation}
where $\epsilon_{\rm energy}$ is a free parameter describing the efficiency
with which the supernova energy couples to the surrounding gas.  As is common 
practice, we model the mass of gas which is reheated by this energy
deposition ($\Delta m_{\rm total}$) as
\begin{equation}
  \Delta m_{\rm total} = \epsilon_{\rm mass} \Delta m_*\ ,
  \label{eqn:reheated_mass}
\end{equation}
where $\epsilon_{\rm mass}$ is a free parameter commonly referred to as the mass
loading factor.

\citet{Croton2006} used constant values of $\epsilon_{\rm energy}$ and
$\epsilon_{\rm mass}$ for all galaxies.  However, we find that we are unable to
replicate the observed shallow low-mass slope of the stellar mass function at
$z{\ge}5$ without adopting a value for these parameters that scale with mass.
We therefore follow \citet{Guo2013} who encountered a similar issue (although
at lower redshifts) leading them to adopt the following parametrizations for
$\epsilon_{\rm energy}$:
\begin{equation}
  \epsilon_{\rm energy} = \alpha_{\rm energy} \left[0.5 + \left(
      \frac{V_{\rm max}}{V_{\rm energy}}\right)^{-\beta_{\rm energy}} \right]\ ,
  \label{eqn:epsilon_energy}
\end{equation}
and similarly for $\epsilon_{\rm mass}$:
\begin{equation}
  \epsilon_{\rm mass} = \min \left\{ \alpha_{\rm mass} \left[0.5 + \left(
      \frac{V_{\rm max}}{V_{\rm mass}}\right)^{-\beta_{\rm mass}} \right], 
  \epsilon_{\rm mass}^{\rm max} \right\}\ ,
  \label{eqn:epsilon_mass}
\end{equation}
where $\alpha_{\rm energy}$, $V_{\rm energy}$, $\beta_{\rm energy}$,
$\alpha_{\rm mass}$, $V_{\rm mass}$, and $\beta_{\rm mass}$ are all free
parameters.  Since $\epsilon_{\rm energy}$ corresponds to an efficiency,
we enforce that it must take a value in the range 0--1 at all times.  We
have also imposed the additional constraint that the mass loading factor,
$\epsilon_{\rm mass}$, cannot exceed an upper limit which we nominally set to
be $\epsilon_{\rm mass}^{\rm max}{=}10$ based on reasonable expectations for
typical high-$z$ galaxies \citep[e.g.][]{Martin1999,Uhlig2012}.  For a standard
energy-driven wind, $\beta_{\rm mass}{=}2$ \citep{Murray2005}.  However, the
value of $\beta_{\rm energy}$ is far less certain and depends on the poorly
understood efficiency with which injected supernova energy is thermalized
\citep{Murray2005} and potential variations in the IMF of stars.

As discussed above, it takes approximately 40Myr for an $8\, {\rm M_{{\sun}}}$
star to go supernova \citep{Portinari1998}.  As a result, the total amount of
supernova energy released by a galaxy at any given snapshot, $\Delta E_{\rm
reheat}$, will be dependent on the mass of stars formed both in the current
and previous snapshots.  We therefore explicitly track the total mass of stars
formed in each galaxy (and all of its progenitors) for the last $N_{\rm SFH}$
snapshots\footnote{SFH $\rightarrow$ star formation history}.  The value of
$N_{\rm SFH}$ is dependent on the input \textit{N}-body simulation and is chosen such
that at least the last $40\,{\rm Myr}$ of star formation is recorded at all
times.  For \Tiamat\ this corresponds to $N_{\rm SFH} = 4$.  At snapshot $j$,
the value of $\Delta E_{\rm reheat}$ is then
\begin{equation}
\Delta E_{{\rm reheat},j} = \sum_{i=j-N_{\rm SFH}}^{i=j} \frac{\Delta
\eta_{i,j}}{\eta_{\rm SNII}} \Delta E_{{\rm total}, i}\
.\label{eqn:delta_e_reheat}
\end{equation}
Similarly, the amount of cold gas reheated by this energy is
\begin{equation}
\Delta m_{{\rm reheat},j} = \sum_{i=j-N_{\rm SFH}}^{i=j} \frac{\Delta
\eta_{i,j}}{\eta_{\rm SNII}} \Delta m_{{\rm total}, i}\
.\label{eqn:delta_m_reheat}
\end{equation}

The term $\Delta \eta_{i,j}$ in the two equations above denotes the fraction of
stars formed during snapshot $i$, that go supernova during snapshot $j$.  This
can be calculated by integrating the stellar IMF ($\phi(m)$) between suitably
chosen mass limits:
\begin{equation}
  \Delta \eta = \int_{m_{\rm low}}^{m_{\rm high}} \phi(m)\,{\rm d}m\ .
\label{eqn:delta_eta}
\end{equation}
The values of $m_{\rm high}$ and $m_{\rm low}$ are set by the range of stellar
masses formed during snapshot $i$ which will have had time to expend their fuel
and go nova during the time spanned by snapshot $j$.  To calculate these we use
a functional form fit to the $Z{=}0.004$ H and He core burning lifetimes
tabulated by \cite{Portinari1998}, under the assumption
that all stars go supernova immediately upon expending their H and He cores,
\begin{equation}
  \log_{10}\left(m(t)\right) = \frac{a}{\log_{10}(t/{\rm Myr})} + b \exp\left(
\frac{c}{\log_{10}(t/{\rm Myr})} \right) + d\ ,
\end{equation}
where $(a,b,c,d) {=} (0.7473, -2.6979, -4.7659, 0.5934)$ and $t$ is the time
since the stars formed.  This fit is accurate to within 6\% for all values of
$t$ appropriate for this work.  For simplicity, we also approximate the star
formation which occurs during any given snapshot by a single coeval burst at the
middle of that snapshot.  Although crude, this approximation results in an error
of at most ${\approx} 15\%$ in $\Delta \eta$ which rapidly becomes negligible as
the time since the burst increases (i.e. $j-i$ increases).  Finally we note
that since we are assuming that all supernova feedback energy is produced by SN-II,
we enforce a minimum $m_{\rm low}$ value of 8${\rm M_{{\sun}}}$ when evaluating
Equations~\ref{eqn:delta_e_reheat} \& \ref{eqn:delta_m_reheat} above.

The eventual fate of the reheated material depends on both its mass, $m_{\rm
  reheat}$, and the amount of energy injected, $\Delta E_{\rm reheat}$.  If we
assume the gas to be adiabatically heated to the virial temperature of its host
halo, the associated change in thermal energy is given by:
\begin{equation}
  \Delta E_{\rm hot} = 0.5 \Delta m_{\rm reheat} V^2_{\rm vir}\ .
\end{equation}
If $\Delta E_{\rm reheat} \ge \Delta E_{\rm hot}$ then there is more energy
injected into the reheated gas than is required to raise it to the virial
temperature of the halo.  We therefore assume the gas to be added to the hot
halo of the host FoF group.  Any excess energy is assumed to then go into
ejecting some fraction of the FoF group hot reservoir from the system entirely:
\begin{equation}
  \Delta m_{\rm eject} = \frac{\Delta E_{\rm reheat} - \Delta E_{\rm
      hot}}{0.5 m_{\rm hot} V_{\rm vir}^2} m_{\rm hot}\ ,
  \label{eqn:delta_m_eject}
\end{equation}
where $V_{\rm vir}$ is the virial velocity of the FoF group.  If instead $\Delta
E_{\rm reheat} < \Delta E_{\rm hot}$, only an energetically feasible fraction
of the total reheated mass is added to the FoF group hot reservoir:
\begin{equation}
  \Delta m_{\rm hot} = \frac{\Delta E_{\rm reheat}}{0.5V^2_{\rm vir}}\ ,
  \label{eqn:delta_m_hot}
\end{equation}
with the rest raining back down on to the galaxy in a galactic fountain.

Any gas and metals which are successfully expelled from the system entirely are
placed into a separate `ejected' reservoir.  Here they are assumed to play no
further role in the evolution of the galaxies in the host FoF group until the
group falls into a more massive system.  At this point, the ejected material is
assumed to be re-accreted into the new group and is added to its hot halo
component.

\subsubsection{Delayed versus contemporaneous feedback}
\label{ssub:delayed_vs_contemporaneous_feedback}

In practice, we apply our supernova scheme in two phases.  First, the amount of 
mass reheated and ejected due to past star formation episodes is calculated as 
described above.  After the masses of the various baryonic reserves have been 
appropriately updated, the amount of new star formation in the current snapshot 
is then determined (cf. Section~\ref{sub:star_formation}).  In this way, ongoing 
energy injection from past star formation episodes is able to prevent new stars 
from forming in the current time step altogether.

After calculating the mass of stars formed, the corresponding reheated and
ejected masses due to any stars with short enough lifetimes to go nova in the
current time step are also calculated.  If the total amount of cold gas removed
from the galaxy due to both star formation and the corresponding contemporaneous
supernova feedback exceeds that which is available, the mass of stars formed in
the current time step is reduced until consistency is achieved.

\subsection{Metal enrichment}
\label{sub:metal_enrichment}

As is common in semi-analytic models
\citep[e.g.][]{De-Lucia2004,Somerville2008,Guo2011}, we implement a simple
metal enrichment scheme whereby a fixed yield, $Y$, of metals is released into
the ISM per unit mass of stars formed.  Again, we assume that these metals are
released predominantly by massive stars which end their lives as SN-II and we
gradually release them over time as these supernov\ae\ occur.  However, since a
more massive star will generally release more metals during its lifetime than a
less massive counterpart, we release these metals in proportion to the {\it
mass} fraction of SN-II (as opposed to the number fraction as was used above).
In other words
\begin{equation}
  \Delta m_{Z, j} = \sum_{i=j-N_{\rm SFH}}^{i=j} \frac{\Delta m_{{\rm
SN};i,j}}{m_{\rm SNII}} Y\Delta m_*\ ,
  \label{eqn:new_metals}
\end{equation}
where $m_{Z, j}$ is the mass of metals released during snapshot $j$, $m_{\rm
SNII}$ is the total fraction of stars with initial masses greater than 8${\rm
M_{{\sun}}}$, and $\Delta m_{{\rm SN};i,j}$ is the fraction of stars formed during
snapshot $i$ that go nova during snapshot $j$.

Analogous to Equations~\ref{eqn:total_SNII} and \ref{eqn:delta_eta} above,
$m_{\rm SN}$ and $\Delta m_{\rm SN}$ are given by
\begin{eqnarray}
  m_{\rm SNII} &=& \int_{8 {\rm M_{{\sun}}}}^{120 {\rm M_{{\sun}}}}m\phi(m)\,{\rm
  d}m =
0.144\ ,\\
\Delta m_{\rm SN} &=& \int_{m_{\rm low}}^{m_{\rm high}} m\phi(m)\,{\rm d}m\
\label{eqn:delta_m_sn},
\end{eqnarray}
where $m_{\rm high}$ and $m_{\rm low}$ are as defined in 
Section~\ref{sub:supernova_feedback}.

The metals released into the ISM in this manner are assumed to be uniformly
mixed with the cold gas of the galaxy.  From here they can be further
distributed to the hot halo or ejected from the system entirely via supernova
feedback.  Metals which do end up in the hot gas reservoir can then enhance
the cooling rate of gas on to the galaxy through metal line emission (see
Equation~\ref{eqn:tcool}).

\subsection{Stellar mass recycling}
\label{sub:stellar_mass_recycling}

A common assumption of many semi-analytic models is the so-called instantaneous
recycling approximation (IRA), in which some fixed fraction of the stellar
mass formed during each time step is instantaneously recycled back into the
ISM.  The precise value of this fraction varies from model to model and is often
left as a free parameter; however, most works employ a value of approximately 
30--40\% \citep[e.g.][]{Cole2000,Croton2006,Somerville2008,Henriques2015}.

Given our choice of IMF (see Section~\ref{sub:supernova_feedback}), a recycle
fraction of 40\% corresponds to all stars more massive than approximately
$1\,{\rm M_{\sun}}$ instantaneously going supernova.  However, the lifetime
of a $1\,{\rm M_{\sun}}$ star is close to the current age of the Universe
\citep[e.g.][]{Portinari1998} and hence the IRA can only be considered valid for
galaxies around $z{=}0$ (i.e. well past the peak of the Universal star formation
rate density).  At $z{\ga}2$, where the majority of galaxies have stellar
populations dominated by recent star formation, this approximation becomes
invalid and we are forced to consider a more realistic alternative.

Our stellar mass recycling prescription is divided into two parts:
\begin{enumerate}
    \item When implementing our delayed supernova feedback prescription, we
      assume that the initial stellar mass of all supernov\ae\ is
      returned to the cold gas reservoir of the galaxy (thus ignoring any mass
      that may be locked up in long lived remnants such as neutron stars and
      black holes).
    \item As discussed in Section~\ref{sub:supernova_feedback},
      we explicitly track the star formation history of each galaxy for the last
      $N_{\rm SFH}$ snapshots.  In order to calculate the recycled mass from
      older stars we approximate these as having formed in a single coeval burst
      which occurred at a time defined by their mass-weighted age.  Equation
      \ref{eqn:delta_m_sn} then allows us to calculate the relevant mass of stars
      which would have gone nova and this is again assumed to be returned
      the cold ISM in its entirety.  Although crude, the approximation of a
      single coeval burst for all older stars provides the correct stellar mass
      loss to within less than 5\% error at all times\footnote{This has been
      confirmed by tests we have performed using a number of idealized star
      formation histories (exponentially increasing, exponentially decreasing,
      constant, multiple burst, and random).} given the snapshot cadence of
      \Tiamat\ and our fiducial value of $N_{\rm SFH}{=}4$.
\end{enumerate}

\subsection{Halo infall and gas stripping}
\label{sub:halo_infall_and_gas_stripping}

As haloes inspiral towards more massive systems, tidal forces experienced
during repeated pericentric passages can lead to the stripping of loosely
bound material from the outer regions.  In \Meraxes, if the mass of an FoF
group drops, then a pro rata fraction of the ejected and/or hot baryonic
content of the halo is also removed.  In practice, the amount of mass which
must be removed is given by the value of $m_{\rm infall}$ (as defined in
Equation~\ref{eqn:fmod_defn}) which will be negative in such systems.  This
material is taken first from the ejected reservoir, with further mass being
removed from the hot halo component if required.  No baryons are ever taken from
the cold gas or stellar mass reservoirs as these are assumed to be protected
from such tidal losses by their position deep in the central potential well of
their haloes.

Further to these long-range tidal forces, galaxies infalling into groups or
clusters are observed to be subjected to a number of dynamical processes which
remove gas from the outskirts of the system, including ram-pressure stripping,
strangulation, and harassment \citep[e.g.][]{van-den-Bosch2008, Peng2015}.  We
model the combined effects of these processes by assuming that all FoF groups
are instantly stripped of their entire hot and ejected gas reservoirs upon
infall into a more massive structure, with their combined mass and metals being
added to the hot component of the new parent.  Although such a rapid stripping
represents the most extreme scenario possible, we note that this approximation
has been made in a number of previous semi-analytic models and defer an improved
and more realistic treatment to future works.

\subsection{Mergers}
\label{sub:mergers}

Mergers play an important role in the build up of galaxy stellar mass, both
through hierarchical mass assembly and induced star formation.  This is
particularly so at high-$z$ where their prevalence is enhanced (cf. Paper I).
In \Meraxes\ (as in almost all semi-analytic models) these galaxy merger
events are triggered by the merging of the corresponding host dark matter
haloes.  Following \citet{Croton2006}, when a dark matter halo is marked as
having merged, we utilize dynamical friction arguments to approximate the time
taken for the orbit of the incoming galaxy to decay and the corresponding
galaxy--galaxy merger to occur \citep{Binney2008}:
\begin{equation}
  t_{\rm merge} = \alpha_{\rm merge} \frac{V_{\rm vir}r_{\rm gal}^2}{G
    m_{\rm gal} \ln (1+M_{\rm vir}/m_{\rm gal})}\ ,
  \label{eqn:t_merge}
\end{equation}
where it is standard to take $\alpha_{\rm merge} {=} 1.17$, $r_{\rm gal}$ is
the distance between the most-bound particle of the parent and the infalling
halo, $m_{\rm gal}$ is the total mass of the infalling galaxy, and $M_{\rm
vir}$ and $V_{\rm vir}$ are the virial properties of the parent.  In \Meraxes,
all of these quantities are evaluated at the last time the infalling halo was
successfully identified in the trees.

The value of $\alpha_{\rm merge}{=}1.17$ is based on the assumption that $t_{\rm
merge}$ is calculated at the moment the infalling halo crosses the virial
radius of the parent.  However, we instead calculate $t_{\rm merge}$ at the
time at which the infalling halo can no longer be identified in the \Tiamat\
simulation and is thus marked as having merged in our input merger trees.
This results in an overestimate of the merger time-scale which worsens the
longer the infalling halo remains identified after crossing the virial radius
of the parent \citep[e.g.][]{Hopkins2010}.  Even in dense environments, the
accurate merger trees produced from \Tiamat\ results in haloes being identified
for extended periods before the merger event occurs.  Furthermore, previous
authors have found that changes to the value of $\alpha_{\rm merge}$ have been
necessary in order to match observational constraints on the luminous end of
the galaxy luminosity function \citep{De-Lucia2007} and idealized \textit{N}-body halo
merger simulations \citep[e.g.][]{Boylan-Kolchin2008}.  On average, we find
that infalling haloes are successfully tracked in our input merger trees until
$r_{\rm gal} {\approx} 0.7R_{\rm vir}$, with a weak trend to be identified to
smaller fractional radii with increasing redshift.  Noting that $t_{\rm merge}
{\propto} r_{\rm gal}^2$, we therefore choose to fix $\alpha_{\rm merge} {=}
0.5$.  We also note that if, after starting the merger clock, the parent galaxy
itself experiences a merger, then we assume that all of its infalling galaxies
also undergo a merger with the same target.

Galaxy mergers can drive strong shocks and turbulence in any participating
cold gas, driving this material towards the inner regions of the parent galaxy
and resulting in an efficient burst of star formation.  We model the fraction
of cold gas consumed by such a burst, $e_{\rm burst}$, using the prescription
introduced by \citet{Somerville2001}:
\begin{equation}
  e_{\rm burst} = \alpha_{\rm burst} (m_{\rm gal}/m_{\rm
    parent})^{\gamma_{\rm burst}}\ ,
  \label{eqn:e_burst}
\end{equation}
where $m_{\rm gal}$ and $m_{\rm parent}$ are the corresponding baryonic masses
(i.e. cold gas + stellar mass), and we follow \citet{Croton2006} by setting
the parameters $\alpha_{\rm burst}{=}0.56$ and $\gamma_{\rm burst}{=}0.7$.
This relation agrees well with the results of numerical simulations of mergers
with baryonic mass ratios in the range 0.1--1.0 \citep{Cox2004}.  For merger
events where the mass ratio is less than 0.1, we suppress any merger-driven star
formation.

For simplicity, we assume that all of the stars formed in a merger-driven burst
do so within a single snapshot.  At $z{\sim}8$, the median dynamical time of a
galaxy disc in \Meraxes\ is ${\sim}$60\% of the time between two consecutive
snapshots of \Tiamat\ (${\sim}$11.2\,Myr).  Hence, this approximation is roughly
equivalent to the assumption that the merger-driven burst occurs on a time-scale
approximately less than one disc dynamical time for the majority of galaxies.
Although the disc dynamical time does increase with decreasing redshift, by
$z{\sim}5$ the median is still only equal to one snapshot and hence this
approximation remains valid for the majority of galaxies.

\subsection{Ghost galaxy evolution}
\label{sub:ghost_galaxy_evolution}

A ghost dark matter halo is one which is temporarily unresolved in our
input merger trees.  This can be due to a number of reasons, but is most
commonly a result of a smaller halo passing through or nearby a much more
massive structure.  The \Tiamat\ merger trees used in this work are carefully
constructed to identify these artefacts, resulting in the skipping of a
potentially large number of snapshots between haloes and their descendants.  In
many semi-analytic models, the galaxies hosted by such haloes are simply ignored
until their halo is later re-identified.  In some cases, re-identification fails
or is not even attempted, resulting in spurious galaxy merger and creation
events.  However, we must ensure that we correctly include these objects at all
snapshots in order to account for their ionizing photon contribution.

Due to the lack of knowledge of the properties of a ghost's host dark matter
halo, we are unable to implement many of the physics prescriptions outlined
above.  We therefore simply allow these galaxies to passively evolve during
the time over which they are identified as ghosts, forming no new stars but
experiencing the delayed supernova feedback from previously formed generations.
When the host halo is eventually re-identified, we assume that any associated
star formation occurred in a single coeval burst at $0.5\Delta t$, where $\Delta
t$ is the time since the halo was last identified.  We then back-fill the
stellar mass history appropriately for use in our delayed supernova feedback
scheme.

\subsection{Reionization}
\label{sub:reionization}

A key goal of \DRAGONS\ is to connect the evolution of the $21\,{\rm cm}$
reionization structure to the formation of the source galaxy population.  As
such, \Meraxes\ has been developed to be the first semi-analytic galaxy
formation model to fully and self-consistently couple the process of
reionization (in particular, the presence of a photo dissociating UVB) to the
evolution of galaxies both temporally and spatially.  To achieve this we have
embedded a specially modified version of the semi-numerical reionization code,
\tocf\ \citep{Mesinger2011}, that includes the calculation of the local
ionizing UVB described by \citet{Sobacchi2013b} and, importantly,
makes full use of the realistic galaxy properties provided by \Meraxes.

\subsubsection{Self-consistent reionization with \tocf}
\label{ssub:self_consistent_reionization_with_21cmfast}

The basic methodology of \tocf\ is to use an excursion set formalism in order
to identify ionized bubbles where the integrated number of ionizing photons is
greater than the number of absorbing atoms and associated recombinations:
\begin{equation}
  N_{\rm b*}(r) N_\gamma f_{\rm esc} \ge (1+\bar N_{\rm rec}) N_{\rm atom}(r)\ .
  \label{eqn:simple_ion_criteria}
\end{equation}
Here $N_{\rm atom}(r)$ is the integrated number of atoms being ionized within
a sphere of radius $r$, $N_{\rm b*}(r)$ is the number of stellar baryons in
the same volume, $N_\gamma$ is the mean number of ionizing photons produced
per stellar baryon, $f_{\rm esc}$ is the escape fraction of these photons, and
$\bar N_{\rm rec}$ is the mean number of recombinations per baryon.  If we
assume that helium is singly ionized at the same rate as cosmic hydrogen, then
expanding Equation~\ref{eqn:simple_ion_criteria} in terms of the integrated
stellar ($m_*(r)$) and total ($M_{\rm tot}(r)$) masses within $r$ gives
\begin{equation}
  {m_*(r) \over m_{\rm p}} N_\gamma f_{\rm esc} \ge 
  (1+\bar N_{\rm rec}){f_{\rm b}(1{-}{3\over 4}Y_{\rm He})M_{\rm tot}(r) \over m_{\rm p}}\
  ,  \label{eqn:ionization_criteria}
\end{equation}
where $Y_{\rm He}$ is the helium mass fraction, $m_{\rm p}$ is the proton mass,
and the term $(1{-}{3 \over 4}Y_{\rm He})$ corresponds to the combined number of
hydrogen and helium atoms per baryon.

It is common for Equation~\ref{eqn:ionization_criteria} to be re-written in
terms of an H\,\textsc{ii} ionizing efficiency, $\xi$:
\begin{equation}
  \xi {m_*(r) \over M_{\rm tot}(r)} \ge 1\ , 
  \label{eqn:xi_criteria}
\end{equation}
where
\begin{equation}
  \xi = 6214 \left( 0.157 \over f_{\rm b}\right) \left( N_\gamma \over
    4000\right) \left( f_{\rm esc} \over 0.2\right) \left( 0.82 \over
    1{-}{3 \over 4}Y_{\rm He}\right)\ ,
  \label{eqn:ionizing_eff}
\end{equation}
and we have excluded the $1{+}\bar N_{\rm rec}$ term based on studies
of the high-redshift Lyman-$\alpha$ forest which suggest $\bar N_{\rm
rec}{\sim}0$ in the diffuse IGM \citep[e.g.][]{Bolton2007,McQuinn2011}.
Despite this simplifying assumption of $\bar N_{\rm rec}{=}0$, we note
that we implicitly include a mean-free path of ionizing photons through
the IGM in our calculation by starting our excursion set calculation at an
appropriate scale \citep{Sobacchi2013b}.  The right-hand side of this equation
includes our fiducial values for each of the physical variables. The number
of ionizing photons per stellar baryon, $N_\gamma$, is set by the assumed
stellar IMF whilst both $f_{\rm b}$ and $Y_{\rm He}$ are well constrained by
cosmology.  Of all of these terms, only $f_{\rm esc}$ is poorly known for
high-redshift galaxies \citep[e.g.]{Wise2009,Raicevic2011,Kuhlen2012}.  Our
fiducial value of 0.2 is primarily chosen to provide a reionization history
which is consistent with the latest {\em Planck} 2015 electron scattering
optical depth measurements \citep[][see Section~\ref{sec:model_calibration}
below]{Planck-Collaboration2015}.

By applying our integrated \tocf\ algorithm to grids of stellar and total
mass within \Meraxes\ we can use Equation~\ref{eqn:xi_criteria} to produce
a neutral hydrogen fraction ($x_{\rm HI}$) grid for the entire simulation
volume.  In order to then determine how this spatially and temporally evolving
ionization structure affects the baryon fraction modifier, $f_{\rm mod}$, of
individual FoF groups we utilize the UVB feedback model of \citet{Sobacchi2013}.
Using idealized 1D hydrodynamical simulations of a static, uniform ionizing UVB
impinging on collapsing dark matter haloes, \citet{Sobacchi2013} found that
$f_{\rm mod}$ was well described by
\begin{equation}
f_{\rm mod} = 2^{-M_{\rm filt}/M_{\rm vir}}\ ,
\label{eqn:fmod}
\end{equation}
where $M_{\rm vir}$ is the mass of the halo and $M_{\rm filt}$ is the `filtering 
mass' representing the mass at which $f_{\rm mod}{=}0.5$:
\begin{equation}
M_{\rm filt} = M_0 J_{21}^a \left(\frac{1{+}z}{10}\right)^b \left[1 -
\left(\frac{1{+}z}{1{+}z_{\rm ion}} \right)^c \right]^d\ .
\label{eqn:Mfilt}
\end{equation}
Here $z_{\rm ion}$ is the redshift at which the collapsing halo was first 
exposed to the UVB and the parameters $(M_0,a,b,c,d) = (2.8{\times} 10^9\, {\rm 
  M_{{\sun}}}, 0.17, -2.1, 2.0, 2.5)$ were found by the authors to provide the
best fit to their results. The $J_{21}$ term represents the local UVB intensity: 
\begin{eqnarray}
J(\nu) &=& J_{21} \left(\frac{\nu}{3.2872{\times} 10^{15}\, {\rm Hz}}\right)^{-\alpha}
\nonumber \\
&&\quad\times 10^{-21} {\rm erg\,s^{-1}\,Hz^{-1}\,(proper\;cm)^{-2}\,sr^{-1}}\ ,
\end{eqnarray}
where $\alpha = 5.0$ for a stellar-driven UV spectrum \citep{Thoul1996}.

In order to calculate $f_{\rm mod}$ using this formalism, we need to know both
the redshift at which the IGM surrounding each halo was first ionized, $z_{\rm
ion}$, and the local ionizing background intensity at this time, $J_{21}$.  The
average UVB intensity which a galaxy is exposed to within an ionized region,
$\bar J_{21}$, is given by
\begin{equation}
  \bar J_{21} = \frac{(1+z)^2}{4 {\rm \pi}}\lambda_{\rm mfp}h \alpha f_{\rm bias} 
\bar\epsilon\ ,
\label{eqn:J21}
\end{equation}
where $\lambda_{\rm mfp}$ is the comoving mean-free path of ionizing photons
(which is assumed here to be equal to the radius of the ionized bubble, $r$)
and $h$ is the Planck constant.  The term $f_{\rm bias}{=}2$ is introduced to
account for the effect of galaxy clustering on boosting the ionizing emissivity
at halo locations relative to the spatial average \citep[][]{Mesinger2008}.
The term $\bar \epsilon$ represents the ionizing emissivity, the time-averaged
number of ionizing photons emitted into the IGM per unit time, per unit comoving
volume.  Approximating the average rate of stellar mass growth of all galaxies
within this region as $m_{*,\rm gross}(r) {/} t_{\rm H}$, where $m_{*,\rm
gross}(r)$ is the gross stellar mass formed within $r$ (i.e. without any
decrement due to stellar evolution) and $t_{\rm H}$ is the Hubble time, $\bar
\epsilon$ can be expressed as
\begin{equation}
  \bar \epsilon = \frac{f_{\rm esc}N_\gamma}{{4 \over 3}{\rm \pi} r^3 m_{\rm 
      p}}{m_{*,\rm gross}(r) \over t_{\rm H}}\ . \label{eqn:ionizing_emissivity}
\end{equation}
Our utilization of $m_{*,\rm gross}(r) {/} t_{\rm H}$, instead of the true
instantaneous star formation rate predicted by \Meraxes, is motivated by the need
to smooth out the often bursty star formation histories of our galaxies.  The
filtering mass formula of Equation~\ref{eqn:Mfilt} assumes that the ionizing
background intensity within a cell remains constant with time.  This is a
reasonable approximation due to the weak sensitivity of $M_{\rm filt}$ on the
UVB intensity ($M_{\rm filt}{\propto}J_{21}^{0.17}$). \citet{Sobacchi2013b} also
find that the $\bar J_{21}$ remains approximately constant within H\,\textsc{ii} regions,
further validating this approximation.  However, fixing the UVB intensity at
the snapshot of ionization results in smaller bubbles having artificially high
or low $\bar J_{21}$ values depending on the current star forming state of the
source galaxies.  This introduces artificial scatter into the calculation which
can become important when comparing the effect of reionization on the evolution
of individual galaxies (see Section~\ref{sub:results-environ}).

In practice, the coupling between galaxy evolution and reionization within 
\Meraxes\ is implemented as follows.
\begin{enumerate}
  \item For a single time step in the simulation, \Meraxes\ evolves all of the 
    galaxies in the entire \Tiamat\ volume.
  \item Our \tocf\ algorithm constructs and processes halo mass, stellar mass,
    and averaged star formation rate grids, along with pre-computed total matter
    density grids.  For this work, and all of the results herein, we use a grid
    resolution of $512^3$. After applying the excursion set formalism and
    equations outlined above, a grid of $\bar J_{\rm 21}$ and $x_{\rm HI}$
    values is generated.
  \item Using the $x_{\rm HI}$ grid, \Meraxes\ keeps track of the redshift at 
    which each cell first became ionized ($z_{\rm ion}$) and the corresponding 
    $\bar J_{\rm 21}$.  It then calculates the baryon fraction modifier of each 
    grid cell following Equation~\ref{eqn:fmod}.
  \item In order to calculate the amount of freshly infalling baryonic material
    it should accrete, each FoF group uses the baryon fraction modifier of the
    grid cell in which it is located in the following simulation time step (see
    Section~\ref{sub:baryonic_infall} above).
  \item This process is then repeated for each of the time steps (of which there 
    are 100 for our input simulation, \Tiamat, between $5{<}z{<}35$).
\end{enumerate}
Through this procedure, the evolution of galaxies and reionization in the simulation
are self-consistently coupled both temporally and spatially.

\subsubsection{Homogeneous model}
\label{ssub:homogeneous_model}

For comparison, we also employ a `homogeneous' reionization prescription using 
$M_{\rm filt}$ values that depend on redshift alone (i.e. no
information about the spatial distribution of the IGM ionization state is
required).  Prescriptions such as this \citep[e.g.][]{Gnedin2000,Kravtsov2004}
have been commonly employed by semi-analytic models for many years 
\citep[e.g.][]{Benson2002,De-Lucia2007,Somerville2008}.

\begin{figure}
  \includegraphics[width=\linewidth]{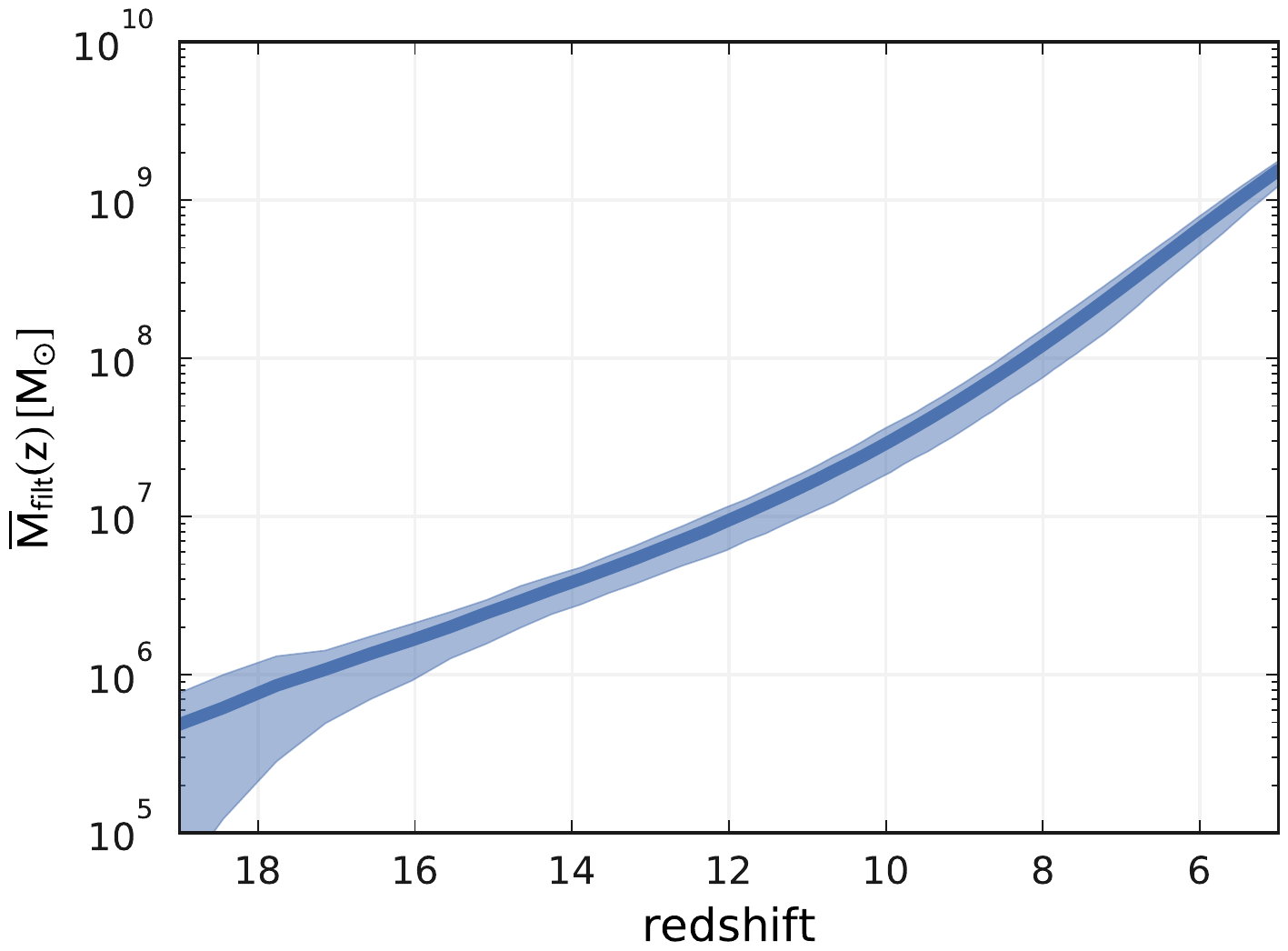}
  \caption{\label{fig:homogeneous_mfilt}The mean filtering mass, $\bar{M}_{\rm
  filt}$, of FoF groups in the fiducial patchy reionization model as a function
  of redshift.  For our homogeneous comparison model, these filtering mass
  values are applied to every FoF group in the simulation, regardless of their
  environment or local reionization history, thus allowing us to directly assess
  the effect of ignoring this information as is commonly done in traditional
  semi-analytic galaxy formation models.  The blue shaded region indicates the
  68\% confidence intervals calculated using the spatial variation in the mean
  from 125 non-overlapping subvolumes comprising the full simulation.}
\end{figure}

At each time step in the simulation, we use our fiducial patchy reionization
prescription described above to calculate the number-weighted mean filtering
mass of all FoF groups in the volume.  We then re-run \Meraxes, applying this
redshift dependent $\bar M_{\rm filt}$ value to all FoF groups when calculating
their baryon fraction modifier as per Equation~\ref{eqn:fmod}.  By generating
$\bar M_{\rm filt}(z)$ from our fiducial model in this way, we are able to use
the homogeneous model result as a baseline with which to assess the detailed
effects of a self-consistent, spatially dependent reionization prescription on
the evolution of the source galaxy population, whilst simultaneously providing
a useful filtering mass relation for $z{\le}5$, that can be applied quickly and
easily.  In Fig.~\ref{fig:homogeneous_mfilt}, we show the evolution of $\bar
M_{\rm filt}$.  The blue shaded region indicates the 68\% confidence intervals
calculated using the spatial variation of the mean in 125 subvolumes.  The
evolution in the filtering mass is well approximated by the following
functional form:
\begin{equation}
  \log_{10}(M_{\rm filt}(z)/{\rm M_{{\sun}}}) = \theta_1 \exp(\theta_2 z) +
  \theta_3\ ,
\end{equation}
where $\theta_1 {=} 7.51^{+0.26}_{-0.18}$, $\theta_2 {=}
-0.090^{+0.01}_{-0.01}$ and $\theta_3 {=} 5.59^{+0.35}_{-0.43}$.  The fitted
parameter values were calculated using Markov chain Monte Carlo (MCMC) methods
with a standard chi-squared likelihood, and provide a fit which is accurate to
within ${\sim}2\%$ of the mean model result across all redshifts.  As we will
demonstrate in Section~\ref{sec:results}, our homogeneous prescription does a
reasonable job of reproducing the mean evolution of the stellar mass functions
and global neutral fractions predicted by the full, patchy reionization
implementation.

\section{Model calibration}
\label{sec:model_calibration}

\begin{table*}
  \begin{minipage}{\textwidth}
  \centering
  \caption{
  \label{tab:fiduical_params}
    The fiducial parameter values used throughout this work.  Values
    were constrained to visually reproduce the observed evolution
    in the galaxy stellar mass function between $z{=}5$ and 7 (see
    Fig.~\ref{fig:smf_calibration}).  The quoted {\em Munich} model values
    represent the range of fiducial values utilized in the following works
    (where appropriate): \citet{Croton2006}, \citet{Guo2011, Guo2013},
    \citet{Mutch2013}, and \citet{Henriques2013, Henriques2015}.  All of
    the parameters in these works were calibrated against $z{<}3$ observations.
    However, they are presented here as a rough guide to the range of plausible
    values.
  }
  \begin{tabular}{c|l|c|l|r@{$\,-\,$}l|r}
  \hline\hline
  {\bf Parameter} & {\bf Prescription} & {\bf Equation} & {\bf Description} & 
  \multicolumn{2}{c}{\bf {\em Munich} model\newline} & {\bf Fiducial} \\
  & & & & \multicolumn{2}{c}{\bf values} & {\bf value}\\
  \hline
  $\Sigma_{\rm norm}$ & Star formation (\S\ref{sub:star_formation}) & 
  \ref{eqn:sigma_crit} & Critical cold gas surface density normalization & 
  \hspace{0.245cm} 0.26 & 0.38 & 0.2 \\
  $\alpha_{\rm SF}$ & -- & \ref{eqn:new_stars} & Star formation efficiency & 
  0.01 & 0.07 & 0.03 \\
  $\alpha_{\rm energy}$ & Supernova feedback 
  (\S\ref{sub:supernova_feedback}) & \ref{eqn:epsilon_energy} & Energy 
  coupling efficiency normalization & 0.18 & 0.7 & 0.5 \\
  $\beta_{\rm energy}$ & -- & \ref{eqn:epsilon_energy} & Coupling efficiency 
  $V_{\rm max}$ scaling & 0 & 3.5 & 2.0 \\
  $V_{\rm energy}$ & -- & \ref{eqn:epsilon_energy} & Coupling efficiency $V_{\rm 
    max}$ normalization & 70 & 336 & 70.0  \\
  $\alpha_{\rm mass}$ & -- & \ref{eqn:epsilon_mass} & Mass loading normalization 
  & 2.1 & 10.3 & 6.0  \\
  $\beta_{\rm mass}$ & -- & \ref{eqn:epsilon_mass} & Mass loading $V_{\rm max}$ 
  scaling & 0 & 3.5 & 0.0 \\
  $V_{\rm mass}$ & -- & \ref{eqn:epsilon_mass} & Mass loading $V_{\rm max}$ 
  normalization & 70 & 430 & 70.0 \\
  $\epsilon_{\rm mass}^{\rm max}$ & -- & \ref{eqn:epsilon_mass} & Maximum mass 
  loading value & \multicolumn{2}{c}{--} & 10.0 \\ 
  $Y$ & Metal enrichment (\S\ref{sub:metal_enrichment}) & 
  \ref{eqn:new_metals} & Mass of metals per unit mass of SN & 0.03 & 0.047 & 0.03 \\
  $f_{\rm esc}$ & Reionization (\S\ref{sub:reionization}) & 
  \ref{eqn:ionizing_eff} & Ionizing photon escape fraction & 
  \multicolumn{2}{c}{--} & 0.2 \\
  \hline\hline
  \end{tabular}
  \end{minipage}
\end{table*}

The free parameters of the model were manually calibrated (by hand) to
replicate the observed evolution of the galaxy stellar mass function between
redshifts 5 and 7, as well as the integrated free electron Thomson scattering
optical depth measurements.  The evolution of the stellar mass function has
been shown by previous statistical investigations of semi-analytic models to
provide a tight constraint on both the star formation efficiency and supernova
feedback parameters \citep[e.g.][]{Henriques2013, Mutch2013}.  By combining this
with the Thomson scattering optical observations, we can additionally put
constraints on the escape fraction of ionizing photons ($f_{\rm esc}$), and
thus all of the free parameters of our model, as listed in
Table~\ref{tab:fiduical_params}.

It could be argued that the luminosity function would provide a more
fundamental constraint on the model, rather the stellar mass function.
However, whilst it is true that converting observed galaxy luminosities to
stellar masses involves a number of assumptions and potentially unreliable
conversions, the same is also true for the inverse procedure of converting
model stellar masses to luminosities.  Stellar masses are intrinsic predictions
of semi-analytic models, whilst luminosities require an extra layer of
modelling.  For example, changing the IMF has a relatively small impact on the
stellar masses predicted by the model, but can have a significant impact on the
resulting luminosity function.  In order to accurately model luminosities in
various bands, one must typically calculate a full spectral energy distribution
(SED) for every object, applying model-dependent Lyman $\alpha$ absorption, sample
selection (e.g. in colour--colour space), and poorly understood dust
corrections.  Furthermore, doing this for all galaxies at multiple redshifts
can take a significant amount of time and memory which prohibits its usefulness
when running a model many times for calibration purposes.  

The redshift range $5{\le}z{\le}7$ corresponds to the highest redshifts for
which reliable observed stellar mass functions are available.  In particular,
we make use of the mass functions estimated by \citet{Gonzalez2011},
\citet{Duncan2014}, \citet{Grazian2014}, and \citet{Song2015}.  Both
\citet{Duncan2014} and \citet{Grazian2014} utilize data collected from the
Cosmic Assembly Near-infrared Deep ExtragaLactic Survey
\citep[CANDELS;][]{Grogin2011,Koekemoer2011} GOODS South field with stellar
masses directly obtained from SED fitting of combined optical and near-infrared
space-based observations, and include the effects of both nebular line and
continuum emission.  In addition, \citet{Grazian2014} include a detailed
treatment of the effects of Eddington bias \citep{Eddington1913} on the
normalization and slope of their derived mass functions. Whilst
\citet{Song2015} also utilize CANDELS infrared data, they instead carry out a
hybrid SED stacking technique to derive a redshift dependent stellar mass--UV
luminosity relation which is then combined with measured UV luminosity
functions to estimate the galaxy stellar mass function. Similarly,
\citet{Gonzalez2011} utilized data combined from \textit{Hubble Space
Telescope} and \textit{Spitzer} observations, but with stellar masses obtained via
mass--UV luminosity relations calibrated at $z{=}4$ alone.

In addition to the stellar mass function which constrains the integrated amount
of star formation driving reionization, the corresponding ionizing photon
budget is also set by the escape fraction.  One of the primary observational
constraints on the timing and duration of reionization comes from the measured
integrated optical depth to Thomson scattering of cosmic microwave background
photons by free electrons, $\tau_e$:
\[
  \tau_e = \int_{z=0}^{\infty}\frac{c\,{\rm d}t}{{\rm d}z}(1+z)^3\sigma_{\rm T}
\]
\begin{equation}
  \hspace{5em}\times\left[Q^{\rm m}_{\rm HII}\left<n_{\rm H}\right> + (Q^{\rm m}_{\rm 
  HeII} + 2Q^{\rm m}_{\rm HeIII})\left<n_{\rm He}\right>\right]\,{\rm d}z\ ,
\end{equation}
where $\sigma_{\rm T}{=}6.652\times 10^{-25}\,{\rm cm^2}$ is the Thomson 
scattering cross-section, $Q^{\rm m}_X$ is the mass-weighted global 
ionized fraction of species $X$, and $\left<n_{\rm H}\right> {=} 1.88\times 
10^{-7} (\Omega_{\rm b}h^2/0.022)\, {\rm cm^{-3}}$ and $\left<n_{\rm He}\right> 
{=} 0.148\times 10^{-7} (\Omega_{\rm b}h^2/0.022)\, {\rm cm^{-3}}$ are the 
average comoving density of hydrogen and helium, respectively \citep{Wyithe2003}.  
For this work, we have assumed that helium is singly ionized at the same rate as 
hydrogen (i.e.  $Q^{\rm m}_{\rm HeII}(z){=}Q^{\rm m}_{\rm HII}(z)$) and only 
becomes doubly ionized at $z{=}4$ \citep[i.e. $Q^{\rm m}_{\rm HIII}{=}1.0$ or 
  0 for $z$ greater than or less than 4 respectively; e.g.][]{Kuhlen2012}.  As 
  we shall demonstrate in the following sections, $\tau_e$ primarily constrains 
  the escape fraction of ionizing photons, $f_{\rm esc}$, in the model.

\begin{figure*}
  \begin{minipage}{\textwidth}
    \begin{center}
      \includegraphics[width=0.85\columnwidth]{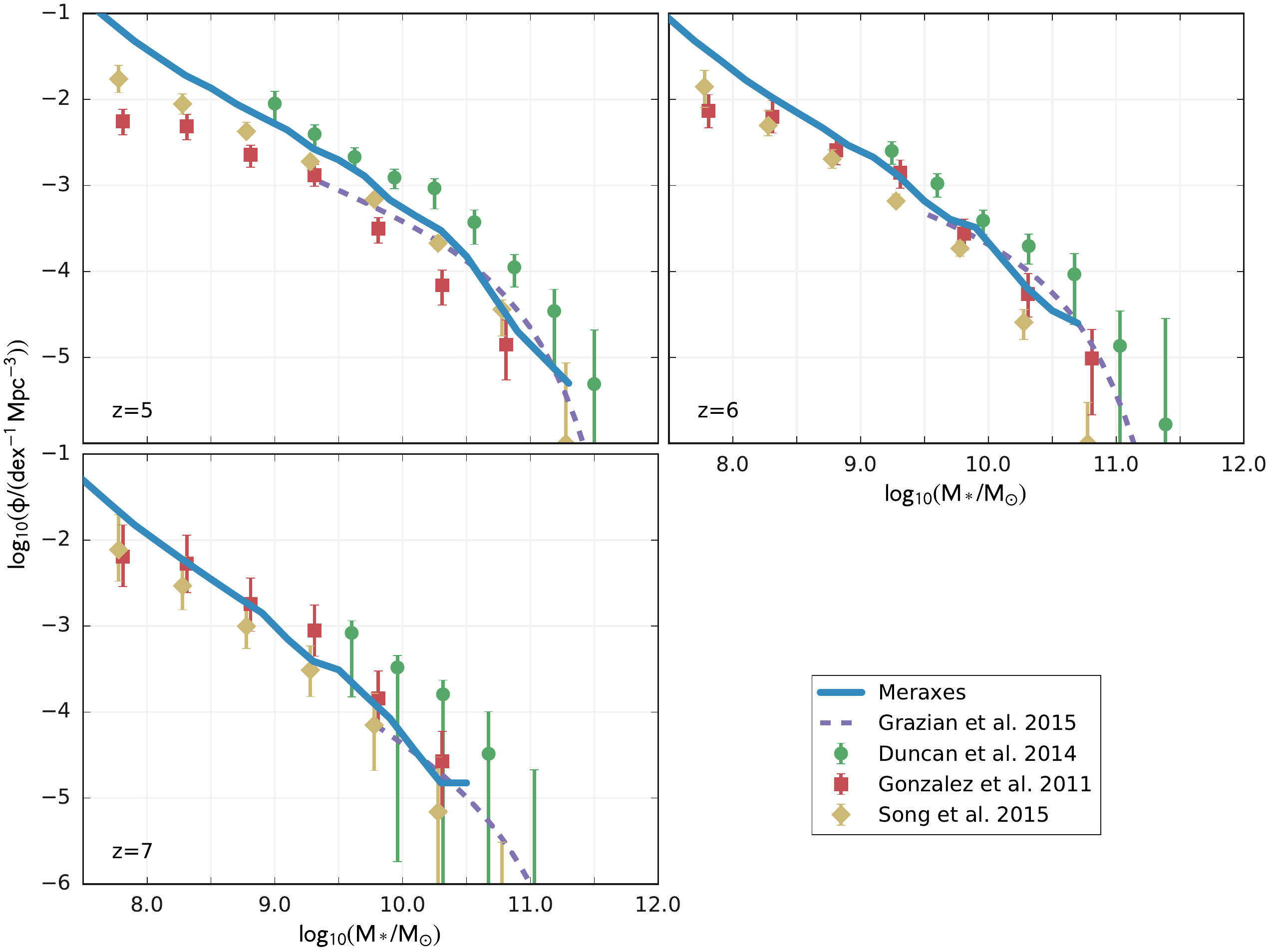}
  \caption{
    \label{fig:smf_calibration} The evolution of the galaxy stellar mass
    function from $z{=}$5--7.  Coupled with the measured Thomson scattering
    optical depth (see Fig.~\ref{fig:tau_e}), these are the only observational
    constraints applied to \Meraxes\ throughout this work unless explicitly
    stated.  Data points show the observations of \citet{Duncan2014},
    \citet{Gonzalez2011} and \citet{Song2015}. The purple dashed lines show the
    best-fitting Schechter functions of \citet{Grazian2014}.  Solid blue lines
    show the self-consistent patchy reionization model using our fiducial
    parameter values.  Best fit low-mass slopes are provided in
    Table~\ref{tab:runs}.  All observations have been corrected to a Salpeter
    IMF where necessary.}
  \end{center}
  \end{minipage}
\end{figure*}

The resulting parameter values for our fiducial model constrained against both
the high-$z$ stellar mass function evolution and integrated electron scattering
optical depth are presented in Table~\ref{tab:fiduical_params}.  It is
important to note that, although these parameter values provide a good match to
the constraining observations and are broadly consistent with comparable works
where appropriate (see second-to-last column), they may not be the only
possible solution.  However, through testing extreme (to the point of being
physically implausible) parameter combinations we are able to ascertain that
supernova feedback is the only feedback mechanism in our model capable of
producing a stellar mass function with a slope consistent with observations.
Regardless, we impress upon the reader that all of the results in this work
must be interpreted within the context of these particular chosen parameter
values alone.  In future work we will carry out a full MCMC analysis to
accurately constrain the free model parameters against a wider range of
observational quantities as well as explore any degeneracies which may exist
between them \citep[e.g.][]{Lu2010,Mutch2013}.

In Fig.~\ref{fig:smf_calibration} we present the fiducial model stellar mass 
functions (blue solid lines) along with the constraining observations.  All 
observations have been converted to a Salpeter IMF and $h{=}0.678$ where 
necessary.  The error bars on observed data points include contributions from 
Poisson noise and uncertainties in photometric redshift determinations.  
However, they neglect the systematic uncertainties associated with the 
estimation of stellar masses from photometric data (e.g. stellar population 
synthesis model variations, and photometric uncertainties).

We are able to achieve an excellent match to the normalization, shape,
and evolution of the observed mass function across all plotted redshifts.  At
$z{=}5$, where there is the largest divergence between different observational
data sets, we chose parameter values which provided a reasonable compromise
between each.  However, at the low-mass end we have chosen to follow the
observations of \citet{Duncan2014} as they use a large data set with stellar
masses obtained from SED fitting and provide actual data points rather than
a Schechter fit.  The quality of the agreement between our model and the
observational data gives us faith that our implemented physical prescriptions
are both reasonable and applicable at the high redshifts of interest in this
work.

Although typically producing fewer ionizing photons than their more massive
counterparts, low mass galaxies with $M{<}M_*$ are expected to contribute
a large fraction of the overall ionizing photon budget due to their high
number density.  For this reason, the low-mass slope of the stellar mass
function, $\alpha$, is of particular importance to reionization.  In
Table~\ref{tab:runs}, we provide the \Meraxes\ best-fit $\alpha$ parameters
at each redshift, obtained by fitting a standard Schechter function to
the model results using MCMC methods\footnote{Fits were carried out using
flat priors in log space and a standard least-squares likelihood function.
Poison uncertainties were used for the model data points.  All MCMC chains
and posterior distributions were visually inspected for convergence.} and
are in good statistical agreement with the corresponding values found by
\citet{Duncan2014} of ${-1.90^{+0.21}_{-0.16}}$, ${-1.91^{+0.91}_{-0.59}}$ and
${-2.31^{+1.31}_{-0.19}}$ for redshifts 5, 6 and 7 respectively.

\begin{figure}
\includegraphics[width=\columnwidth]{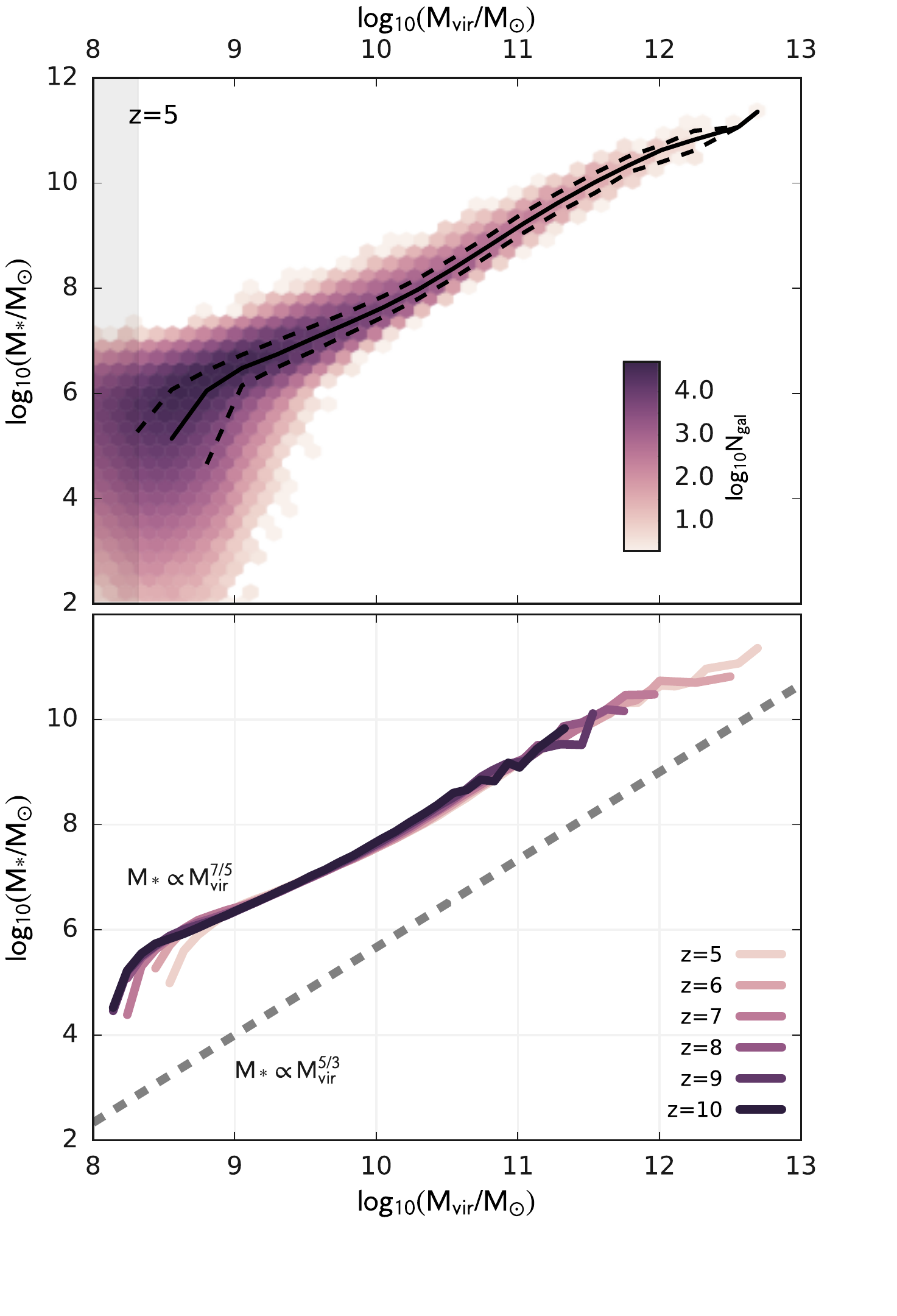}
\caption{\label{fig:stellar_vs_halo} Upper: the distribution of total
  stellar mass as a function of FoF group virial mass for the \fiducial\ model
  at $z{=}5$.  The solid and dashed black lines show the median and 68\%
  confidence intervals of the distribution.  The grey shaded region indicates
  halo masses below the atomic cooling mass threshold. Lower: the
  evolution of the \fiducial\ model median $M_*$--$M_{\rm vir}$ relation
  between redshifts 10--5.  There is no significant evolution in either
  the slope or normalization with redshift.  The grey dashed line indicates
  the theoretically motivated slope of $5/3$ (${\approx} 1.7$)
  suggested by \citet{Wyithe2013} for supernova regulated galaxy growth.  The
  normalization of this line has been arbitrarily chosen to allow a comparison
  with the slope of 1.4 predicted by \Meraxes.}
\end{figure}

In the top panel of Fig.~\ref{fig:stellar_vs_halo} we present the $z{=}5$
distribution of total stellar mass (i.e. summed over all galaxies) in each
FoF group as a function of the group virial mass in our fiducial model.  The
median relation (solid black line) is well described by a power law with a slope
of ${\sim} 1.4$.  This shows good agreement with simple energy conservation
arguments which suggest a slope of ${\sim} 1.7$ for supernova feedback-regulated
galaxy growth and a fixed cold gas mass fraction \citep{Wyithe2013}.  However,
at FoF group masses below $M_{\rm vir}{\approx}10^{9.5}\,{\rm M_{{\sun}}}$ there
is a rapid increase in the spread of stellar mass values.  This is due to
a combination of supernova and reionization feedback effects, as well as a
low star formation efficiency in these small, often diffuse haloes.  In the
lower panel of Fig.~\ref{fig:stellar_vs_halo} we show the evolution of the
median FoF group $M_*$--$M_{\rm vir}$ relation as a function of redshift.
Interestingly, there is no statistically significant evolution in either
the slope or normalization of the relation with redshift.  However, the same
simple energy conservation arguments which provided a good agreement for
the slope of the relation suggest that the normalization should evolve with
redshift \citep{Wyithe2003}.  Despite this, our model indicates that in order to
reproduce the observed evolution of the galaxy stellar mass function over the
redshifts considered in this work, the efficiency of galaxy formation and the
associated feedback processes must conspire to provide a constant star formation
efficieny in haloes of a fixed mass.  This agrees with similar findings from 
subhalo abundance-matching (SHAM) studies at lower redshifts 
\citep[e.g.][]{Behroozi2013b}. 

\begin{figure}
\includegraphics[width=\columnwidth]{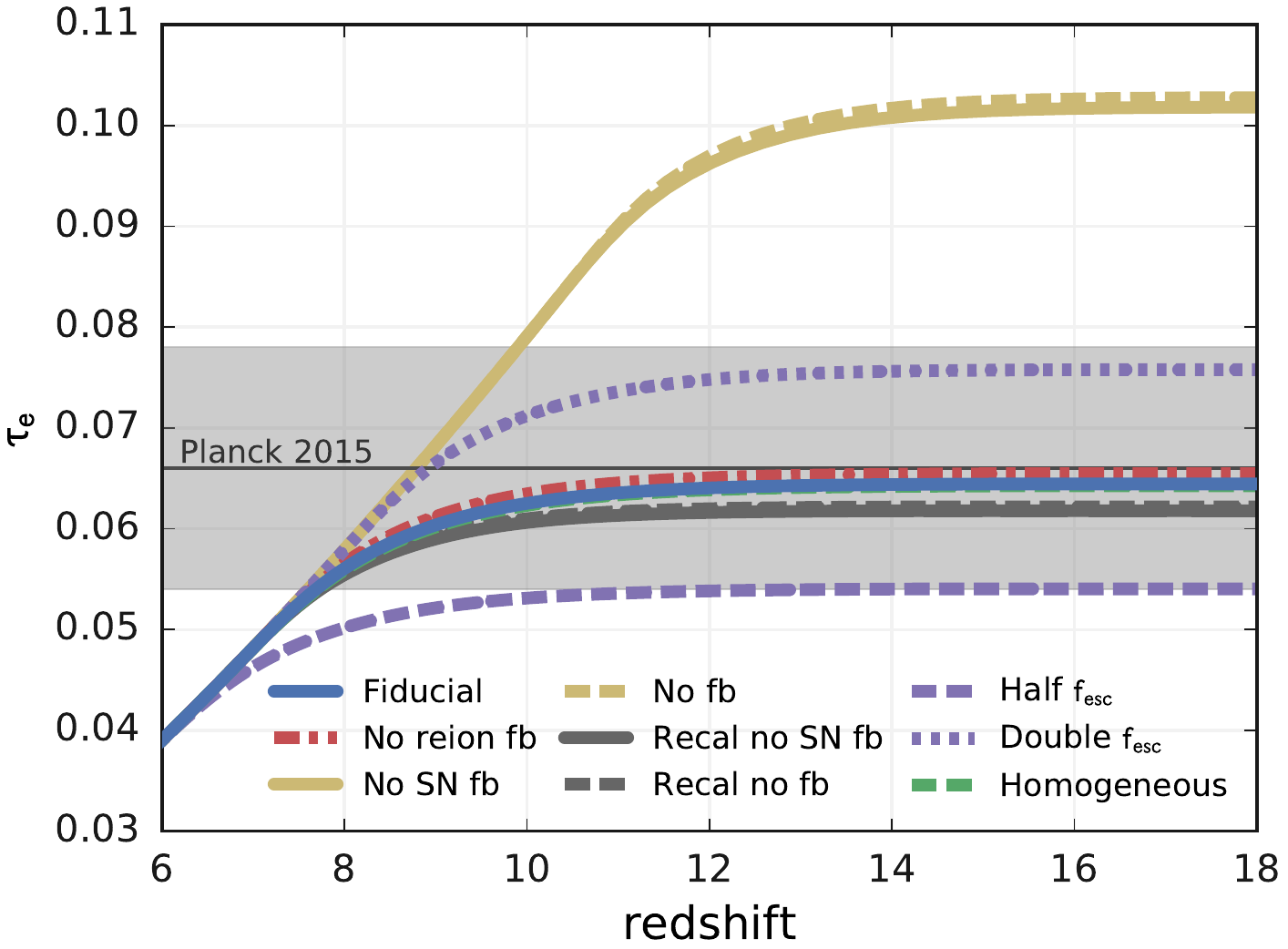}
\caption{\label{fig:tau_e} The integrated free electron scattering optical 
  depth, $\tau_e$, as a function of redshift.  The grey horizontal line and
  shaded region indicate the constraints on $\tau_e$ to $z\sim 1100$ from the
  {\em Planck} 2015 data release \citep{Planck-Collaboration2015}.  The blue solid
  line shows the fiducial model which is constrained to reproduce the {\em 
    Planck} result.  The \homogeneous\ model result is obscured by the \fiducial\ line.}
\end{figure}

In Fig.~\ref{fig:tau_e}, we present the electron scattering optical depth of 
our fiducial model (blue line) against the current best observational 
measurements provided by the {\em Planck} satellite 
\citep[][]{Planck-Collaboration2015}.  The other model variations shown in this 
plot will be discussed in detail below.  However, as can be seen, our fiducial 
model provides an integrated optical depth which is in excellent agreement with 
the {\em Planck} results.

An important consideration when assessing the results of any cosmological
simulation is the potential loss of stellar mass (and therefore ionizing
photon contribution) due to finite mass resolution.  This issue is
explored in detail for \Meraxes\ and the \Tiamat\ suite of simulations in
Appendix~\ref{sec:appendix-resolution}.  In summary, we find that the halo mass
function of the full \Tiamat\ simulation is complete down to approximately
$0.5\,{\rm dex}$ above the atomic cooling mass threshold at $z{=}5$, whilst
the lower volume but higher mass resolution \MediTiamat\ is fully complete
down to this mass limit.  Using \MediTiamat\ to quantify the fraction of total
stellar mass missing from the full simulation, we find that at $z{=}10$ we
miss approximately 25\% of stellar mass from low mass, unresolved systems.  By
$z{=}6$ this fraction falls to ${\la}5\%$.

\begin{figure*}
\begin{minipage}{\textwidth}
  \begin{center}
    \includegraphics[width=0.86\columnwidth]{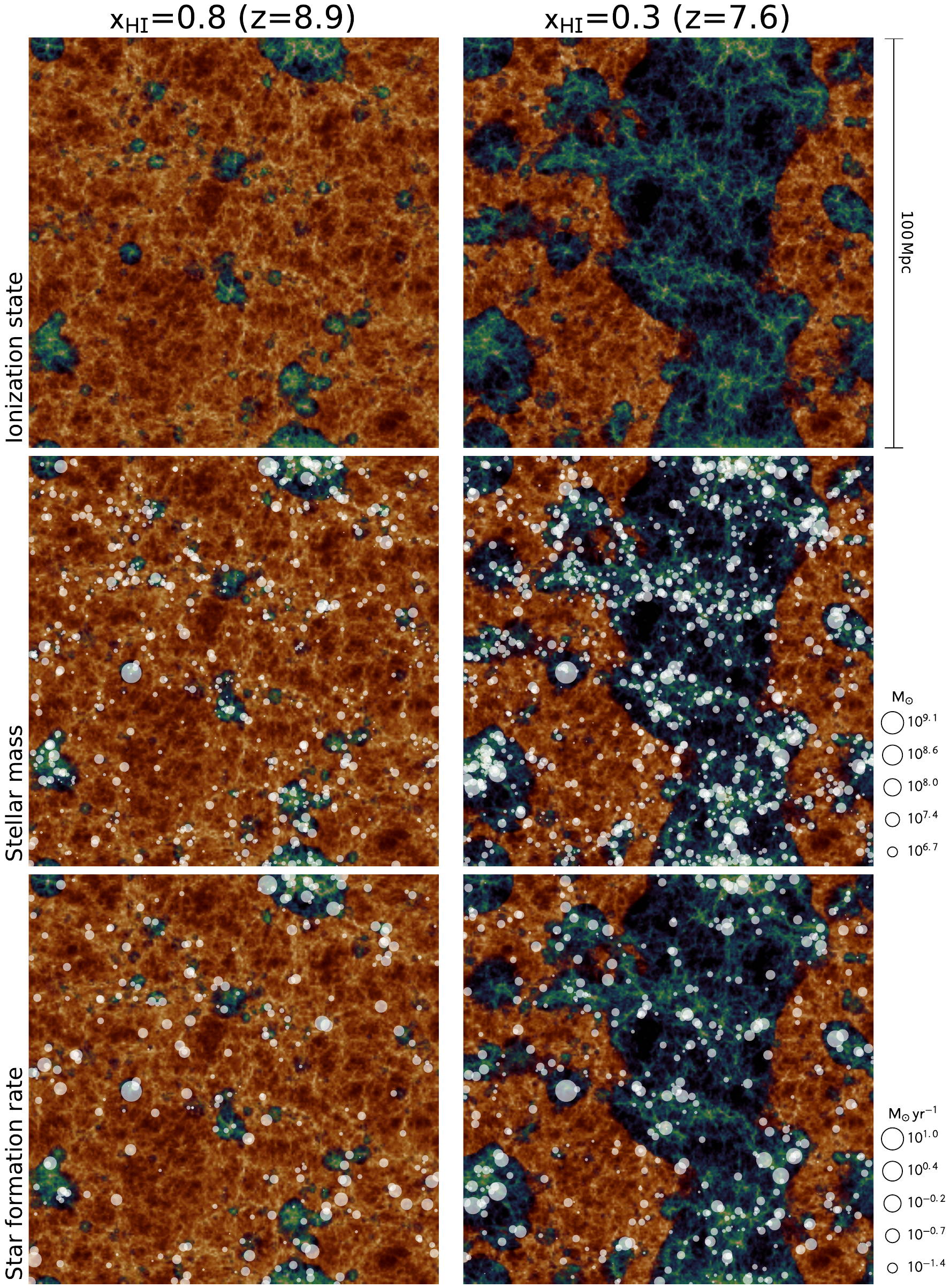}
    \caption{\label{fig:ionization_state_maps} The ionization structure of
    a $4\,{\rm Mpc}$ thick slab produced by \Meraxes.  The left- and right-hand
    columns correspond to a volume-weighted global neutral fraction of
    $\left<x_{\rm HI}\right>{=} 0.8$ and 0.3, respectively. Top row:
    the ionization state of the slab.  Orange indicates regions of the volume
    which are neutral, whilst the green structure shows the underlying
    matter distribution inside the ionized bubbles surrounding the densest
    structures.  At $\left<x_{\rm HI}\right>{=} 0.3$ (right), there is a large
    `lane' of ionized IGM approximately $50\,{\rm Mpc}$ wide extending from
    top to bottom where multiple bubbles overlap. Middle row: the same
    ionization structure with a random 1/50th of the ionizing galaxy
    population (selected at $\left<x_{\rm HI}\right>{=} 0.3$) overlaid as
    points scaled by stellar mass.
    Bottom row: the same galaxy population as above but with point size
    scaled by instantaneous star formation rate.  There are fewer
    galaxies in this row, compared to above, due to the bursty nature of star
    formation resulting in some galaxies forming no new stars in the current
    snapshot.}
  \end{center}
\end{minipage}
\end{figure*}

Finally, in Fig.~\ref{fig:ionization_state_maps}, we present a $4\,{\rm Mpc}$
deep slab extracted from the \fiducial\ \Meraxes\ model (see
Section~\ref{sec:results} below) at a volume-averaged neutral fraction of 0.8
(left) and 0.3 (right).  The orange regions indicate the neutral portions of
the simulation volume, whilst the green structure shows the underlying matter
distribution within the ionized bubbles.  At early times, these ionized bubbles
surround the peaks in the density field where the first galaxies form.  As
reionization progresses the bubbles begin to overlap, and by $x_{\rm HI}{=}0.3$
a large `lane' of ionized hydrogen, extending the entire length of the
simulation volume is formed.  The ability to investigate the distribution and
evolution of bubble morphologies and the associated observable $21\,{\rm cm}$
power spectra is a key feature of our \Meraxes\ framework and is investigated
in detail in Paper V.  In the middle and bottom rows of
Fig.~\ref{fig:ionization_state_maps}, we overplot the positions of a random
1/50th of the full galaxy population in the slab, selected at $x_{\rm
HI}{=}0.3$.  The galaxies plotted in the $x_{\rm HI}{=}0.8$ panels are the main
progenitors of this subsampled population.  In the middle row, the points are
scaled by the stellar mass of each galaxy, whilst in the bottom row they are
scaled by their instantaneous star formation rates.  The fact that fewer points
are present in the star formation rate panels is a result of the bursty nature
of star formation in the model meaning some fraction of galaxies have no
ongoing star formation at any particular redshift.


\section{The interplay between reionization and galaxy growth}
\label{sec:results}

In this section, we use \Meraxes\ to investigate the relative importance of 
reionization feedback for regulating both the growth of galaxy stellar mass and 
the timing and duration of the EoR.  We also quantify how 
simple variations to the physics of reionization affect the galaxy stellar mass 
function, before finally elucidating the importance of local environment for 
determining the stellar mass of galaxies affected by photoionization 
suppression.  We again note that the results of this work should be interpreted
within the context of our chosen parameter values, constrained to match the
evolution of the galaxy stellar mass function and the optical depth to electron
scattering as described in Section~\ref{sec:model_calibration}.

\begin{table*}
  \begin{minipage}{\textwidth}
    \centering

    \caption{
      \label{tab:runs}
      Summary table of the different model runs explored in this work (see
      Section~\ref{sec:results}).  $\Delta z_{x_{\rm HI}{=}0.8\rightarrow 0.2}$
      indicates the redshift spanned between a global neutral fraction of 80\%
      and 20\%. The $\alpha_{\rm SF}$ and $f_{\rm esc}$ columns indicate the
      star formation efficiency and ionizing photon escape fraction of each run
      respectively.  The $z{=}5$, 6, and 7 SMF $\alpha$ parameters (last
      three columns) refer to the low mass slope of the corresponding
      stellar mass functions fit with a standard Schechter function.  For
      reference, the $\alpha$ parameters measured by \citet{Duncan2014}
      are ${-1.90^{+0.21}_{-0.16}}$, ${-1.91^{+0.91}_{-0.59}}$ and
      ${-2.31^{+1.31}_{-0.19}}$ for redshifts 5, 6, and 7, respectively.
    }

    \begin{tabular}{r|l|c|c|c|c|c|c|c}
      \hline \hline
      &&&&&&\multicolumn{3}{c}{SMF $\alpha$}\\
      \cline{7-9} &{\bf Model} & $\alpha_{\rm SF}$ & $f_{\rm esc}$ & $z_{x_{\rm 
      HI}{=}0.5}$ & $\Delta z_{x_{\rm HI}{=}0.8\rightarrow 0.2}$ & $z{=}5$ &    
      $z{=}6$ & $z{=}7$                                                         
      \\
      \hline
      \fiducialLine & Fiducial & 0.03 & 0.2 & 7.96 & 1.47 & $-1.84^{+0.01}_{-0.01}$ & 
      $-1.92^{+0.02}_{-0.02}$ & $-2.05^{+0.03}_{-0.03}$ \\
      \noreionfbLine & No reionization feedback & 0.03 & 0.2 & 8.08 & 1.42 & 
      $-1.88^{+0.01}_{-0.01}$ & $-1.96^{+0.02}_{-0.02}$ & 
      $-2.06^{+0.03}_{-0.03}$ \\
      \nosnfbLine & No SN feedback & 0.03 & 0.2 & 11.41 & 1.52 & $-1.69^{+0.01}_{-0.00}$ & 
      $-1.86^{+0.01}_{-0.01}$ & $-2.04^{+0.01}_{-0.01}$ \\
      \nofbLine & No feedback & 0.03 & 0.2 & 11.50 & 1.47 & $-2.06^{+0.00}_{-0.00}$ & 
      $-2.14^{+0.01}_{-0.01}$ & $-2.23^{+0.01}_{-0.01}$ \\
      \recalibnosnfbLine & Recalibrated no SN feedback & 0.00106 & 0.2 & 7.75 & 1.29
      & $-2.25^{+0.02}_{-0.02}$ & $-2.51^{+0.03}_{-0.03}$ & $-2.69^{+0.07}_{-0.06}$ \\
      \recalibnofbLine & Recalibrated no feedback & 0.00106 & 0.2 & 7.8 & 1.26 & $-2.29^{+0.02}_{-0.02}$ 
      & $-2.52^{+0.03}_{-0.03}$ & $-2.66^{+0.07}_{-0.07}$ \\
      \halffescLine & Half $f_{\rm esc}$ & 0.03 & 0.1 & 6.89 & 1.38 &
      $-1.85^{+0.01}_{-0.01}$ & $-1.94^{+0.02}_{-0.02}$ &
      $-2.05^{+0.03}_{-0.03}$ \\
      \doublefescLine & Double $f_{\rm esc}$ & 0.03 & 0.4 & 9.04 & 1.50 & 
      $-1.83^{+0.01}_{-0.01}$ & $-1.91^{+0.02}_{-0.02}$ & 
      $-2.03^{+0.03}_{-0.03}$ \\
      \homogeneousLine & Homogeneous & 0.03 & 0.2 & 7.94 & 1.50 & $-1.84^{+0.01}_{-0.01}$ & 
      $-1.94^{+0.02}_{-0.02}$ & $-2.06^{+0.03}_{-0.03}$ \\
      \hline \hline
    \end{tabular}

  \end{minipage}
\end{table*}

Throughout we focus on the following model variations.
\begin{description}

  \item[\textit{Fiducial (\fiducialLine)}:] both the spatial and temporal 
    evolution of the reionization structure and UVB intensity are
    fully coupled to the growth of the source galaxy population.  This is the
    model calibrated in Section~\ref{sec:model_calibration} above to match the
    evolution of stellar mass function and Thompson scattering optical depth
    measurements.

  \item[\textit{No feedback (\nofbLine)}:] the \fiducial\ model with no 
    supernova feedback or reionization feedback included, resulting in runaway
    star formation.

  \item[\textit{No supernova feedback (\nosnfbLine)}:] the \fiducial\ model with 
    supernova feedback turned off.  All remaining physical processes, including 
    reionization feedback, remain unchanged.

  \item[\textit{No reionization feedback (\noreionfbLine)}:] the \fiducial\ 
    model with reionization feedback removed by setting $f_{\rm mod}{=}1$ for 
    all galaxies at all times.  All remaining physical processes, including 
    supernova feedback, remain unchanged.
  
  \item[\textit{Recalibrated no supernova feedback (\recalibnosnfbLine)}:] the
    \nosnfb\ model with a lower star formation efficiency, $\alpha_{\rm
    SF}{=}1.06{\times}10^{-3}$, chosen to replicate the total $z$=5 stellar
    mass density of the \fiducial\ model.

  \item[\textit{Recalibrated no feedback (\recalibnofbLine)}:] 
    the \nofb\ model with the same lower star formation efficiency
    ($\alpha_{\rm SF}{=}1.06{\times}10^{-3}$).
  
  \item[\textit{Half $f_{\rm esc}$ (\halffescLine)}:] the fully coupled 
    \fiducial\ model but with a lower escape fraction of $f_{\rm esc}{=}0.1$.

  \item[\textit{Double $f_{\rm esc}$ (\doublefescLine)}:] the fully coupled 
    \fiducial\ model but with a higher escape fraction of $f_{\rm esc}{=}0.4$.

  \item[\textit{Homogeneous (\homogeneousLine)}:] the evolution of reionization 
    is decoupled from that of the growth of galaxies.  The baryon
    fraction modifier of each halo, $f_{\rm mod}$, is determined using the halo
    number-weighted average $M_{\rm filt}$ as a function of redshift calculated
    using the fiducial model (see Section~\ref{ssub:homogeneous_model} above).  All
    remaining physical processes remain unchanged from the \fiducial\ case.

\end{description}

\subsection{The relative importance of reionization feedback}
\label{sub:results-relative_eff}

In Section~\ref{sec:model_calibration} above, we demonstrated that our 
\fiducial\ model successfully reproduces the evolution of the high-$z$ galaxy 
stellar mass function, as well as the most recent electron scattering optical 
depth measurements.  In this section, we utilize the resulting realistic 
population of galaxies to investigate how important photoionization suppression 
of baryonic infall is for regulating the stellar mass content of dark matter 
haloes when compared to galactic feedback processes such as supernova feedback.

\subsubsection{The stellar mass function}
\label{ssub:the_stellar_mass_function}

\begin{figure*}
  \begin{minipage}{\textwidth}
    \centering
    \includegraphics[width=\textwidth]{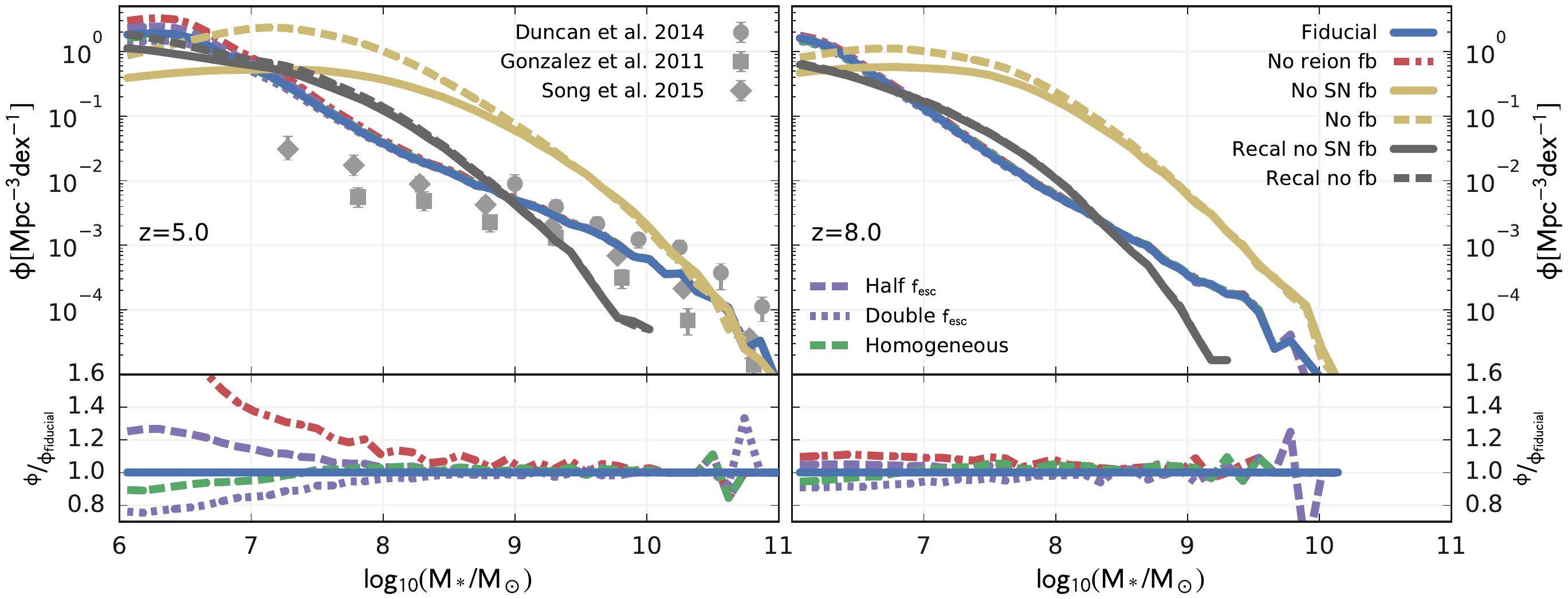}
    \caption{\label{fig:smf_comparison} The $z{=}5$ and 8 galaxy stellar 
      mass functions.  The solid blue line indicates the result of the
      \fiducial\ patchy reionization model which has been calibrated to
      reproduce the observed $z$=5--7 mass functions.  Grey points in the
      left-hand panel show the relevant observational data.  The small
      difference introduced by omitting reionization feedback (red
      dash--dotted) demonstrates the minor role which this mechanism plays in
      regulating stellar mass growth.  Conversely, the removal of supernova
      feedback (gold solid) produces a much larger effect.  However, comparison
      between the \nofb\ (gold dashed) and \nosnfb\ (grey solid) models shows
      that, in the absence of the dominant supernova feedback, reionization
      does suppress the number density of $M_*{\la}10^9\,{\rm M_{\sun}}$
      galaxies if star formation is gas supply limited in a large fraction of
    haloes.}
  \end{minipage}
\end{figure*}

In Fig.~\ref{fig:smf_comparison} we show the $z{=}5$ (left) and $z{=}8$
(right) galaxy stellar mass functions from each of our model variations.  The
latter redshift value corresponds to a volume-averaged neutral hydrogen fraction
of 50\% in the \fiducial\ model (\fiducialLine; cf. Fig.~\ref{fig:xHI_frac}
and Table~\ref{tab:runs}).  The bottom panels also indicate the fractional
differences of a subset of the models with respect to the fiducial result.
Immediately apparent is that models without supernova feedback (\nofbLine,
\nosnfbLine, \recalibnofbLine ,\recalibnosnfbLine) produce the most significant
change to the $z{=}5$ stellar mass function and are the only models which are
not consistent with the observational data (grey points).  The gold solid line
(\nosnfbLine) shows the predicted stellar mass content of haloes in the \nosnfb\
model.  Here, reionization feedback is still included using the fiducial
escape fraction of $f_{\rm esc}{=}0.2$; however, it is unable to counter the
runaway star formation which occurs in the absence of supernova feedback.
The net result is a large boost in the number densities of galaxies with
respect to the \fiducial\ model (\fiducialLine) for $M_*{\la}10^{10.5}\,{\rm
M_{{\sun}}}$.  At higher masses, supernova feedback becomes inefficient and
the mass function converges to the \fiducial\ result.  This is because
these large galaxies preferentially reside in the most massive haloes where
supernov\ae\ are unable to provide the required energy to heat gas to/beyond the
virial temperature, thus preventing it from being used for further star formation
(cf. Section~\ref{sub:supernova_feedback} and equations therein).

The red dash--dotted line (\noreionfbLine) in Fig.~\ref{fig:smf_comparison}
shows the results of our \noreionfb\ model.  Here we use the fiducial escape
fraction of $f_{\rm esc}{=}0.2$ in order to calculate the ionization state of
the IGM; however, we decouple the effect of photoionization suppression from
the infall of baryons into each FoF group by setting $f_{\rm mod}{=}1$ (cf.
Equation~\ref{eqn:fmod_defn}) for all groups at all times.  Hence, in this
model variation, galaxy evolution proceeds independently of the ionization
state of the IGM.  By turning off reionization feedback in this manner whilst
still leaving the strong supernova feedback required to reproduce the observed
high-$z$ stellar mass functions in place, we find little change to the stellar
mass function during the EoR (right panel) with respect to the \fiducial\ model
(\fiducialLine).  However, the effects of reionization are cumulative over time
(see Equation~\ref{eqn:Mfilt}) and by $z{=}5$ (left panel) we find the space
density of galaxies with stellar masses less than $10^{7}\, {\rm M_{\sun}}$ is
increased by up to ${\sim}40\%$ relative to the \fiducial\ result.  However,
this effect is small compared to that found in the absence of supernova
feedback.  Furthermore, as we move to higher masses, the relative differences
rapidly decrease.  This is a reflection of the fact that reionization feedback
is only effective in low-mass haloes (hosting typically low-mass galaxies) with
shallow potential wells susceptible to accretion suppression from the UVB (cf.
Section~\ref{sub:results-relative_eff}).

By comparing the \nofb\ (\nofbLine) and \nosnfb\ (\nosnfbLine) model lines, we
can investigate the isolated impact of reionization on the growth of galaxy
stellar mass.  In the former model, there is no reionization or supernova
feedback included; however, in the latter we turn on the photoionization
suppression of baryon accretion.  This results in a clear decrement in the
number density of low mass galaxies with $M_* \la 10^9\,{\rm M_{\sun}}$ at
$z{=}5$.  Higher mass galaxies are largely unaffected since, by the time
their massive host haloes were exposed to the UVB, they already
provided a potential well deep enough to accrete gas despite the presence of the
photoionizing UVB.  During reionization (right-hand panel), we see that
the effects of reionization feedback are more modest due to a smaller fraction
of galaxies being exposed to the UVB as well as the typically shorter
exposure times for those which have.

Simply turning off supernova feedback whilst leaving the remaining model
parameters constant (\nosnfb; \nosnfbLine) leads to an over-prediction of the
total $z$=5 stellar mass density by a factor of 5 relative to the \fiducial\
model as well as a mass function slope which is too steep to be consistent with
observations.  This enhanced star formation may lead to an under-estimate in the
importance of reionization feedback.  In order to test this hypothesis we have
run a \recalibnosnfb\ model (\recalibnosnfbLine) with a reduced star formation
efficiency parameter, $\alpha_{\rm SF}$, chosen to provide the same total
$z$=5 stellar mass density as the \fiducial\ model.  The \recalibnofb\ model
(\recalibnofbLine) additionally shows the result of this altered $\alpha_{\rm
SF}$ with reionization feedback also omitted.  Again, these models predict a
stellar mass function which is too steep to be consistent with observations.
However, we find that there is no combination of remaining parameters in our
model which can reproduce the slope of the observed stellar mass function in the
absence of supernova feedback.

The small relative difference between the \recalibnosnfb\ and \recalibnofb\
models in Fig.~\ref{fig:smf_comparison} indicates that reionization feedback
is even less effective at regulating galaxy growth than was the case with the
original \nosnfb\ and \nofb\ variations.  This is because our fiducial star
formation efficiency parameter results in stellar mass growth in the majority
of small haloes being gas supply limited.  By reducing the star formation
efficiency parameter in the recalibrated models, this is no longer the case and
there is more gas available in these systems than can be converted into stars
in a single time step.  Reionization-driven photosuppression of accretion
into these gas-rich systems therefore has little impact on the growth of
stellar mass, resulting in a quantitatively similar change to the stellar mass
function as is found in the presence of supernova feedback (i.e. comparing the
\fiducial\ and \noreionfb\ model results).

Together, these results highlight an over-estimate of the importance of
reionization feedback for regulating star formation in models which do not
employ the galactic feedback processes necessary to reproduce the observed
stellar mass functions at high-$z$.  Alternatively, at the very least, a star
formation efficiency low enough to result in a gas-rich star formation scenario
is needed.

\subsubsection{The stellar mass content of haloes}
\label{ssub:the_stellar_mass_content_of_haloes}

\begin{figure*}
  \centering
  \subfloat{\includegraphics[width=1.0\textwidth]{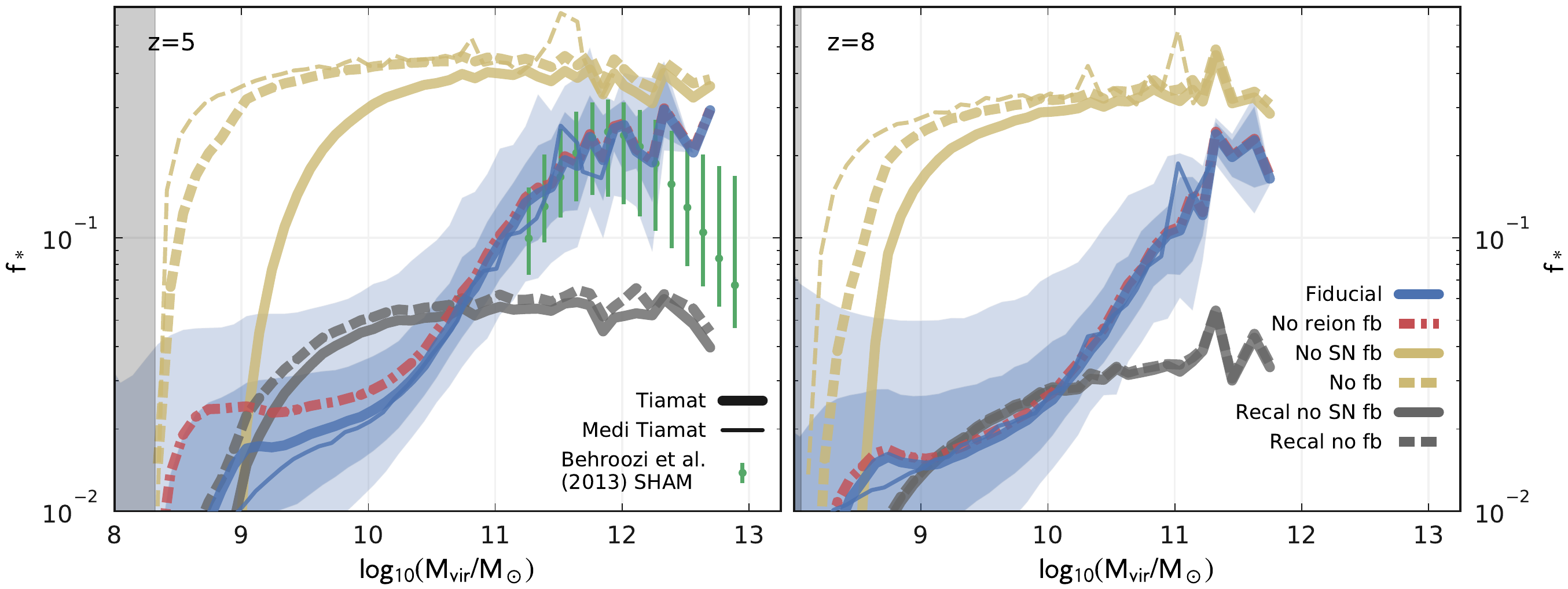}}
  \caption{\label{fig:fstar_eff} The median fraction of baryons in the form of 
    stars, $f_*{=}M_*/(f_{\rm b}M_{\rm vir})$, as a function of FoF group mass,
    at $z{=}5$ (left) and $z{=}7.8$ (right).  Thick lines show the results of
    \Meraxes\ when run on the full \Tiamat\ simulation merger trees, whilst
    thin lines indicate the results of running on the higher resolution
    \MediTiamat\ trees.  The blue solid lines and surrounding shaded regions
    show the median result of the \fiducial\ (\Tiamat) model and associated 68
    and 95\% confidence intervals.  The magnitudes of these statistical
    uncertainties are representative of those of all of the models shown in
    each panel.  The grey shaded region at the left of each panel denotes halo
    masses below the atomic cooling mass threshold, $M_{\rm cool}(z)$,
    corresponding to a virial temperature of $10^4\, {\rm K}$.  The green data
    points in the $z{=}5$ panel display the SHAM results of
    \citet{Behroozi2013} which are in excellent agreement with our \fiducial\
    model.  Comparison between the \nofb\ (gold dashed) and \nosnfb\ (grey
    solid) lines demonstrates that reionization feedback is most effective in
    low mass haloes.  However, supernova feedback of the level required to
    reproduce the observed high-$z$ stellar mass functions dominates the
    suppression of star formation across all halo masses.  The strong halo mass
    dependence of $f_*$ contrasts the constant value assumed by the majority of
  reionization structure studies.}
\end{figure*}

In Section~\ref{ssub:the_stellar_mass_function}, we demonstrated that the impact
of reionization on regulating the growth of galaxies and the production of
ionizing photons in \Meraxes\ is minimal owing to the importance of supernova
feedback.  In this section, we explore this topic further by investigating the
stellar mass content of haloes as predicted by a subset of our model variations,
both subsequent to and during reionization.  In Fig.~\ref{fig:fstar_eff},
we present the fraction of baryons in the form of stars, $f_*{=}M_*/(f_{\rm
b}M_{\rm vir})$, as a function of FoF group virial mass, at $z{=}5$ and 8 (the
latter redshift corresponding to $x_{\rm HI}{\sim}0.5$ in the fiducial model).
Thick lines indicate the results from running \Meraxes\ on the full \Tiamat\
simulation.  Thin lines show a subset of the same models run on the higher
resolution \MediTiamat\ trees which are complete down to the atomic cooling mass
threshold at both redshifts shown (cf. Appendix~\ref{sec:appendix-resolution}).

At low halo masses, there are minor differences between the \fiducial\ model
results of each simulation.  These are predominantly driven by an increased
prevalence of merger-driven halo mass growth in \MediTiamat\ which is
unresolved in the lower resolution \Tiamat\ trees.  These merger events occur
less frequently than \textit{in situ} star formation episodes, but they are more
efficient, giving rise to more energetic supernova feedback episodes capable of
ejecting significant amounts of material from haloes and temporarily halting
star formation.  The net result is a reduction in the stellar mass growth of
low mass haloes in the higher resolution trees.  During reionization
(right-hand panel), the lower atomic cooling mass threshold results in \Tiamat\
missing a larger fraction of the lowest mass haloes than is the case at
$z{=}5$.  This can be seen by comparing the \nofb\ (\nofbLine) model lines from
each simulation.  However, when feedback is included, the stellar mass content
of these haloes is greatly reduced and the results of the different simulations
again come to a good agreement.  We also note that despite the minor
discrepancies at low masses, there is excellent overall agreement between the
\Tiamat\ and \MediTiamat\ results across all other masses in both panels.
Hence, Fig.~\ref{fig:fstar_eff} demonstrates that our full \Tiamat\ volume has
sufficient mass resolution to correctly model the growth of galaxies across the
range of masses and redshifts relevant for this work.

Our \fiducial\ model (\fiducialLine) predicts a strong decline in the stellar
mass content of haloes with $M_{\rm vir}{\la}10^{12}\,{\rm M_{\sun}}$.  For
comparison, a constant stellar baryon fraction of $f_{*}{\sim}0.05$ is commonly
employed by previous studies utilizing \tocf\ \citep[e.g.][]{Mesinger2011,
Sobacchi2013b}, as well as previous \textit{N}-body-based radiative transfer calculations
\citep[e.g.][]{Iliev2007}.  Whilst it is important to remember that, for the
purposes of reionization, the precise value of $f_*$ is degenerate with other
quantities such as the escape fraction of ionizing photons, it is clear that the
approximation of a constant value with halo mass is poor.  Correctly predicting
and self-consistently utilizing this relation is a key feature of \Meraxes, as
is investigating and predicting further potential contributing variables such as
environmental density (see Section~\ref{sub:results-environ} below) and ionizing
escape fraction.

For comparison, we have also plotted the $z{=}5$ $f_*$--$M_{\rm vir}$ relation
and 1-$\sigma$ scatter found by the SHAM study of
\citet[][green error bars]{Behroozi2013}.  Our \fiducial\ model (\fiducialLine)
shows excellent agreement with these results.\footnote{We note that the
SHAM results of \citet{Behroozi2013} actually provide
the $f_*$--$M_{\rm vir}$ relation for central subhaloes, not for FoF groups as
we have presented in Fig.~\ref{fig:fstar_eff}.  However, at the high halo
masses probed by the abundance matching data, the central subhalo and its galaxy
dominate the mass of the FoF group in the vast majority of cases.  Therefore
this comparison is fair.  We have also explicitly confirmed that the subhalo
$f_*$--$M_{\rm vir}$ relation of our fiducial model, as well as its scatter,
remain consistent with the above results over the relevant mass range.}  This is
perhaps not unexpected given that both studies have utilized \textit{N}-body simulations
with free model parameters constrained to match observed high-$z$ stellar mass
functions.  However, the agreement is still noteworthy given that we have
utilized different N-body simulations, halo finders, observational data sets and 
methodologies.  Less obvious is the high level of agreement between the
scatter in $f_*$ values at a fixed $M_{\rm vir}$ which is an unconstrained
prediction of our \Meraxes\ results.  We also highlight that our model provides
predictions down to masses at least two orders of magnitude lower than can be
directly probed by current observations and SHAM studies.

A notable feature of the \nofb\ model (\nofbLine) is the sharp turn-over in the
$f_*$--$M_{\rm vir}$ relation at halo masses approaching, but greater than, the
atomic cooling mass threshold.  This occurs in both the full \Tiamat\ (thick
lines) and higher resolution \MediTiamat\ volumes (thin lines), indicating
that this is a robust prediction of our model.  Further investigation suggests
that stellar mass growth in these low mass haloes is almost entirely dominated
by \textit{in situ} star formation (as opposed to merger driven growth).  However, the
fraction of their lifetimes over which the galaxies hosted by these objects have
been actively forming stars is small.  This is due to an inability to obtain
enough gas to exceed the critical surface density required for star formation
(cf.  Equation~\ref{eqn:sigma_crit}).

By comparing the \nofb\ (\nofbLine) and \nosnfb\ (\nosnfbLine) model lines we
can again assess the importance of reionization-driven photosuppression in the
absence of the effects of supernova feedback.  In these models, the star
formation efficiency in a significant number of low-mass haloes remains gas
supply limited and hence reducing the infall of fresh baryons via
photoionization suppression has a significant effect.  Comparing the
\recalibnofb\ (\recalibnofbLine) and \recalibnosnfb\ (\recalibnosnfbLine) model
lines, where the lowered star formation efficiency results in galaxy growth
which is no longer limited by the availability of cold gas, the effect of
reionization is far less pronounced, just as is the case in the model
variations with supernova feedback included (i.e. the \fiducial\ and
\noreionfb\ models).

\begin{figure}
  \centering
  \includegraphics[width=\columnwidth]{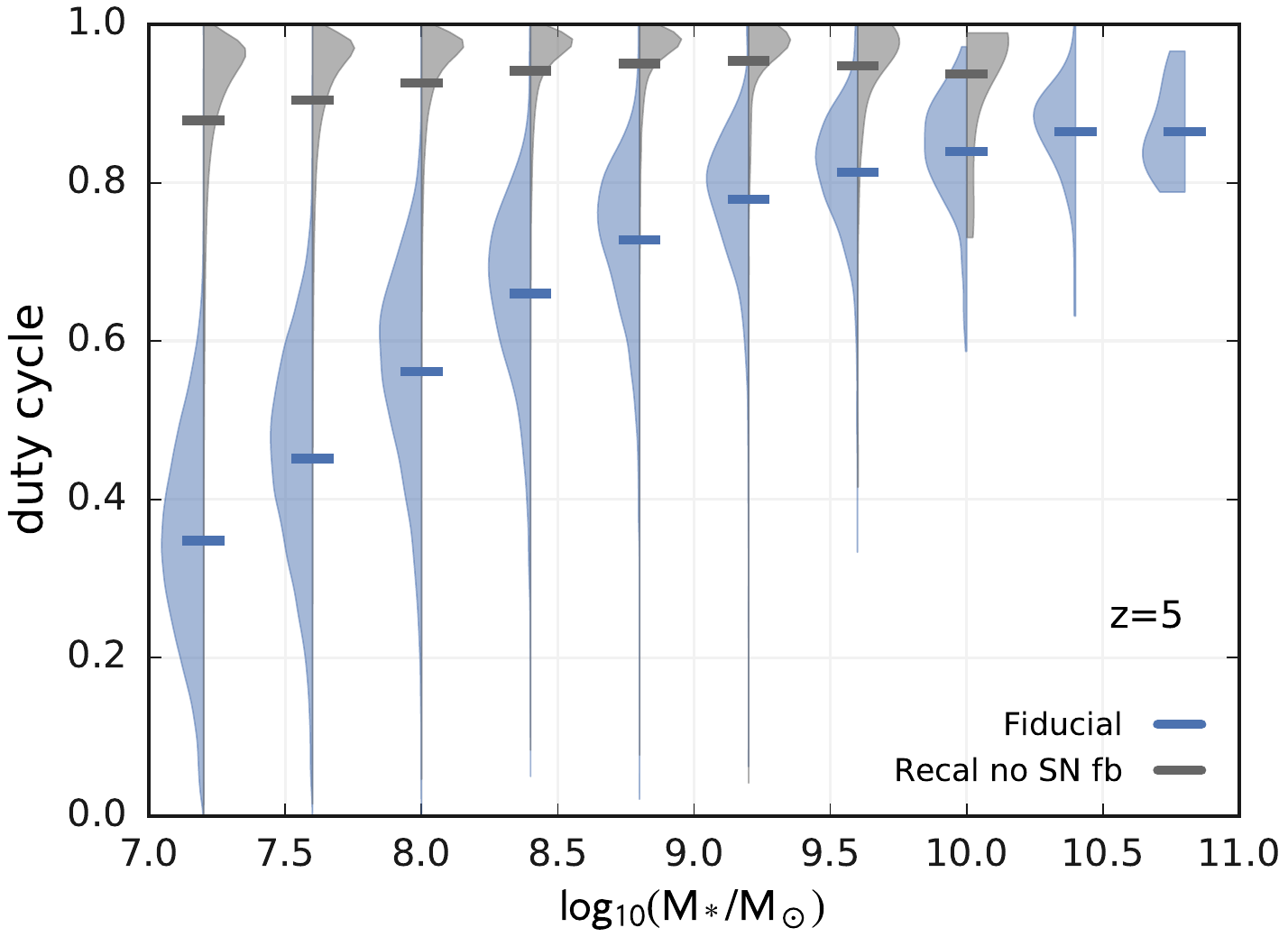}
  \caption{\label{fig:duty_cycle}%
    The fraction of time over which galaxies have been actively forming stars
    during their lifetimes (duty cycle) as a function of their $z{=}5$ stellar
    mass in both the \fiducial{} (blue) and \recalibnosnfb{} (grey) models.
    The shaded regions show a kernel-density estimate of the distribution of
    duty cycles for each stellar mass bin whilst horizontal bars indicate the
    mean values.  Supernova feedback in the \fiducial{} model
    blows out cold gas in low mass systems, stifling star formation and
    reducing the fraction of time over which they are actively forming
    stars.  In the \recalibnosnfb{} model, the lack of supernova feedback
    coupled with a low star formation efficiency results in the ample
    availability of cold gas at all times and hence a duty cycle of ${\sim}$1
    at all masses. 
}
\end{figure}

To demonstrate the effect of gas supply limited star formation more explicitly,
in Fig.~\ref{fig:duty_cycle} we present the fraction of their lifetime over
which galaxies have been actively forming stars (i.e. their duty cycle) as a
function of stellar mass in both the \fiducial{} (blue) and \recalibnosnfb{}
(grey) models.  The shaded regions show a kernel-density estimate of the
distribution of duty cycles for each stellar mass bin whilst horizontal bars
indicate the mean values.  In the \fiducial{} model, supernova feedback reduces
the availability of cold gas for star formation, such that only a small
fraction of the infalling material is converted to stars in low mass systems.
In the \recalibnosnfb{} model, the lack of supernova feedback, coupled with a
low star formation efficiency, means that there is an ample supply of cold gas
for star formation at all times and hence a duty cycle of ${\sim}$1 at all
masses.

\subsection{Quantifying the effects of reionization on the galaxy stellar mass 
  function}
\label{sub:results-smf}

Having demonstrated the relatively small importance of reionization feedback for
regulating galaxy growth, in this section we move on to quantify the effects of
varying the escape fraction of ionizing photons on the stellar mass function.
The ability to self-consistently and quantitatively investigate such outcomes
is a key feature of \Meraxes.

In Fig.~\ref{fig:smf_comparison} we present the stellar mass functions
predicted by our \doublefesc (\doublefescLine) and \halffesc (\halffescLine)
models.  By doubling the escape fraction of ionizing photons, we see a
suppression in the stellar mass function by around 20\% for masses of around
$10^{7}\, {\rm M_{\sun}}$.  The reason is that by increasing the ionizing
emissivity of all galaxies, reionization occurs earlier than in the \fiducial\
model (see Fig.~\ref{fig:xHI_frac}).  This results in an increased time
over which haloes are exposed to ionizing radiation.  Coupled with the fact
that haloes are typically less massive and more susceptible to the effects of
reionization at higher redshifts, the final result is a suppression in the
number of low mass galaxies relative to the \fiducial\ model.  The space density
of more massive objects is again largely unaffected since, even with an earlier
exposure to the ionizing UVB, their haloes were already massive
enough to accrete gas from the reionized IGM.  Any reduction
in the stellar mass of these objects is hence mainly driven by the accretion
of less stellar mass through mergers.  By halving the escape fraction of
ionizing photons (\halffesc\ model; \halffescLine), we obtain a mirror effect.
Reionization occurs later, and hence haloes are typically exposed to the UVB for
less time and have larger masses when this occurs.

Current observations are only able to probe the $z{=}5$ stellar mass function
down to $M_*{\ga}10^{7.5}\, {\rm M_{\sun}}$, and even then only through the
use of uncertain mass-to-light ratios \citep[e.g.][]{Gonzalez2011, Song2015}
with relatively large systematic and statistical uncertainties.  Thus, even
in the case of no reionization feedback, the relative change between the
mass functions of Fig.~\ref{fig:smf_comparison} is well below the level
of observational uncertainties.  In Table~\ref{tab:runs}, we indicate the
low-mass slope, $\alpha$, for each run, where we have only fitted to galaxies
with stellar masses greater than the observational limit of $10^{7.5}\,
{\rm M_{\sun}}$.  The differences between the resulting $\alpha$ values are
again too small to be detected observationally without extremely precise
measurements beyond what is currently achievable.  On the right-hand side of
Fig.~\ref{fig:smf_comparison}, we present the equivalent galaxy stellar mass
functions at $z{=}8$, corresponding to a global neutral hydrogen fraction of
approximately $0.5$ in the \fiducial\ model (Table~\ref{tab:runs}).  Here,
even at the resolved masses, the variations in space density with respect to
the \fiducial\ model are typically less than 10\%.  This suggests that the
statistics of low redshift ($z{\leq}5$) galaxy populations, as opposed to
galaxies during the EoR itself, provide the stronger potential
constraints on reionization.

On the other hand, the insensitivity of the observable portion of the $z{=}5$
mass function to the details of reionization suggests that we can calibrate our
galaxy formation models to high-redshift stellar mass/luminosity functions and
then use additional observations to constrain other unknowns in high-$z$ galaxy
formation.  These include the dependence of the ionizing escape fraction on
properties such as mass, redshift and star formation rate.

\subsection{The evolution of ionized hydrogen in the IGM}
\label{sub:results-xHI}

\begin{figure}
  \centering
  \includegraphics[width=\columnwidth]{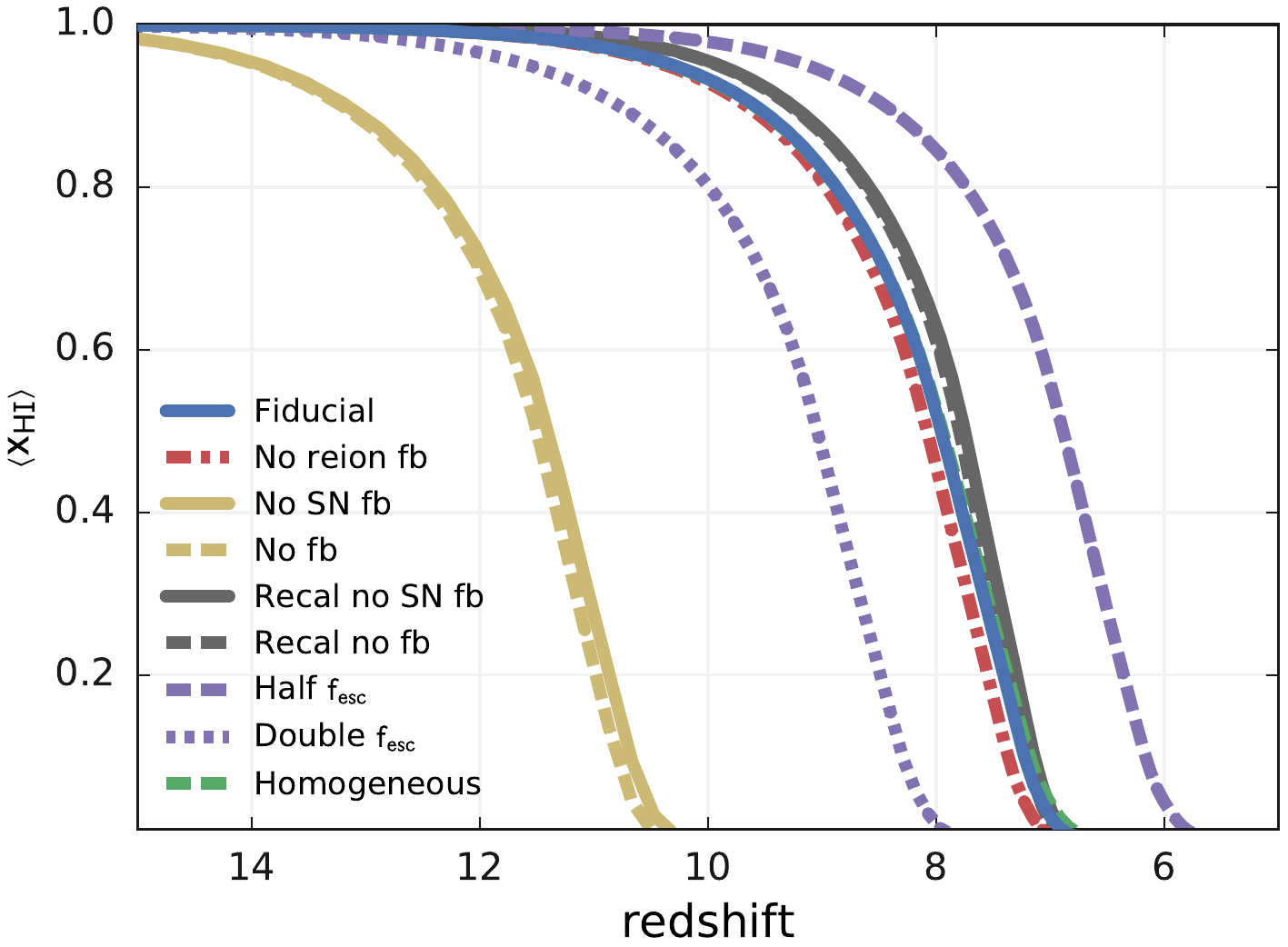}
  \caption{\label{fig:xHI_frac}The evolution of the volume-weighted global 
    neutral fraction, $\left<x_{\rm HI}\right>$, as a function of redshift.
    The blue solid line indicates the result of our \fiducial\ model which
    matches both the observed evolution of the galaxy stellar mass function
    (Fig.~\ref{fig:smf_calibration}) and the most recent electron scattering
    optical depth measurements (Fig.~\ref{fig:tau_e}).  The small perturbation
    to the neutral fraction history caused by the omission of reionization
    feedback (red dash--dotted) indicates that the EoR is not self-regulated.
    Instead, supernova feedback is almost entirely responsible for regulating
    the production of ionizing photons and the progression of reionization.  The
    \homogeneous\ and \recalibnofb\ lines are obscured by the \fiducial\ and
    \recalibnosnfb\ results, respectively.}
\end{figure}

In Fig.~\ref{fig:xHI_frac}, we present the evolution of the volume-averaged
neutral fraction of our \fiducial\ patchy reionization model (\fiducialLine)
and all eight model variations.  The second and third columns
of Table~\ref{tab:runs} also provide some basic statistics on the timing
and duration of reionization in each model in the form of the redshift at
which the volume becomes 50\% ionized ($z_{x_{\rm HI}{=}0.5}$) and the
redshift spanned between neutral fractions of 80 and 20\% ($z_{x_{\rm
HI}{=}0.8\rightarrow 0.2}$).

The purple lines in Fig.~\ref{fig:xHI_frac} demonstrate the effect of
either doubling (\doublefescLine) or halving (\halffescLine) the ionizing
escape fraction from its fiducial value of $f_{\rm esc}{=}0.2$.  This shifts
reionization to earlier and later times, respectively, but produces no
significant change to its duration (Table~\ref{tab:runs}).  Interestingly,
Fig.~\ref{fig:tau_e} shows that, despite a relative shift in the midpoint of
reionization with respect to the \fiducial\ model of $\Delta z{\approx}{\pm}
1$, both the double and half ionizing escape fraction models provide
reionization histories which are marginally consistent with the {\em
Planck} optical depth constraints.  Furthermore, the relative change in
$\tau_e$ with respect to the \fiducial\ model is approximately symmetric
(Table~\ref{tab:runs}).  As demonstrated in Section~\ref{sub:results-smf},
the effect of reionization feedback on the growth of stellar mass is weak,
with both of these model variations predicting stellar mass functions which
remain in good agreement with observational data.  Hence, $f_\textrm{esc}$ is
effectively decoupled from the stellar mass function in our model and the
electron scattering optical depth measurements can be directly translated to
an approximate constraint on this parameter.\footnote{We emphasize that the
ability of the electron scattering depth to fully constrain $f_{\rm esc}$ is
only true within the framework of our model where other confounding processes,
such as the intrinsic ionizing photon production rate of stellar populations,
are fixed.} In the case of an ionizing escape fraction which is constant with
both mass and redshift, $0.1{\la}f_{\rm esc}{\la}0.4$ for our fiducial model.

Changing the escape fraction of ionizing photons has two competing outcomes for 
reionization.  First, the ionizing efficiency parameter, $\xi$, is directly 
proportional to $f_{\rm esc}$ (cf. Equation~\ref{eqn:ionizing_eff}), and so doubling the 
escape fraction results in a doubling in the efficiency of reionization for a 
fixed mass of stars.  As a consequence, the first galaxies to form have a larger 
impact, moving the start of reionization to earlier times.  To balance this, the 
increased ionizing emissivity also leads to a more efficient photoionization 
suppression of baryonic infall, reducing star formation rates and regulating the 
production of further ionizing photons.  The converse is also true for the case 
of halving the escape fraction.  Here, more stellar mass is required to produce the 
same number of ionizing photons escaping into the IGM; hence, reionization moves 
to later times.  To counter this, the UVB photosuppression of baryonic 
infall is also reduced.  This `self-regulation' mechanism has been proposed as 
a potentially important effect for modulating the timing and duration of 
reionization \citep{Iliev2007}.

We can exploit our framework to quantify just how effective photoionization
suppression is in self-regulating reionization for a realistic population
of galaxies which simultaneously reproduces both the growth of stellar
mass in the early Universe and current optical depth constraints.  In
Fig.~\ref{fig:xHI_frac} we again show the results of our \noreionfb\ model
(\noreionfbLine) in which galaxy evolution proceeds independently of the
ionization state of the IGM.  As can be seen in Fig.~\ref{fig:xHI_frac}, the
resulting change to the evolution of the global neutral fraction with respect
to the \fiducial\ model is minimal, with a shift in the midpoint of reionization
of less than $0.1$ in redshift (cf. Table~\ref{tab:runs}).  This is easily
understood by considering the minor role which reionization suppression plays in
modulating star formation, and hence the production of ionizing photons, during
the EoR itself (cf. Section~\ref{sub:results-relative_eff}
above).  Our \Meraxes\ framework thus predicts that reionization is
not self-regulated, supporting similar claims made by other authors
\citep[e.g.][]{Kim2013b, Sobacchi2013b, Wyithe2013}.

Instead, the strong importance of galactic feedback processes (such as supernova
feedback) for regulating the growth of stellar mass leads to them being
dominant in controlling the timing and duration of reionization along with
$f_{\rm esc}$.  The gold solid line (\nosnfbLine) in Fig.~\ref{fig:xHI_frac}
shows the predicted neutral fraction evolution in the absence of supernova
feedback processes (\nosnfb\ model).  The resulting runaway star formation,
in tandem with the fiducial escape fraction of $f_{\rm esc}{=}0.2$, results
in an early and rapid reionization process with a midpoint at $z{=}11.2$ and
a time of just $71\,{\rm Myr}$ between volume averaged neutral fractions of
80 and 20\%.  We also show in Fig.~\ref{fig:xHI_frac} the \nofb\ model,
where neither reionization nor supernova feedback is included.  The small
perturbation to the reionization history with respect to the \nosnfb\ model
(where only reionization feedback is included) is again extremely small (see
also Table~\ref{tab:runs}).  The same is also true of the \recalibnofb\
(\recalibnofbLine) and \recalibnosnfb\ (\recalibnosnfbLine) models, further
reinforcing the negligible contribution of photosuppression in self-regulating
reionization.

In summary, the level of supernova feedback required to reproduce observed
high-$z$ stellar mass functions dominates over photoionization suppression with
regard to modulating the progression of reionization.  This agrees with the
results of previous works \citep[e.g.][]{Kim2013b, Wyithe2013}.  However, we do
note that if reionization were more extended than is predicted by our models,
the role of reionization feedback may be enhanced.  For example, inhomogeneous
IGM recombinations or a redshift-varying escape fraction of ionizing photons
(see Section~\ref{sub:results-emissivity} below) could delay the end stages of
reionization and lead to an enhanced suppression of the large-scale reionization
structure \citep{Sobacchi2014}.

Finally, for reference, the green dashed line (\homogeneousLine) of
Fig.~\ref{fig:xHI_frac} shows the neutral fraction evolution of the
\homogeneous\ model variation.  This model produces a reionization history which
is in excellent agreement with the fiducial self-consistent, spatially dependent
scenario (\fiducialLine) by construction.  This is due to the fact that the
\homogeneous\ model was calibrated to reproduce the same mean filtering mass
as the \fiducial\ model at all redshifts.  However, we will go on to show in
Section~\ref{sub:results-environ} that the use of a parametrized homogeneous
model such as this has important, environmentally dependent implications for the
predicted properties of individual galaxies.

\subsection{The evolution of the instantaneous ionizing emissivity}
\label{sub:results-emissivity}

\begin{figure}
  \includegraphics[width=\columnwidth]{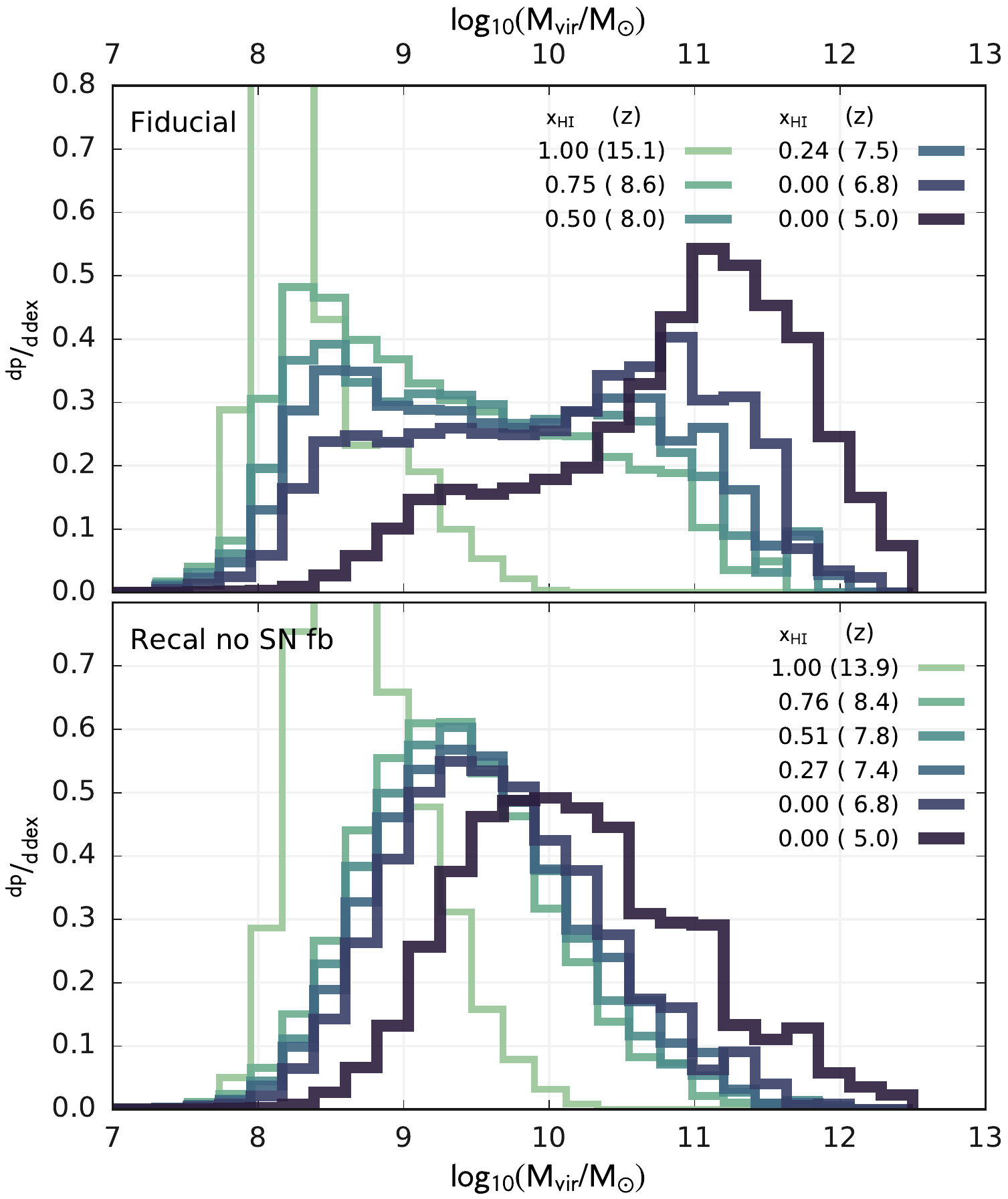}
    \caption{\label{fig:mvir_contrib} The probability distributions of
    instantaneous ionizing photon contribution as a function of FoF group
    mass for various global neutral fractions in the \fiducial\ (top) and
    \recalibnosnfb\ (bottom) models.  The inclusion of supernova feedback in
    the \fiducial\ model results in a broad distribution of halo masses which
    contribute to reionization.}
\end{figure}

Having demonstrated that reionization feedback plays only a minor role in 
regulating the evolution of the global neutral fraction, we now focus on 
elucidating the dominant effects of supernova feedback.  In 
Fig.~\ref{fig:mvir_contrib}, we present the probability distribution function 
of instantaneous ionizing photon contributions as a function of FoF group virial 
mass for both the \fiducial\ and \recalibnosnfb\ models at various neutral 
fractions.  At high $x_{\rm HI}$ (high-$z$), massive haloes have not had time to 
form and so the ionizing emissivity is dominated by masses near the atomic 
cooling mass threshold in both models.  After reionization has begun, the 
distributions predicted by the two models rapidly begin to diverge, with our 
\fiducial\ model predicting a broad range of contributing halo masses.  This is 
due to the inclusion of supernova feedback which suppresses the relative 
ionizing photon contribution of low-mass systems.  In the \recalibnosnfb\ model 
variation, the absence of supernova feedback means that low-mass haloes instead 
remain the dominant source of ionizing photons at all times.

\begin{figure}
  \includegraphics[width=\columnwidth]{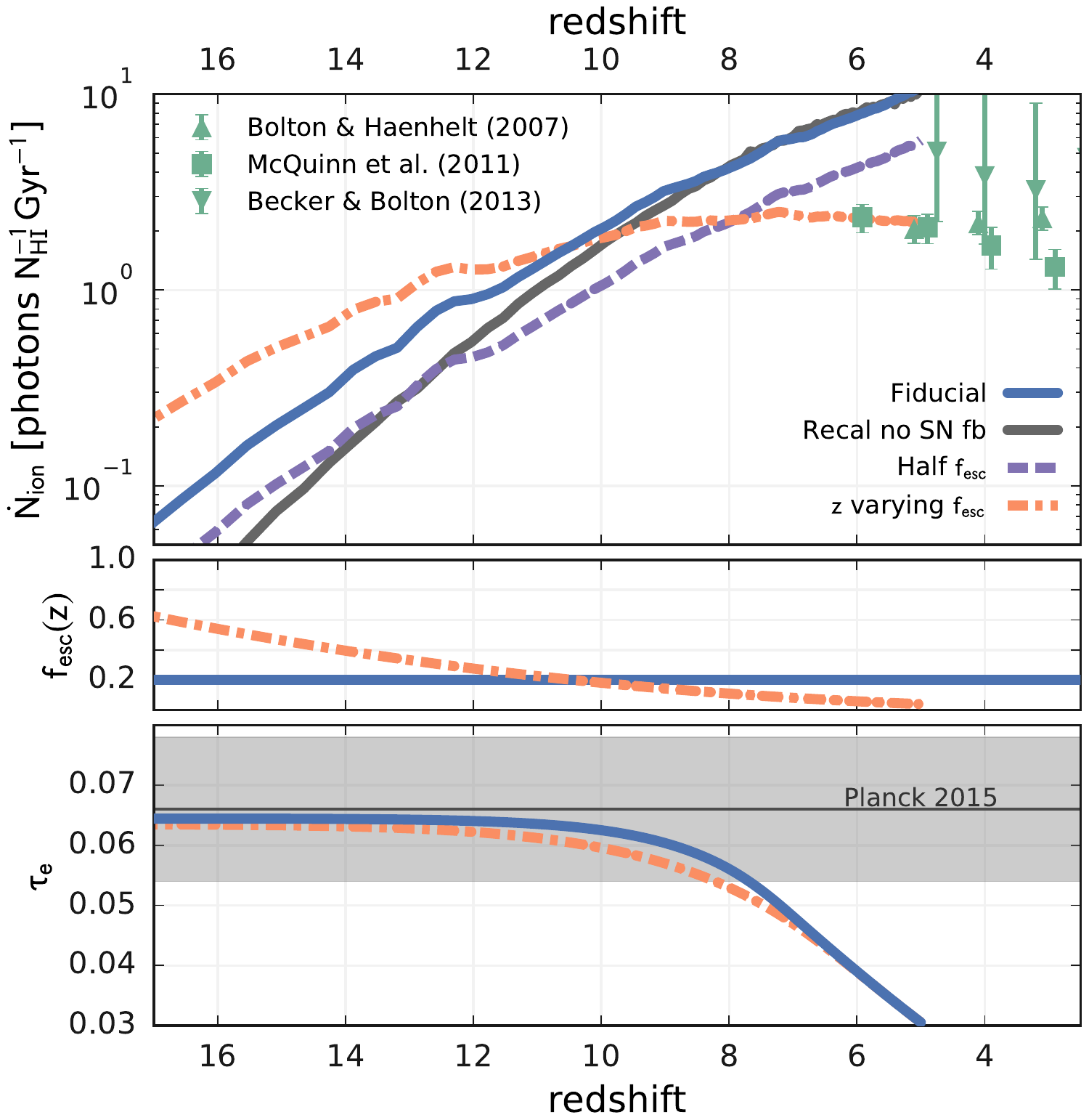}
  \caption{%
    \label{fig:emissivity_evo} Top panel: the evolution of the ionizing
    emissivity in units of photons per hydrogen atom per Gyr.  The green data
    points are taken from works combining IGM Lyman $\alpha$ opacities with
    detailed hydrodynamical simulations
    \citep{Bolton2007,McQuinn2011,Becker2013}.  Both the \fiducial\ (solid
    blue) and \recalibnosnfb\ (solid grey) models overpredict the
    instantaneous ionizing efficiencies by a factor of $\sim$5 at $z$=5.  The
    \halffesc\ model (purple dashed), which is only marginally consistent with
    the {\em Planck} optical depth measurements, still fails to reproduce the
    observational data.  However, an evolving escape fraction (orange
    dash--dotted) can simultaneously reproduce the observed flat emissivity and
    its normalization at $z{\le}6$. Middle panel: the escape fraction as
    a function of redshift for the \zvaryingfesc\ model shown in the panel
    above.  Bottom panel: the integrated free electron scattering optical
    depth, $\tau_{\rm e}$, as a function of redshift for the \fiducial\ and
    \zvaryingfesc\ models.  The grey horizontal line and shaded region indicate
    the constraints on $\tau_{\rm e}$ to $z\sim$1100 from the {\em Planck} 2015
    data release \citep{Planck-Collaboration2015}.
}
\end{figure}

In Fig.~\ref{fig:emissivity_evo}, we show the evolution of the
instantaneous ionizing emissivity, $\dot N_{\rm ion}$, in units of ionizing
photons per hydrogen atom per Gyr:
\begin{equation}
  \dot N_{\rm ion} = \frac{\Delta m_*}{\Delta t} \frac{N_{\gamma}f_{\rm
  esc}}{f_{\rm b}M_{\rm tot}(1-0.75 Y_{\rm He})} = \frac{\Delta m_*}{\Delta 
t}\frac{\xi}{M_{\rm tot}}\ ,
\label{eqn:dNiondt}
\end{equation}
where $\xi$ is the ionizing efficiency as defined in
Equation~\ref{eqn:ionizing_eff} and $\Delta m_*$ is the change in gross stellar
mass (excluding losses due to stellar evolution) between two consecutive
snapshots separated by a time $\Delta t$.  The data points present recent
measurements derived by combining observed IGM Lyman $\alpha$ opacities
with detailed hydrodynamical simulations including radiative transfer
\citep{Bolton2007,McQuinn2011,Becker2013}.

Despite correctly reproducing the electron scattering optical depth (see
Section~\ref{sec:model_calibration}), our \fiducial\ model (\fiducialLine)
predicts a high and steeply increasing ionizing emissivity at $z{=}5$
that is inconsistent with the observational data.  By construction, the
\recalibnosnfb\ model (\recalibnosnfbLine) predicts a $z$=5 global emissivity
in agreement with the \fiducial\ result.  However, the absence of supernova
feedback leads to a steeper evolution with redshift in the recalibrated model.
This is due to the lower star formation efficiency causing a delay in the
build-up of significant stellar mass, and therefore the onset of reionization.
As shown in Fig.~\ref{fig:mvir_contrib}, as galaxies grow with decreasing
redshift, the absence of supernova feedback ensures that low-mass, low-bias
haloes continue to dominate ionizing photon production.  This leads to a more
rapid growth in the global emissivity and a faster reionization process for
$\left< x_{\rm HI}\right> \la 0.5$ (cf. Fig.~\ref{fig:xHI_frac}).

One possible modification to reduce the ionizing emissivity of the \fiducial\
model, and thus bring it into qualitative agreement with the observational
constraints from the Lyman $\alpha$ forest, would be to reduce the escape
fraction of ionizing photons.  As shown in Fig.~\ref{fig:tau_e}, the
\halffesc\ model (\halffescLine) provides the lowest escape fraction
that is consistent with the {\em Planck} optical depth constraints.
However, from Fig.~\ref{fig:emissivity_evo} we see that this model still
results in a steep emissivity evolution which fails to reproduce the
trend of the $z{\le}6$ Lyman $\alpha$ forest constraints.  We can easily
understand this result by noting that $\dot N_{\rm ion}{\propto}f_{\rm esc}$
(Equation~\ref{eqn:dNiondt}).  Therefore, halving $f_{\rm esc}$ simply results
in a halving of $\dot N_{\rm ion}$.

\subsubsection{An evolving escape fraction}
\label{ssub:results-evolving_escape_fraction}

We now investigate how we can modify our model in order to simultaneously
match the normalization and flat slope of the observed ionizing emissivity at
$z{\le}6$, as well as our fiducial constraints of the {\em Planck} $\tau_{\rm
e}$ measurements and high-$z$ galaxy stellar mass functions.  To fully address
this question would require a full statistical investigation of the model's free
parameter space, a task which is beyond the scope of this work.  However, the
value of $f_{\rm esc}$ for an individual galaxy is strongly dependent on its
chemical, structural, and kinematic properties, and there exists considerable
theoretical and observational evidence to suggest that the average $f_{\rm
esc}$ does indeed vary with redshift, mass, and/or star formation rate
\citep[e.g.][]{Gnedin2008, Wise2009, Paardekooper2011, Yajima2011, Kuhlen2012,
Mitra2013, Paardekooper2015}.

\begin{figure}
  \centering
  \includegraphics[width=\columnwidth]{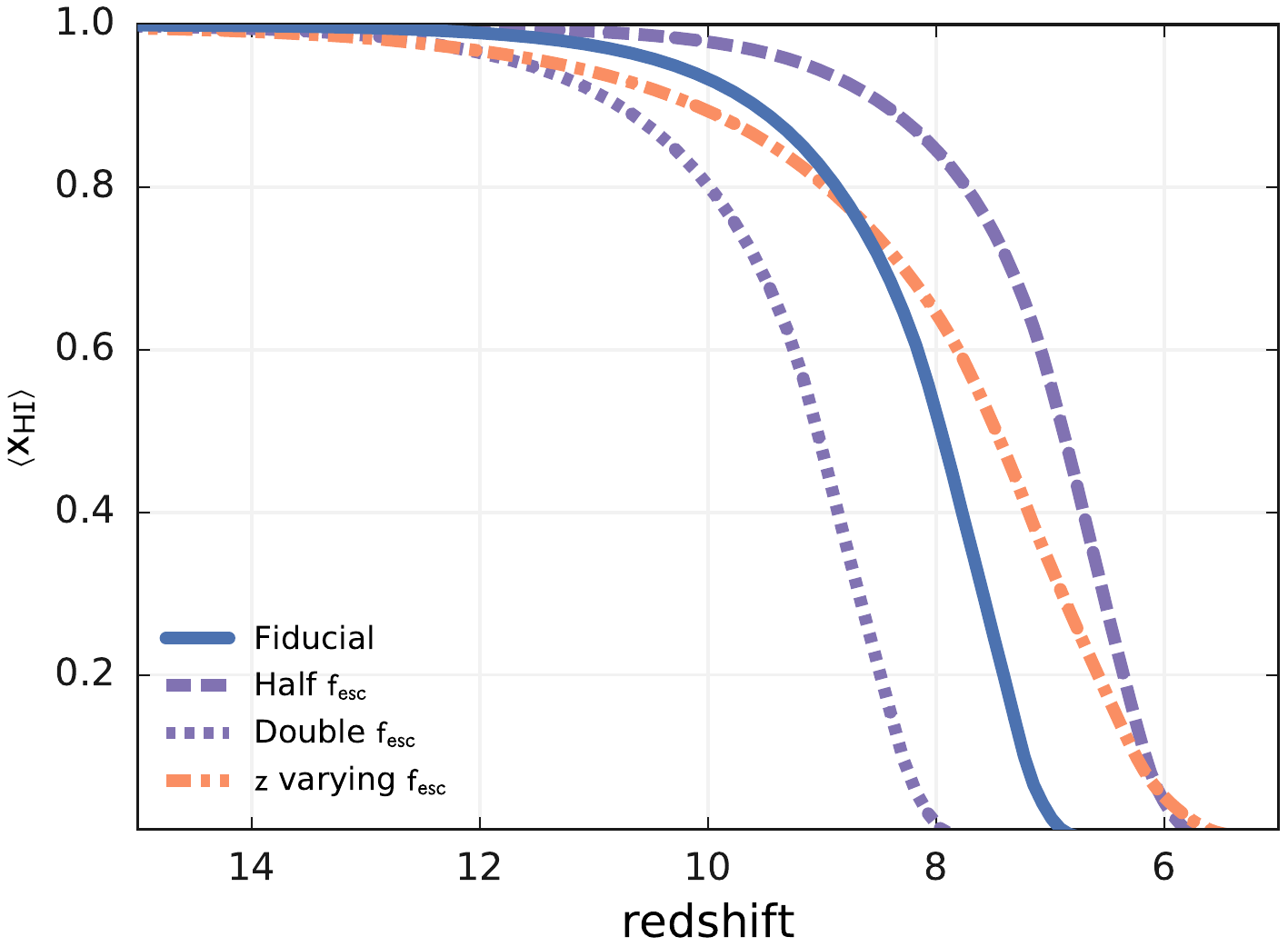}
  \caption{\label{fig:zvarying_xHI_frac}The evolution of the volume-weighted
    global neutral fraction, $\left<x_{\rm HI}\right>$, as a function
    of redshift for the \zvaryingfesc\ model (orange dashed).  Also shown for
    comparison are the results of the \fiducial\ (blue solid), \halffesc\
    (purple dashed), and \doublefesc\ (purple dotted) models.  The high escape
    fraction at high redshift in the \zvaryingfesc\ model leads to the early
    onset of reionization, whilst the declining $f_{\rm esc}$ with decreasing
    redshift also prolongs its duration.}
\end{figure}

\begin{figure*}
  \begin{minipage}{\textwidth}
    \includegraphics[width=\columnwidth]{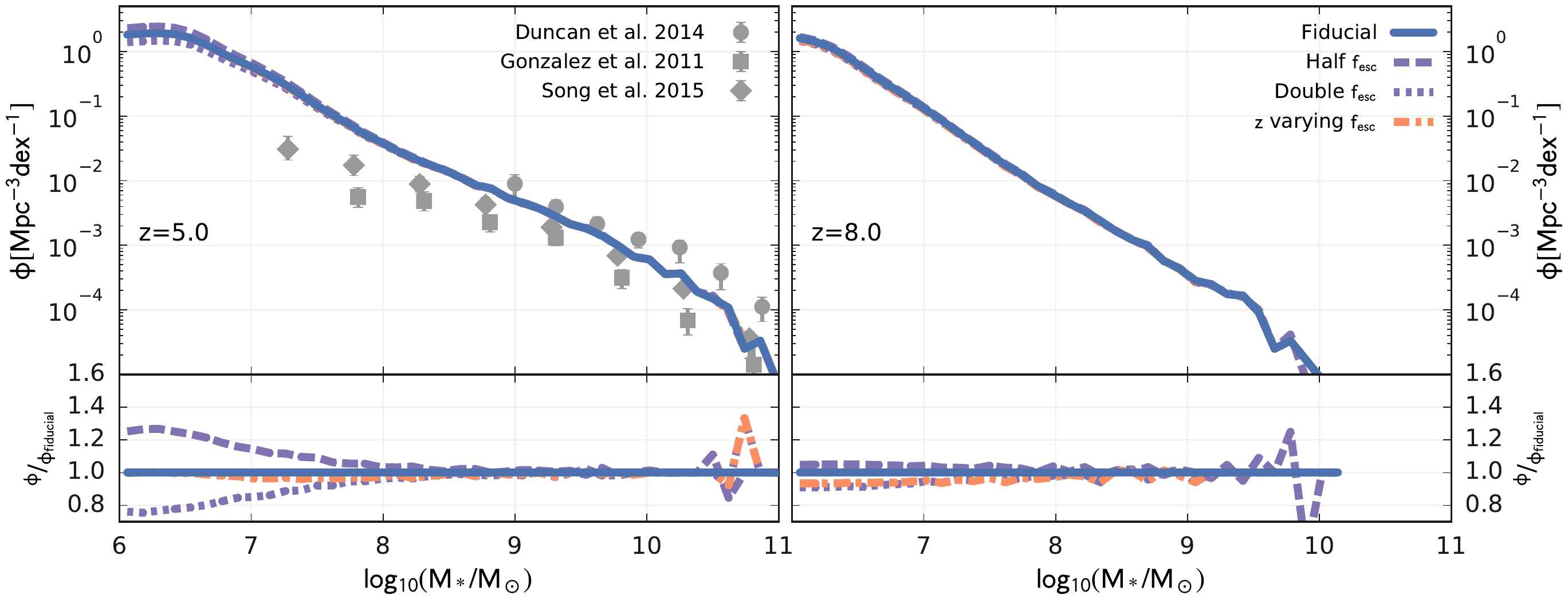}
    \caption{\label{fig:zvarying_smf} The galaxy stellar mass functions produced
    by the \zvaryingfesc\ model (orange dash-dotted) at $z{=}7$ and 8.  Also
    shown for comparison are the results of the \fiducial\ (solid blue),
    \halffesc\ (purple dashed) and \doublefesc\ (purple dotted) models.  Grey
    points in the left-hand panel show the observational data used to constrain
    the \fiducial\ model.  The early onset of reionization in the \zvaryingfesc\
    model is largely compensated for by its extended duration, leading to an
    evolution in the stellar mass function which shows good agreement with our
    \fiducial\ result.}
  \end{minipage}
\end{figure*}

In Fig.~\ref{fig:emissivity_evo}, we show the results of a \zvaryingfesc\ model
(\zvaryingfescLine) which utilizes a modified version of the single power-law
relation proposed by \citet{Kuhlen2012}:
\begin{equation}
  f_{\rm esc}(z)={\rm min}\left[ f_{\rm 
      esc}|_{z=5}\left(\frac{1+z}{6}\right)^\kappa, 1 \right]\ .
\end{equation}
The free parameters of this relation, $f_{\rm esc}|_{z=5}{=}0.04$ and
$\kappa{=}2.5$, have been chosen to reproduce the \citet{McQuinn2011} ionizing
emissivity and {\em Planck} $\tau_{\rm e}$ measurements.  We note that these
values agree well with the range of values found by \citet{Kuhlen2012} to
simultaneously reproduce the observed UV luminosity function\footnote{Assuming
a low luminosity cut-off of $M_{\rm UV}{\la}-14$, equivalent to the resolution
limit of our simulation (Paper IV).}, ionizing emissivity, and electron
scattering optical depth.  The middle panel of Fig.~\ref{fig:emissivity_evo}
shows the corresponding evolution of $f_{\rm esc}(z)$.

In order to self-consistently implement a varying escape fraction in our
model, we remove $f_{\rm esc}$ from the H\,\textsc{ii} ionizing emissivity and efficiency
equations (Equation~\ref{eqn:ionizing_eff} and \ref{eqn:ionizing_emissivity}
respectively) and instead track the build-up of the $f_{\rm esc}(z)$-weighted
gross stellar mass of each galaxy.  This weighted mass is then used in all
21cmFAST calculations.  All other semi-analytic model parameters remain fixed to
their \fiducial\ values.  In the bottom panel of Fig.~\ref{fig:emissivity_evo}
we plot the resulting evolution of the integrated free electron scattering
optical depth for this model, demonstrating that it is consistent with the
{\em Planck} constraints.

At $z{\ga}17$ the \zvaryingfesc\ model escape fraction is large, and the slope
of the ionizing emissivity is similar to that of the fiducial model.  At lower
redshifts, $f_{\rm esc}$ begins to decline, leading to a flattening of the
$\dot N_{\rm ion}$ evolution at $z{\la}9$.  The presence of this flattening
indicates that the escape fraction is declining at a rate which is approximately
equal to the growth of stellar mass in the simulation.  The resulting slow-down
in the rate of ionizing photon emission prolongs the latter stages of
reionization, causing the slower evolution of $\tau_{\rm e}$ seen in the bottom
panel of Fig.~\ref{fig:emissivity_evo}.

In Fig.~\ref{fig:zvarying_xHI_frac}, we present the evolution of the
volume-weighted global neutral fraction of the \zvaryingfesc\ model
(\zvaryingfescLine).  The large escape fraction in this model at high-$z$
leads to an early onset of reionization.  However, since $f_{\rm esc}$
decreases over time, the speed at which reionization progresses declines with
decreasing neutral fraction, resulting in an extended EoR.  In
Fig.~\ref{fig:zvarying_smf}, we also present the galaxy stellar mass functions
of the \zvaryingfesc\ model at $z{=}5$ and 8.  From this, we see that the early
onset of reionization is largely compensated for by its extended duration,
leading to an evolution in the stellar mass function which shows good agreement
with our \fiducial\ result.

In summary, we have demonstrated that, for a realistic population of galaxies
which match current observational measurements of the growth of stellar mass
and electron scattering optical depth, the additional constraining power of
the observed post-reionization ionizing emissivity provides potential evidence
for the requirement of a varying escape fraction of ionizing photons.  This 
agrees well with the findings of other works using both $\tau_{\rm e}$ and $\dot 
N_{\rm ion}$ as constraints \citep[e.g.][]{Kuhlen2012}.

\subsection{The importance of environment}
\label{sub:results-environ}

As demonstrated in Figs~\ref{fig:smf_comparison} and \ref{fig:xHI_frac},
there is excellent agreement between the stellar mass functions and neutral
fraction evolutions predicted by the \fiducial\ (\fiducialLine) and
\homogeneous\ (\homogeneousLine) reionization prescriptions.  Despite this, we
show in this section that there can be significant differences in the stellar
mass predictions for individual galaxies.  We also explore these differences in
detail in order to elucidate the important role of environment in modulating the
UVB suppression of galaxy growth.

\begin{figure}
  \includegraphics[width=\columnwidth]{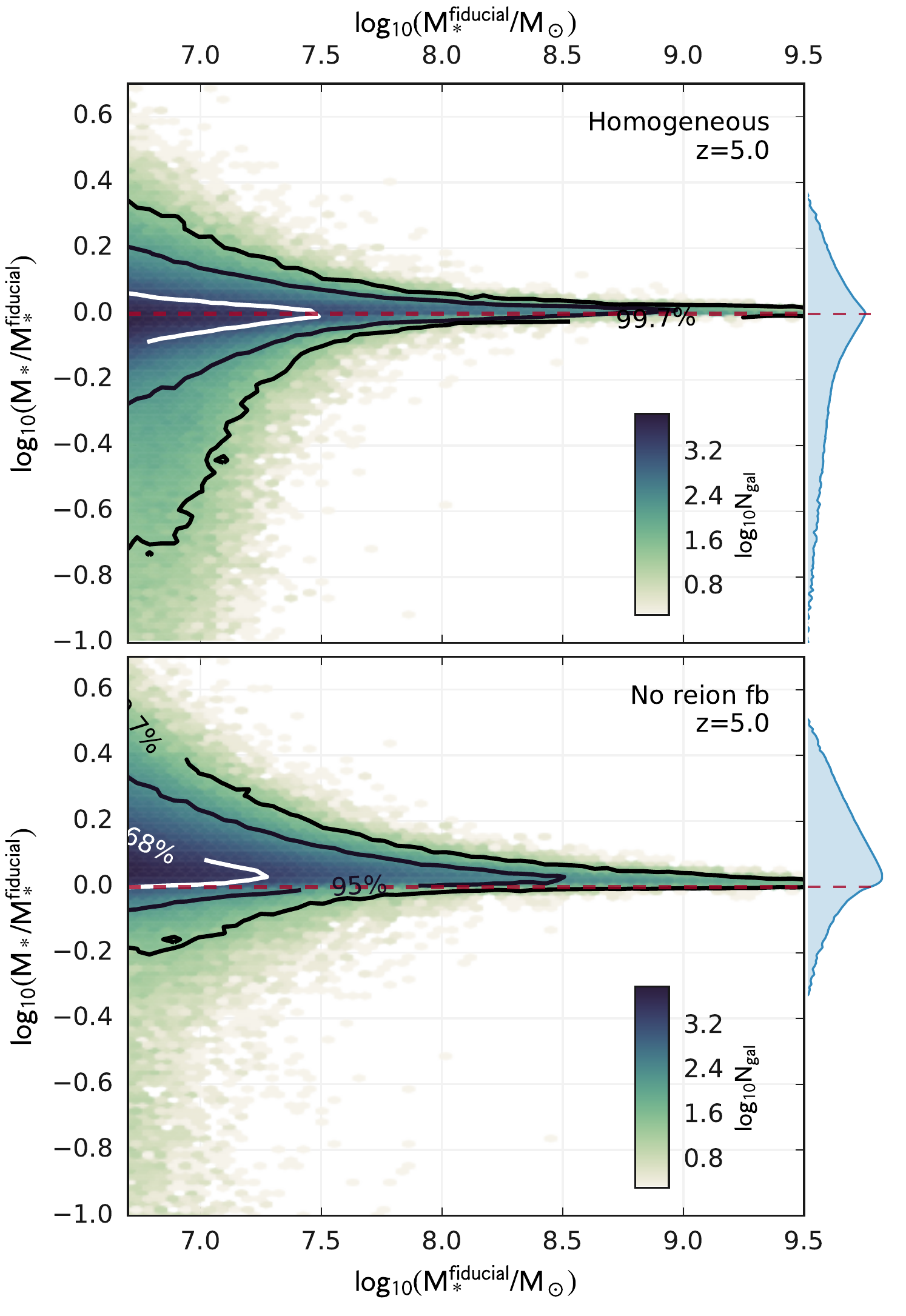}
  \caption{\label{fig:oto_sm} The fractional change in stellar masses of
  individual galaxies with respect to the \fiducial\ model at $z{=}5$.
  Contours indicate the 1$\sigma$, 2$\sigma$ and 3$\sigma$ confidence intervals
  of each distribution.  Blue distributions on the right axes show a log
  normalized kernel density estimation of the marginalized distribution of
  stellar mass fractions.  The spread in the galaxy masses predicted by the
  \homogeneous\ model compared to the self-consistent, patchy \fiducial\ model
  (top panel) is largely a result of environmental effects which are not fully
  encoded by the former model.  Furthermore, there are a number of galaxies
  with stellar masses in the \noreionfb\ model which are less than their
  counterparts in the \fiducial\ feedback case (bottom panel).  This highlights
  the potentially complicated nature of photoionization suppression on the
  growth of stellar mass in the early Universe.}
\end{figure}

In the top panel of Fig.~\ref{fig:oto_sm}, we show the distribution of
fractional changes in individual galaxy stellar masses between the \fiducial\
and \homogeneous\ models.  Contours indicate the 68, 95 and 99\% confidence
intervals of the distribution, whilst the histograms on the right-hand side show
the distribution\footnote{The plotted curves are a log normalized kernel-density
estimate of full marginalized distribution.} of all stellar mass ratios for
galaxies in the \fiducial\ model with stellar masses larger than $10^{6.7}\,{\rm
M_{{\sun}}}$.  As expected from the close agreement between the stellar mass
functions of these two prescriptions, the distribution is peaked around
$M_*^{\rm homog}{/}M_*^{\rm fiducial} {=} 1$.  However, some galaxies exhibit
significant variations, especially at lower stellar masses.  Below $10^{7}\,
{\rm M_{\sun}}$, it is common for galaxies to vary in mass by factors of 2 or
more, with order-of-magnitude differences possible for lower masses.  The
distribution in galaxy mass ratios at a fixed stellar mass is largely symmetric,
meaning that the total stellar functions produced by these two prescriptions
remain in good agreement.  However, the different predictions for individual
galaxy masses may have important consequences for galaxy clustering statistics
and $21$cm power spectra owing to the Poisson noise that this scatter adds
between the halo and galaxy clustering.  This highlights that despite the good
agreement in stellar mass functions, there can be important consequences of
using a self-consistent, spatially dependent model of reionization that is not
fully encoded in a parametrized homogeneous description.

In the bottom panel of Fig.~\ref{fig:oto_sm} we plot the same distribution of
variations in individual galaxy masses between the \fiducial\ and \noreionfb\
runs.  As expected, the bulk of galaxies experience a boost in their masses
in the absence of reionization feedback.  This is because each galaxy attains
access to the full Universal fraction of baryons, even after their surroundings
have been fully ionized.  There are, however, a small number of galaxies for
which the absence of photoionization results in a decreased stellar mass.  In
the majority of cases, this can be explained by an increase in the number of
galaxies populating low-mass haloes.  In the \fiducial\ model, infall into these
low-mass haloes is suppressed due to the ionizing background.  Thus, when they
are accreted into larger systems, with deeper potential wells, their new parents
are able to accrete the previously photoheated to fuel new star formation.
However, in the \noreionfb\ model, the absence of the ionizing background allows
many small haloes to accrete baryons which then cool and condense down into a
galaxy.  When these haloes are eventually accreted into larger systems, their
baryons remain locked up in the infalling satellite, making them unavailable to
fuel the growth of the central galaxy.  In addition, small satellite galaxies
often have extremely long merger times (cf.  Equation~\ref{eqn:t_merge}),
further compounding their ability to keep baryons locked up in the form of stars
and cold gas.  The net effect is that the stellar mass of their host halo's
central galaxy can actually go down in the absence of reionization feedback,
again highlighting the potentially complicated effects of photoionization
suppression on the growth of stellar mass in the early Universe.

\begin{figure}
  \includegraphics[width=1.0\columnwidth]{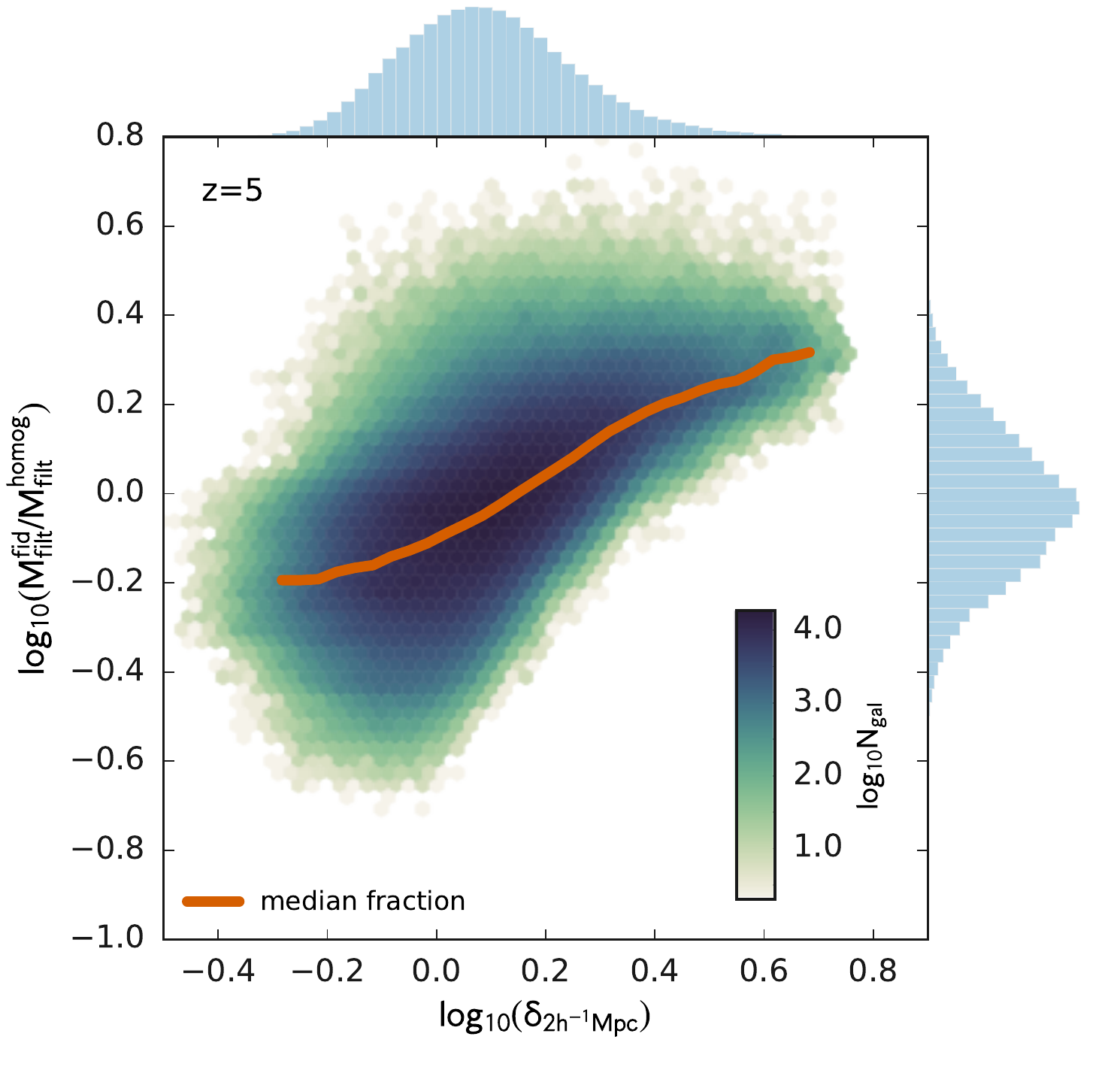}
  \caption{The fractional difference in filtering mass ($M_{\rm filt}$) values 
    of individual haloes at $z{=}5$ predicted by the \fiducial\ patchy
    reionization model and the homogeneous parametrized model as a function
    of total matter density smoothed over a $2h^{-1}\, {\rm Mpc}$ scale.
    Histograms on the right and top axes indicate the marginalized distributions
    of filtering mass ratios and densities, respectively.  There is a clear
    trend of increasing $M_{\rm filt}$ (i.e. more effective photoionization
    suppression) with increasing environmental density in the \fiducial\ model
    which is not captured in the simple homogeneous case.  This can be easily
    understood as the effect of UVB suppression from ionizing photons produced
    by nearby galaxies in over-dense environments, an effect which is not fully
    included in the parametrized \homogeneous\ model.}
  \label{fig:fmod_env_corr}
\end{figure}

In Fig.~\ref{fig:fmod_env_corr}, we plot a 2D histogram of the ratio of 
filtering mass values ($M_{\rm filt}$; cf. Equation~\ref{eqn:Mfilt}) predicted 
by the \fiducial\ and \homogeneous\ models for each FoF group at $z{=}5$, as a 
function of the local total matter density smoothed on a $2h^{-1}\, {\rm Mpc}$ 
scale.  The histogram on the right-hand axes indicates the distribution of 
filtering mass ratios marginalized over all environmental densities.  We find 
that the patchy \fiducial\ model predicts a photoionization suppression which 
depends on environment with up to an order of magnitude difference in the 
filtering mass between over- and underdense regions.  The trend of increasing 
filtering mass with increasing over-density can be seen more clearly by 
considering the median filtering mass ratio as a function of environment (solid 
orange line) and is due to the contribution of ionizing flux from nearby 
galaxies which is not self-consistently included in the \homogeneous\ model.  At 
higher local densities, there is an increase in the average number and mass of 
nearby sources contributing ionizing photons which in the \fiducial, patchy 
reionization model results in an increased $M_{\rm filt}$ value.

The inability of parametrized global reionization prescriptions to capture the
density-dependent spread in filtering masses, as well as the trend of
increasing filtering mass with density, could have important consequences for
the galaxy clustering and cross-correlation statistics.  It is also important
to note that we were only able to calibrate the homogeneous variation through
the use of our full fiducial run with self-consistently coupled reionization.
Any change to the free model parameters listed in
Table~\ref{tab:fiduical_params}, or underlying cosmological model, would require
the mean $M_{\rm filt}$--$z$ relationship to be recalculated.


\section{Conclusions}
\label{sec:conclusions}

In this work we have presented a new semi-analytic galaxy formation model, 
\Meraxes, developed as part of the \DRAGONS\ programme.  \Meraxes\ has been 
designed to investigate the growth of the first galaxies and their 
role in driving the EoR.  It possesses the following key 
features.

\begin{itemize}
  \item A temporally and spatially coupled treatment of reionization 
    provided by the integration of the semi-numerical reionization model \tocf\ 
    (Section~\ref{sub:reionization}).
  \item The use of accurate hierarchical merger trees extracted from a custom 
    \textit{N}-body simulation with a large volume and high mass resolution, as
    well as the temporal resolution required to resolve star formation and
    supernova feedback in the early Universe
    (Section~\ref{sub:n_body_simulation}).
  \item Galaxy formation physics based on the \citet{Croton2006} semi-analytic 
    model, with updates to deal with the high temporal resolution provided
    by our input merger trees such as a time-delayed supernova feedback
    prescription and the removal of the instantaneous recycling approximation
    (Sections~\ref{sub:baryonic_infall}--\ref{sub:ghost_galaxy_evolution}).
\end{itemize}

We have calibrated the free parameters of \Meraxes\ against the observed
evolution of the galaxy stellar mass function between $5{\le}z{\le}7$,
the latest {\em Planck} \citep{Planck-Collaboration2015} constraints on
the electron optical scattering depth (Section~\ref{sec:model_calibration}),
and the Lyman $\alpha$ forest constraints on the $z{\la}6$ ionizing emissivity
(Section~\ref{ssub:results-evolving_escape_fraction}).  Through the subsequent
investigation of a number of model runs with varying reionization and galaxy
feedback parameters, we find the following key results.

\begin{itemize}
  \item Supernova feedback is the dominant physical mechanism for regulating the 
    growth of galaxy stellar mass both during and immediately prior to 
    reionization.  We find that it is only in the absence of supernova feedback 
    that suppression of star formation due to the presence of an ionizing UVB 
    reaches significant levels, and even then, only if star formation is
    efficient enough so as to be gas supply limited.
  \item We further find that the reionization history of the IGM is similarly
    insensitive to reionization feedback.  This implies that the process of
    reionization is not self-regulated
    (Section~\ref{sub:results-xHI})\footnote{However, we note that if the
      process of reionization is more extended than is predicted by our model
      (for example, due to inhomogeneous recombinations in the IGM) then
    reionization feedback may play a more effective role.}.
  \item The \fiducial\ model also predicts that a broad range of halo masses
    contribute to ionizing photon production during reionization
    (Section~\ref{sub:results-xHI}).
  \item Models employing a constant escape fraction of ionizing photons are
    unable to simultaneously match the observed growth of galaxy stellar mass,
    Thomson scattering optical depth measurements, and the Lyman $\alpha$
    forest constraints on the $z{\la}6$ global ionizing emissivity.  However,
    we show that an evolving escape fraction, with $0.05{\la}f_{\rm
    esc}{\la}0.1$ at $z{\sim}6$ and a redshift scaling proportional to
    $\left(1{+}z\right)^{\kappa}$, where $\kappa{\sim}3$, does allow our model
    to simultaneously satisfy these constraints.
    (Section~\ref{sub:results-emissivity}).
  \item Using our framework we are able to quantify the effect of reionization 
    on the shape and normalization of the stellar mass function, finding no
    observationally detectable imprint for stellar masses of $M_{*}{>}10^{7.5}$
    at $z{\ge}5$ (Section~\ref{sub:results-smf}).
  \item By comparing to a simple, homogeneous reionization prescription of the 
    kind traditionally used in semi-analytic models, we find that the inclusion of 
    a self-consistent patchy reionization model can result in significant, 
    environmentally dependent variations in the stellar masses of individual 
    galaxies by factors of 2--3 (Section~\ref{sub:results-environ}).
\end{itemize}

The scatter and density-dependent effects of UVB suppression could
have important consequences for galaxy clustering and cross-correlation
statistics.  These could in turn lead to detectable signatures in the $21\,{\rm
cm}$ power spectrum measurements of current and upcoming radio surveys.  The
ability to self-consistently include and quantify these effects is a key feature
of \Meraxes\ and the \DRAGONS\ programme and is investigated in detail in Paper
V.

In this work, we have chosen to calibrate our model against the high-$z$ stellar
mass function.  However, a more directly observable quantity is UV luminosity.
In Paper IV, we demonstrate the excellent agreement of our fiducial
model with the observed UV luminosity function over a broad range in redshift
($5{<}z{<}10$), allowing us to further investigate the validity of the commonly
used Kennicutt relation \citep{Kennicutt1998,Madau1998} to approximate galaxy
star formation rates from UV luminosities at high-$z$.

Finally, we note that we have chosen to base the galaxy physics of our initial
\Meraxes\ implementation on the \citet{Croton2006} model in order to provide
confidence in our newly developed framework and to allow us to test the
relative effects of supernova and reionization feedback for a set of
well-understood and tested physics prescriptions.  However, in future work we
will expand upon \Meraxes, adapting and improving the model to provide a
better match with the results of our suite of hydrodynamical simulations,
{\em Smaug} \citep[][]{Duffy2014}, run as part of the \DRAGONS\ programme.

\section*{Acknowledgements}
\label{sec:Acknowledgements}

This research was supported by the Victorian Life Sciences Computation
Initiative (VLSCI), grant ref. UOM0005, on its Peak Computing Facility hosted
at the University of Melbourne, an initiative of the Victorian Government,
Australia. Part of this work was performed on the gSTAR national facility at
Swinburne University of Technology. gSTAR is funded by Swinburne and the
Australian Governments Education Investment Fund.  AM acknowledges support from
the European Research Council (ERC) under the European Unions Horizon 2020
research and innovation programme (grant agreement no. 638809 AIDA). This
research programme is funded by the Australian Research Council through the ARC
Laureate Fellowship FL110100072 awarded to JSBW.

\bibliographystyle{mn2e}
\bibliography{meraxes}

\begin{thebibliography}{117}
\expandafter\ifx\csname natexlab\endcsname\relax\def\natexlab#1{#1}\fi

\bibitem[{{Angel} {et~al}\mbox{.}(2016){Angel}, {Poole}, {Ludlow}, {Duffy},
  {Geil}, {Mutch}, {Mesinger}, \& {Wyithe}}]{Angel2016}
{Angel} P.~W., {Poole} G.~B., {Ludlow} A.~D., {Duffy} A.~R., {Geil} P.~M.,
  {Mutch} S.~J., {Mesinger} A., {Wyithe} J.~S.~B., 2016, \mnras, 459, 2106

\bibitem[{{Barkana} \& {Loeb}(2001)}]{Barkana2001}
{Barkana} R., {Loeb} A., 2001, \physrep, 349, 125

\bibitem[{{Baugh}(2006)}]{Baugh2006}
{Baugh} C.~M., 2006, Rep. Prog. Phys., 69, 3101

\bibitem[{{Becker} \& {Bolton}(2013)}]{Becker2013}
{Becker} G.~D., {Bolton} J.~S., 2013, \mnras, 436, 1023

\bibitem[{{Behroozi} {et~al}\mbox{.}(2013{\natexlab{a}}){Behroozi}, {Wechsler},
  \& {Conroy}}]{Behroozi2013b}
{Behroozi} P.~S., {Wechsler} R.~H., {Conroy} C., 2013{\natexlab{a}}, \apjl,
  762, L31

\bibitem[{{Behroozi} {et~al}\mbox{.}(2013{\natexlab{b}}){Behroozi}, {Wechsler},
  \& {Conroy}}]{Behroozi2013}
{Behroozi} P.~S., {Wechsler} R.~H., {Conroy} C., 2013{\natexlab{b}}, \apj, 770,
  57

\bibitem[{{Benson} {et~al}\mbox{.}(2002){Benson}, {Lacey}, {Baugh}, {Cole}, \&
  {Frenk}}]{Benson2002}
{Benson} A.~J., {Lacey} C.~G., {Baugh} C.~M., {Cole} S., {Frenk} C.~S., 2002,
  \mnras, 333, 156

\bibitem[{{Benson} {et~al}\mbox{.}(2006){Benson}, {Sugiyama}, {Nusser}, \&
  {Lacey}}]{Benson2006}
{Benson} A.~J., {Sugiyama} N., {Nusser} A., {Lacey} C.~G., 2006, \mnras, 369,
  1055

\bibitem[{{Binney} \& {Tremaine}(2008)}]{Binney2008}
{Binney} J., {Tremaine} S., 2008, {Galactic Dynamics: Second Edition}.
  Princeton University Press, Princeton, NJ

\bibitem[{{Bolton} \& {Haehnelt}(2007)}]{Bolton2007}
{Bolton} J.~S., {Haehnelt} M.~G., 2007, \mnras, 382, 325

\bibitem[{{Bournaud} {et~al}\mbox{.}(2014){Bournaud}, {Perret}, {Renaud},
  {Dekel}, {Elmegreen}, {Elmegreen}, {Teyssier}, {Amram}, {Daddi}, {Duc},
  {Elbaz}, {Epinat}, {Gabor}, {Juneau}, {Kraljic}, \& {Le
  Floch'}}]{Bournaud2014}
{Bournaud} F. {et~al.}, 2014, \apj, 780, 57

\bibitem[{{Bouwens} {et~al}\mbox{.}(2011){Bouwens}, {Illingworth}, {Oesch},
  {Labb{\'e}}, {Trenti}, {van Dokkum}, {Franx}, {Stiavelli}, {Carollo},
  {Magee}, \& {Gonzalez}}]{Bouwens2011}
{Bouwens} R.~J. {et~al.}, 2011, \apj, 737, 90

\bibitem[{{Bower} {et~al}\mbox{.}(2006){Bower}, {Benson}, {Malbon}, {Helly},
  {Frenk}, {Baugh}, {Cole}, \& {Lacey}}]{Bower2006}
{Bower} R.~G., {Benson} A.~J., {Malbon} R., {Helly} J.~C., {Frenk} C.~S.,
  {Baugh} C.~M., {Cole} S., {Lacey} C.~G., 2006, \mnras, 370, 645

\bibitem[{{Boylan-Kolchin} {et~al}\mbox{.}(2008){Boylan-Kolchin}, {Ma}, \&
  {Quataert}}]{Boylan-Kolchin2008}
{Boylan-Kolchin} M., {Ma} C.-P., {Quataert} E., 2008, \mnras, 383, 93

\bibitem[{Bullock {et~al}\mbox{.}(2001)Bullock, Dekel, Kolatt, Kravtsov,
  Klypin, Porciani, \& Primack}]{Bullock2001}
Bullock J.~S., Dekel A., Kolatt T.~S., Kravtsov A., Klypin A., Porciani C.,
  Primack J.~R., 2001, ApJ, 555, 240

\bibitem[{{Chen} {et~al}\mbox{.}(2014){Chen}, {Wise}, {Norman}, {Xu}, \&
  {O'Shea}}]{Chen2014}
{Chen} P., {Wise} J.~H., {Norman} M.~L., {Xu} H., {O'Shea} B.~W., 2014, \apj,
  795, 144

\bibitem[{{Ciardi} {et~al}\mbox{.}(2003){Ciardi}, {Stoehr}, \&
  {White}}]{Ciardi2003}
{Ciardi} B., {Stoehr} F., {White} S.~D.~M., 2003, \mnras, 343, 1101

\bibitem[{{Cole} {et~al}\mbox{.}(2000){Cole}, {Lacey}, {Baugh}, \&
  {Frenk}}]{Cole2000}
{Cole} S., {Lacey} C.~G., {Baugh} C.~M., {Frenk} C.~S., 2000, \mnras, 319, 168

\bibitem[{{Cox} {et~al}\mbox{.}(2004){Cox}, {Primack}, {Jonsson}, \&
  {Somerville}}]{Cox2004}
{Cox} T.~J., {Primack} J., {Jonsson} P., {Somerville} R.~S., 2004, \apjl, 607,
  L87

\bibitem[{Croton {et~al}\mbox{.}(2006)Croton, Springel, White, De~Lucia, Frenk,
  Gao, Jenkins, Kauffmann, Navarro, \& Yoshida}]{Croton2006}
Croton D.~J. {et~al.}, 2006, MNRAS, 365, 11

\bibitem[{{De Lucia} \& {Blaizot}(2007)}]{De-Lucia2007}
{De Lucia} G., {Blaizot} J., 2007, \mnras, 375, 2

\bibitem[{{De Lucia} {et~al}\mbox{.}(2004){De Lucia}, {Kauffmann}, \&
  {White}}]{De-Lucia2004}
{De Lucia} G., {Kauffmann} G., {White} S.~D.~M., 2004, \mnras, 349, 1101

\bibitem[{{D{\'{\i}}az} {et~al}\mbox{.}(2014){D{\'{\i}}az}, {Koyama},
  {Ryan-Weber}, {Cooke}, {Ouchi}, {Shimasaku}, \& {Nakata}}]{Diaz2014}
{D{\'{\i}}az} C.~G., {Koyama} Y., {Ryan-Weber} E.~V., {Cooke} J., {Ouchi} M.,
  {Shimasaku} K., {Nakata} F., 2014, \mnras, 442, 946

\bibitem[{{Dijkstra} {et~al}\mbox{.}(2004){Dijkstra}, {Haiman}, {Rees}, \&
  {Weinberg}}]{Dijkstra2004}
{Dijkstra} M., {Haiman} Z., {Rees} M.~J., {Weinberg} D.~H., 2004, \apj, 601,
  666

\bibitem[{Duffy {et~al}\mbox{.}(2014)Duffy, Wyithe, Mutch, \&
  Poole}]{Duffy2014}
Duffy A.~R., Wyithe J. S.~B., Mutch S.~J., Poole G.~B., 2014, MNRAS, 443, 3435

\bibitem[{Duncan {et~al}\mbox{.}(2014)Duncan, Conselice, Mortlock, Hartley,
  Guo, Ferguson, Dav{\'e}, Lu, Ownsworth, Ashby, Dekel, Dickinson, Faber,
  Giavalisco, Grogin, Kocevski, Koekemoer, Somerville, \& White}]{Duncan2014}
Duncan K. {et~al.}, 2014, MNRAS, 444, 2960

\bibitem[{{Eddington}(1913)}]{Eddington1913}
{Eddington} A.~S., 1913, \mnras, 73, 359

\bibitem[{{Feng} {et~al}\mbox{.}(2016){Feng}, {Di-Matteo}, {Croft}, {Bird},
  {Battaglia}, \& {Wilkins}}]{Feng2015}
{Feng} Y., {Di-Matteo} T., {Croft} R.~A., {Bird} S., {Battaglia} N., {Wilkins}
  S., 2016, \mnras, 455, 2778

\bibitem[{{Finlator} {et~al}\mbox{.}(2011){Finlator}, {Dav{\'e}}, \&
  {{\"O}zel}}]{Finlator2011}
{Finlator} K., {Dav{\'e}} R., {{\"O}zel} F., 2011, \apj, 743, 169

\bibitem[{{Furlanetto}(2006)}]{Furlanetto2006}
{Furlanetto} S.~R., 2006, New Astron. Rev., 50, 157

\bibitem[{{Furlanetto} {et~al}\mbox{.}(2004){Furlanetto}, {Zaldarriaga}, \&
  {Hernquist}}]{Furlanetto2004}
{Furlanetto} S.~R., {Zaldarriaga} M., {Hernquist} L., 2004, \apj, 613, 1

\bibitem[{{Geil} {et~al}\mbox{.}(2015){Geil}, {Mutch}, {Poole}, {Angel},
  {Duffy}, {Mesinger}, \& {Wyithe}}]{Geil2015}
{Geil} P.~M., {Mutch} S.~J., {Poole} G.~B., {Angel} P.~W., {Duffy} A.~R.,
  {Mesinger} A., {Wyithe} J.~S.~B., 2015, preprint (arXiv: e-prints) (Paper V)

\bibitem[{{Geil} \& {Wyithe}(2008)}]{Geil2008}
{Geil} P.~M., {Wyithe} J.~S.~B., 2008, \mnras, 386, 1683

\bibitem[{{Genel} {et~al}\mbox{.}(2014){Genel}, {Vogelsberger}, {Springel},
  {Sijacki}, {Nelson}, {Snyder}, {Rodriguez-Gomez}, {Torrey}, \&
  {Hernquist}}]{Genel2014}
{Genel} S. {et~al.}, 2014, \mnras, 445, 175

\bibitem[{{Glazebrook}(2013)}]{Glazebrook2013}
{Glazebrook} K., 2013, \pasa, 30, 56

\bibitem[{{Gnedin}(2000)}]{Gnedin2000}
{Gnedin} N.~Y., 2000, \apj, 542, 535

\bibitem[{{Gnedin} {et~al}\mbox{.}(2008){Gnedin}, {Kravtsov}, \&
  {Chen}}]{Gnedin2008}
{Gnedin} N.~Y., {Kravtsov} A.~V., {Chen} H.-W., 2008, \apj, 672, 765

\bibitem[{Gonz{\'a}lez {et~al}\mbox{.}(2011)Gonz{\'a}lez, Labbe, Bouwens,
  Illingworth, Franx, \& Kriek}]{Gonzalez2011}
Gonz{\'a}lez V., Labbe I., Bouwens R.~J., Illingworth G., Franx M., Kriek M.,
  2011, ApJ Letters, 735, L34

\bibitem[{{Grazian} {et~al}\mbox{.}(2015){Grazian}, {Fontana}, {Santini},
  {Dunlop}, {Ferguson}, {Castellano}, {Amorin}, {Ashby}, {Barro}, {Behroozi},
  {Boutsia}, {Caputi}, {Chary}, {Dekel}, {Dickinson}, {Faber}, {Fazio},
  {Finkelstein}, {Galametz}, {Giallongo}, {Giavalisco}, {Grogin}, {Guo},
  {Kocevski}, {Koekemoer}, {Koo}, {Lee}, {Lu}, {Merlin}, {Mobasher}, {Nonino},
  {Papovich}, {Paris}, {Pentericci}, {Reddy}, {Renzini}, {Salmon}, {Salvato},
  {Sommariva}, {Song}, \& {Vanzella}}]{Grazian2014}
{Grazian} A. {et~al.}, 2015, \aap, 575, A96

\bibitem[{{Grogin} {et~al}\mbox{.}(2011){Grogin}, {Kocevski}, {Faber},
  {Ferguson}, {Koekemoer}, {Riess}, {Acquaviva}, {Alexander}, {Almaini},
  {Ashby}, {Barden}, {Bell}, {Bournaud}, {Brown}, {Caputi}, {Casertano},
  {Cassata}, {Castellano}, {Challis}, {Chary}, {Cheung}, {Cirasuolo},
  {Conselice}, {Roshan Cooray}, {Croton}, {Daddi}, {Dahlen}, {Dav{\'e}}, {de
  Mello}, {Dekel}, {Dickinson}, {Dolch}, {Donley}, {Dunlop}, {Dutton}, {Elbaz},
  {Fazio}, {Filippenko}, {Finkelstein}, {Fontana}, {Gardner}, {Garnavich},
  {Gawiser}, {Giavalisco}, {Grazian}, {Guo}, {Hathi}, {H{\"a}ussler},
  {Hopkins}, {Huang}, {Huang}, {Jha}, {Kartaltepe}, {Kirshner}, {Koo}, {Lai},
  {Lee}, {Li}, {Lotz}, {Lucas}, {Madau}, {McCarthy}, {McGrath}, {McIntosh},
  {McLure}, {Mobasher}, {Moustakas}, {Mozena}, {Nandra}, {Newman}, {Niemi},
  {Noeske}, {Papovich}, {Pentericci}, {Pope}, {Primack}, {Rajan},
  {Ravindranath}, {Reddy}, {Renzini}, {Rix}, {Robaina}, {Rodney}, {Rosario},
  {Rosati}, {Salimbeni}, {Scarlata}, {Siana}, {Simard}, {Smidt}, {Somerville},
  {Spinrad}, {Straughn}, {Strolger}, {Telford}, {Teplitz}, {Trump}, {van der
  Wel}, {Villforth}, {Wechsler}, {Weiner}, {Wiklind}, {Wild}, {Wilson},
  {Wuyts}, {Yan}, \& {Yun}}]{Grogin2011}
{Grogin} N.~A. {et~al.}, 2011, \apjs, 197, 35

\bibitem[{Guo {et~al}\mbox{.}(2013)Guo, White, Angulo, Henriques, Lemson,
  Boylan-Kolchin, Thomas, \& Short}]{Guo2013}
Guo Q., White S., Angulo R.~E., Henriques B., Lemson G., Boylan-Kolchin M.,
  Thomas P., Short C., 2013, MNRAS, 428, 1351

\bibitem[{Guo {et~al}\mbox{.}(2011)Guo, White, Boylan-Kolchin, De~Lucia,
  Kauffmann, Lemson, Li, Springel, \& Weinmann}]{Guo2011}
Guo Q. {et~al.}, 2011, MNRAS, 413, 101

\bibitem[{{Harker} {et~al}\mbox{.}(2006){Harker}, {Cole}, {Helly}, {Frenk}, \&
  {Jenkins}}]{Harker2006}
{Harker} G., {Cole} S., {Helly} J., {Frenk} C., {Jenkins} A., 2006, \mnras,
  367, 1039

\bibitem[{{Henriques} {et~al}\mbox{.}(2015){Henriques}, {White}, {Thomas},
  {Angulo}, {Guo}, {Lemson}, {Springel}, \& {Overzier}}]{Henriques2015}
{Henriques} B.~M.~B., {White} S.~D.~M., {Thomas} P.~A., {Angulo} R., {Guo} Q.,
  {Lemson} G., {Springel} V., {Overzier} R., 2015, \mnras, 451, 2663

\bibitem[{{Henriques} {et~al}\mbox{.}(2013){Henriques}, {White}, {Thomas},
  {Angulo}, {Guo}, {Lemson}, \& {Springel}}]{Henriques2013}
{Henriques} B.~M.~B., {White} S.~D.~M., {Thomas} P.~A., {Angulo} R.~E., {Guo}
  Q., {Lemson} G., {Springel} V., 2013, \mnras, 431, 3373

\bibitem[{Hopkins {et~al}\mbox{.}(2010)Hopkins, Croton, Bundy, Khochfar,
  van~den Bosch, Somerville, Wetzel, Keres, Hernquist, Stewart, Younger, Genel,
  \& Ma}]{Hopkins2010}
Hopkins P.~F. {et~al.}, 2010, ApJ, 724, 915

\bibitem[{{Iliev} {et~al}\mbox{.}(2008){Iliev}, {Mellema}, {Pen}, {Bond}, \&
  {Shapiro}}]{Iliev2008}
{Iliev} I.~T., {Mellema} G., {Pen} U.-L., {Bond} J.~R., {Shapiro} P.~R., 2008,
  \mnras, 384, 863

\bibitem[{{Iliev} {et~al}\mbox{.}(2007){Iliev}, {Pen}, {Bond}, {Mellema}, \&
  {Shapiro}}]{Iliev2007}
{Iliev} I.~T., {Pen} U.-L., {Bond} J.~R., {Mellema} G., {Shapiro} P.~R., 2007,
  \apj, 660, 933

\bibitem[{{Jaacks} {et~al}\mbox{.}(2012){Jaacks}, {Nagamine}, \&
  {Choi}}]{Jaacks2012}
{Jaacks} J., {Nagamine} K., {Choi} J.~H., 2012, \mnras, 427, 403

\bibitem[{{Jeon} {et~al}\mbox{.}(2014){Jeon}, {Pawlik}, {Bromm}, \&
  {Milosavljevi{\'c}}}]{Jeon2014}
{Jeon} M., {Pawlik} A.~H., {Bromm} V., {Milosavljevi{\'c}} M., 2014, \mnras,
  444, 3288

\bibitem[{Kauffmann(1996)}]{Kauffmann1996}
Kauffmann G., 1996, MNRAS, 281, 475

\bibitem[{Kennicutt(1989)}]{Kennicutt1989}
Kennicutt R.~C., 1989, ApJ, 344, 685

\bibitem[{Kennicutt(1998)}]{Kennicutt1998}
Kennicutt R. C.~J., 1998, ApJ, 498, 541

\bibitem[{{Kim} {et~al}\mbox{.}(2013{\natexlab{a}}){Kim}, {Wyithe}, {Park}, \&
  {Lacey}}]{Kim2013}
{Kim} H.-S., {Wyithe} J.~S.~B., {Park} J., {Lacey} C.~G., 2013{\natexlab{a}},
  \mnras, 433, 2476

\bibitem[{{Kim} {et~al}\mbox{.}(2013{\natexlab{b}}){Kim}, {Wyithe}, {Raskutti},
  {Lacey}, \& {Helly}}]{Kim2013b}
{Kim} H.-S., {Wyithe} J.~S.~B., {Raskutti} S., {Lacey} C.~G., {Helly} J.~C.,
  2013{\natexlab{b}}, \mnras, 428, 2467

\bibitem[{{Koekemoer} {et~al}\mbox{.}(2011){Koekemoer}, {Faber}, {Ferguson},
  {Grogin}, {Kocevski}, {Koo}, {Lai}, {Lotz}, {Lucas}, {McGrath}, {Ogaz},
  {Rajan}, {Riess}, {Rodney}, {Strolger}, {Casertano}, {Castellano}, {Dahlen},
  {Dickinson}, {Dolch}, {Fontana}, {Giavalisco}, {Grazian}, {Guo}, {Hathi},
  {Huang}, {van der Wel}, {Yan}, {Acquaviva}, {Alexander}, {Almaini}, {Ashby},
  {Barden}, {Bell}, {Bournaud}, {Brown}, {Caputi}, {Cassata}, {Challis},
  {Chary}, {Cheung}, {Cirasuolo}, {Conselice}, {Roshan Cooray}, {Croton},
  {Daddi}, {Dav{\'e}}, {de Mello}, {de Ravel}, {Dekel}, {Donley}, {Dunlop},
  {Dutton}, {Elbaz}, {Fazio}, {Filippenko}, {Finkelstein}, {Frazer}, {Gardner},
  {Garnavich}, {Gawiser}, {Gruetzbauch}, {Hartley}, {H{\"a}ussler},
  {Herrington}, {Hopkins}, {Huang}, {Jha}, {Johnson}, {Kartaltepe},
  {Khostovan}, {Kirshner}, {Lani}, {Lee}, {Li}, {Madau}, {McCarthy},
  {McIntosh}, {McLure}, {McPartland}, {Mobasher}, {Moreira}, {Mortlock},
  {Moustakas}, {Mozena}, {Nandra}, {Newman}, {Nielsen}, {Niemi}, {Noeske},
  {Papovich}, {Pentericci}, {Pope}, {Primack}, {Ravindranath}, {Reddy},
  {Renzini}, {Rix}, {Robaina}, {Rosario}, {Rosati}, {Salimbeni}, {Scarlata},
  {Siana}, {Simard}, {Smidt}, {Snyder}, {Somerville}, {Spinrad}, {Straughn},
  {Telford}, {Teplitz}, {Trump}, {Vargas}, {Villforth}, {Wagner}, {Wandro},
  {Wechsler}, {Weiner}, {Wiklind}, {Wild}, {Wilson}, {Wuyts}, \&
  {Yun}}]{Koekemoer2011}
{Koekemoer} A.~M. {et~al.}, 2011, \apjs, 197, 36

\bibitem[{{Kravtsov} {et~al}\mbox{.}(2004){Kravtsov}, {Gnedin}, \&
  {Klypin}}]{Kravtsov2004}
{Kravtsov} A.~V., {Gnedin} O.~Y., {Klypin} A.~A., 2004, \apj, 609, 482

\bibitem[{{Kuhlen} \& {Faucher-Gigu{\`e}re}(2012)}]{Kuhlen2012}
{Kuhlen} M., {Faucher-Gigu{\`e}re} C.-A., 2012, \mnras, 423, 862

\bibitem[{{Lacey} {et~al}\mbox{.}(2011){Lacey}, {Baugh}, {Frenk}, \&
  {Benson}}]{Lacey2011}
{Lacey} C.~G., {Baugh} C.~M., {Frenk} C.~S., {Benson} A.~J., 2011, \mnras, 412,
  1828

\bibitem[{{Lagos} {et~al}\mbox{.}(2012){Lagos}, {Bayet}, {Baugh}, {Lacey},
  {Bell}, {Fanidakis}, \& {Geach}}]{Lagos2012}
{Lagos} C.~d.~P., {Bayet} E., {Baugh} C.~M., {Lacey} C.~G., {Bell} T.~A.,
  {Fanidakis} N., {Geach} J.~E., 2012, \mnras, 426, 2142

\bibitem[{{Leroy} {et~al}\mbox{.}(2008){Leroy}, {Walter}, {Brinks}, {Bigiel},
  {de Blok}, {Madore}, \& {Thornley}}]{Leroy2008}
{Leroy} A.~K., {Walter} F., {Brinks} E., {Bigiel} F., {de Blok} W.~J.~G.,
  {Madore} B., {Thornley} M.~D., 2008, \aj, 136, 2782

\bibitem[{{Liu} {et~al}\mbox{.}(2015){Liu}, {Mutch}, {Angel}, {Duffy}, {Geil},
  {Poole}, {Mesinger}, \& {Wyithe}}]{Liu2015}
{Liu} C., {Mutch} S.~J., {Angel} P.~W., {Duffy} A.~R., {Geil} P.~M., {Poole}
  G.~B., {Mesinger} A., {Wyithe} J.~S.~B., 2015, preprint (arXiv: e-prints)
  (Paper IV)

\bibitem[{{Lu} {et~al}\mbox{.}(2011{\natexlab{a}}){Lu}, {Kere{\v s}}, {Katz},
  {Mo}, {Fardal}, \& {Weinberg}}]{Lu2011}
{Lu} Y., {Kere{\v s}} D., {Katz} N., {Mo} H.~J., {Fardal} M., {Weinberg} M.~D.,
  2011{\natexlab{a}}, \mnras, 416, 660

\bibitem[{{Lu} {et~al}\mbox{.}(2011{\natexlab{b}}){Lu}, {Mo}, {Weinberg}, \&
  {Katz}}]{Lu2010}
{Lu} Y., {Mo} H.~J., {Weinberg} M.~D., {Katz} N., 2011{\natexlab{b}}, \mnras,
  416, 1949

\bibitem[{{Madau} {et~al}\mbox{.}(1998){Madau}, {Pozzetti}, \&
  {Dickinson}}]{Madau1998}
{Madau} P., {Pozzetti} L., {Dickinson} M., 1998, \apj, 498, 106

\bibitem[{{Martin}(1999)}]{Martin1999}
{Martin} C.~L., 1999, \apj, 513, 156

\bibitem[{{McLure} {et~al}\mbox{.}(2013){McLure}, {Dunlop}, {Bowler},
  {Curtis-Lake}, {Schenker}, {Ellis}, {Robertson}, {Koekemoer}, {Rogers},
  {Ono}, {Ouchi}, {Charlot}, {Wild}, {Stark}, {Furlanetto}, {Cirasuolo}, \&
  {Targett}}]{McLure2013}
{McLure} R.~J. {et~al.}, 2013, \mnras, 432, 2696

\bibitem[{{McQuinn} {et~al}\mbox{.}(2011){McQuinn}, {Oh}, \&
  {Faucher-Gigu{\`e}re}}]{McQuinn2011}
{McQuinn} M., {Oh} S.~P., {Faucher-Gigu{\`e}re} C.-A., 2011, \apj, 743, 82

\bibitem[{Mesinger \& Dijkstra(2008)}]{Mesinger2008}
Mesinger A., Dijkstra M., 2008, MNRAS, 390, 1071

\bibitem[{{Mesinger} \& {Furlanetto}(2007)}]{Mesinger2007}
{Mesinger} A., {Furlanetto} S., 2007, \apj, 669, 663

\bibitem[{Mesinger {et~al}\mbox{.}(2011)Mesinger, Furlanetto, \&
  Cen}]{Mesinger2011}
Mesinger A., Furlanetto S., Cen R., 2011, MNRAS, 411, 955

\bibitem[{{Mitra} {et~al}\mbox{.}(2013){Mitra}, {Ferrara}, \&
  {Choudhury}}]{Mitra2013}
{Mitra} S., {Ferrara} A., {Choudhury} T.~R., 2013, \mnras, 428, L1

\bibitem[{{Morales} \& {Wyithe}(2010)}]{Morales2010}
{Morales} M.~F., {Wyithe} J.~S.~B., 2010, \araa, 48, 127

\bibitem[{{Murray} {et~al}\mbox{.}(2005){Murray}, {Quataert}, \&
  {Thompson}}]{Murray2005}
{Murray} N., {Quataert} E., {Thompson} T.~A., 2005, \apj, 618, 569

\bibitem[{Mutch {et~al}\mbox{.}(2013)Mutch, Poole, \& Croton}]{Mutch2013}
Mutch S.~J., Poole G.~B., Croton D.~J., 2013, MNRAS, 428, 2001

\bibitem[{{Norman} {et~al}\mbox{.}(2015){Norman}, {Reynolds}, {So}, {Harkness},
  \& {Wise}}]{Norman2015}
{Norman} M.~L., {Reynolds} D.~R., {So} G.~C., {Harkness} R.~P., {Wise} J.~H.,
  2015, \apjs, 216, 16

\bibitem[{{Ocvirk} {et~al}\mbox{.}(2015){Ocvirk}, {Gillet}, {Shapiro},
  {Aubert}, {Iliev}, {Teyssier}, {Yepes}, {Choi}, {Sullivan}, {Knebe},
  {Gottloeber}, {D'Aloisio}, {Park}, {Hoffman}, \& {Stranex}}]{Ocvirk2015}
{Ocvirk} P. {et~al.}, 2015, preprint (arXiv: e-prints)

\bibitem[{{Paardekooper} {et~al}\mbox{.}(2015){Paardekooper}, {Khochfar}, \&
  {Dalla Vecchia}}]{Paardekooper2015}
{Paardekooper} J.-P., {Khochfar} S., {Dalla Vecchia} C., 2015, \mnras, 451,
  2544

\bibitem[{{Paardekooper} {et~al}\mbox{.}(2011){Paardekooper}, {Pelupessy},
  {Altay}, \& {Kruip}}]{Paardekooper2011}
{Paardekooper} J.-P., {Pelupessy} F.~I., {Altay} G., {Kruip} C.~J.~H., 2011,
  \aap, 530, A87

\bibitem[{{Pawlik} {et~al}\mbox{.}(2015){Pawlik}, {Schaye}, \& {Dalla
  Vecchia}}]{Pawlik2015}
{Pawlik} A.~H., {Schaye} J., {Dalla Vecchia} C., 2015, \mnras, 451, 1586

\bibitem[{{Peng} {et~al}\mbox{.}(2015){Peng}, {Maiolino}, \&
  {Cochrane}}]{Peng2015}
{Peng} Y., {Maiolino} R., {Cochrane} R., 2015, Nat, 521, 192

\bibitem[{{Planck Collaboration}(2015)}]{Planck-Collaboration2015}
{Planck Collaboration}, 2015, preprint (arXiv: e-prints)

\bibitem[{{Poole} {et~al}\mbox{.}(2016){Poole}, {Angel}, {Mutch}, {Power},
  {Duffy}, {Geil}, {Mesinger}, \& {Wyithe}}]{Poole2016}
{Poole} G.~B., {Angel} P.~W., {Mutch} S.~J., {Power} C., {Duffy} A.~R., {Geil}
  P.~M., {Mesinger} A., {Wyithe} S.~B., 2016, \mnras, 459, 3025 (Paper I)

\bibitem[{Portinari {et~al}\mbox{.}(1998)Portinari, Chiosi, \&
  Bressan}]{Portinari1998}
Portinari L., Chiosi C., Bressan A., 1998, A\&A, 334, 505

\bibitem[{{Rai{\v c}evi{\'c}} {et~al}\mbox{.}(2011){Rai{\v c}evi{\'c}},
  {Theuns}, \& {Lacey}}]{Raicevic2011}
{Rai{\v c}evi{\'c}} M., {Theuns} T., {Lacey} C., 2011, \mnras, 410, 775

\bibitem[{{Robertson} {et~al}\mbox{.}(2013){Robertson}, {Furlanetto},
  {Schneider}, {Charlot}, {Ellis}, {Stark}, {McLure}, {Dunlop}, {Koekemoer},
  {Schenker}, {Ouchi}, {Ono}, {Curtis-Lake}, {Rogers}, {Bowler}, \&
  {Cirasuolo}}]{Robertson2013}
{Robertson} B.~E. {et~al.}, 2013, \apj, 768, 71

\bibitem[{{Salpeter}(1955)}]{Salpeter1955}
{Salpeter} E.~E., 1955, \apj, 121, 161

\bibitem[{{Salvaterra} {et~al}\mbox{.}(2011){Salvaterra}, {Ferrara}, \&
  {Dayal}}]{Salvaterra2011}
{Salvaterra} R., {Ferrara} A., {Dayal} P., 2011, \mnras, 414, 847

\bibitem[{{Schaye} {et~al}\mbox{.}(2015){Schaye}, {Crain}, {Bower}, {Furlong},
  {Schaller}, {Theuns}, {Dalla Vecchia}, {Frenk}, {McCarthy}, {Helly},
  {Jenkins}, {Rosas-Guevara}, {White}, {Baes}, {Booth}, {Camps}, {Navarro},
  {Qu}, {Rahmati}, {Sawala}, {Thomas}, \& {Trayford}}]{Schaye2015}
{Schaye} J. {et~al.}, 2015, \mnras, 446, 521

\bibitem[{{Schenker} {et~al}\mbox{.}(2013){Schenker}, {Robertson}, {Ellis},
  {Ono}, {McLure}, {Dunlop}, {Koekemoer}, {Bowler}, {Ouchi}, {Curtis-Lake},
  {Rogers}, {Schneider}, {Charlot}, {Stark}, {Furlanetto}, \&
  {Cirasuolo}}]{Schenker2013}
{Schenker} M.~A. {et~al.}, 2013, \apj, 768, 196

\bibitem[{{Shin} {et~al}\mbox{.}(2008){Shin}, {Trac}, \& {Cen}}]{Shin2008}
{Shin} M.-S., {Trac} H., {Cen} R., 2008, \apj, 681, 756

\bibitem[{{So} {et~al}\mbox{.}(2014){So}, {Norman}, {Reynolds}, \&
  {Wise}}]{So2014}
{So} G.~C., {Norman} M.~L., {Reynolds} D.~R., {Wise} J.~H., 2014, \apj, 789,
  149

\bibitem[{Sobacchi \& Mesinger(2013)}]{Sobacchi2013b}
Sobacchi E., Mesinger A., 2013, MNRAS, 432, 3340

\bibitem[{{Sobacchi} \& {Mesinger}(2013)}]{Sobacchi2013}
{Sobacchi} E., {Mesinger} A., 2013, \mnras, 432, 51

\bibitem[{{Sobacchi} \& {Mesinger}(2014)}]{Sobacchi2014}
{Sobacchi} E., {Mesinger} A., 2014, \mnras, 440, 1662

\bibitem[{{Sokasian} {et~al}\mbox{.}(2003){Sokasian}, {Abel}, \&
  {Hernquist}}]{Sokasian2003}
{Sokasian} A., {Abel} T., {Hernquist} L., 2003, \mnras, 340, 473

\bibitem[{{Somerville} {et~al}\mbox{.}(2008){Somerville}, {Hopkins}, {Cox},
  {Robertson}, \& {Hernquist}}]{Somerville2008}
{Somerville} R.~S., {Hopkins} P.~F., {Cox} T.~J., {Robertson} B.~E.,
  {Hernquist} L., 2008, \mnras, 391, 481

\bibitem[{Somerville {et~al}\mbox{.}(2001)Somerville, Primack, \&
  Faber}]{Somerville2001}
Somerville R.~S., Primack J.~R., Faber S.~M., 2001, MNRAS, 320, 504

\bibitem[{{Song} {et~al}\mbox{.}(2016){Song}, {Finkelstein}, {Ashby},
  {Grazian}, {Lu}, {Papovich}, {Salmon}, {Somerville}, {Dickinson}, {Duncan},
  {Faber}, {Fazio}, {Ferguson}, {Fontana}, {Guo}, {Hathi}, {Lee}, {Merlin}, \&
  {Willner}}]{Song2015}
{Song} M. {et~al.}, 2016, \apj, 825, 5

\bibitem[{{Springel} {et~al}\mbox{.}(2005){Springel}, {White}, {Jenkins},
  {Frenk}, {Yoshida}, {Gao}, {Navarro}, {Thacker}, {Croton}, {Helly},
  {Peacock}, {Cole}, {Thomas}, {Couchman}, {Evrard}, {Colberg}, \&
  {Pearce}}]{Springel2005}
{Springel} V. {et~al.}, 2005, Nat, 435, 629

\bibitem[{{Springel} {et~al}\mbox{.}(2001){Springel}, {White}, {Tormen}, \&
  {Kauffmann}}]{Springel2001}
{Springel} V., {White} S.~D.~M., {Tormen} G., {Kauffmann} G., 2001, \mnras,
  328, 726

\bibitem[{Sutherland \& Dopita(1993)}]{Sutherland1993}
Sutherland R.~S., Dopita M.~A., 1993, ApJS, 88, 253

\bibitem[{{Thoul} \& {Weinberg}(1996)}]{Thoul1996}
{Thoul} A.~A., {Weinberg} D.~H., 1996, \apj, 465, 608

\bibitem[{{Trac} \& {Cen}(2007)}]{Trac2007}
{Trac} H., {Cen} R., 2007, \apj, 671, 1

\bibitem[{{Trac} {et~al}\mbox{.}(2008){Trac}, {Cen}, \& {Loeb}}]{Trac2008}
{Trac} H., {Cen} R., {Loeb} A., 2008, \apjl, 689, L81

\bibitem[{{Trenti} {et~al}\mbox{.}(2015){Trenti}, {Perna}, \&
  {Jimenez}}]{Trenti2015}
{Trenti} M., {Perna} R., {Jimenez} R., 2015, \apj, 802, 103

\bibitem[{{Uhlig} {et~al}\mbox{.}(2012){Uhlig}, {Pfrommer}, {Sharma}, {Nath},
  {En{\ss}lin}, \& {Springel}}]{Uhlig2012}
{Uhlig} M., {Pfrommer} C., {Sharma} M., {Nath} B.~B., {En{\ss}lin} T.~A.,
  {Springel} V., 2012, \mnras, 423, 2374

\bibitem[{{van den Bosch} {et~al}\mbox{.}(2008){van den Bosch}, {Aquino},
  {Yang}, {Mo}, {Pasquali}, {McIntosh}, {Weinmann}, \&
  {Kang}}]{van-den-Bosch2008}
{van den Bosch} F.~C., {Aquino} D., {Yang} X., {Mo} H.~J., {Pasquali} A.,
  {McIntosh} D.~H., {Weinmann} S.~M., {Kang} X., 2008, \mnras, 387, 79

\bibitem[{White \& Frenk(1991)}]{White1991}
White S. D.~M., Frenk C.~S., 1991, ApJ, 379, 52

\bibitem[{{Wise} \& {Cen}(2009)}]{Wise2009}
{Wise} J.~H., {Cen} R., 2009, \apj, 693, 984

\bibitem[{{Wisnioski} {et~al}\mbox{.}(2011){Wisnioski}, {Glazebrook}, {Blake},
  {Wyder}, {Martin}, {Poole}, {Sharp}, {Couch}, {Kacprzak}, {Brough},
  {Colless}, {Contreras}, {Croom}, {Croton}, {Davis}, {Drinkwater}, {Forster},
  {Gilbank}, {Gladders}, {Jelliffe}, {Jurek}, {Li}, {Madore}, {Pimbblet},
  {Pracy}, {Woods}, \& {Yee}}]{Wisnioski2011}
{Wisnioski} E. {et~al.}, 2011, \mnras, 417, 2601

\bibitem[{{Wyithe} \& {Loeb}(2003)}]{Wyithe2003}
{Wyithe} J.~S.~B., {Loeb} A., 2003, \apj, 586, 693

\bibitem[{{Wyithe} \& {Loeb}(2004)}]{Wyithe2004}
{Wyithe} J.~S.~B., {Loeb} A., 2004, Nat, 432, 194

\bibitem[{{Wyithe} \& {Loeb}(2013)}]{Wyithe2013}
{Wyithe} J.~S.~B., {Loeb} A., 2013, \mnras, 428, 2741

\bibitem[{{Yajima} {et~al}\mbox{.}(2011){Yajima}, {Choi}, \&
  {Nagamine}}]{Yajima2011}
{Yajima} H., {Choi} J.-H., {Nagamine} K., 2011, \mnras, 412, 411

\bibitem[{{Zahn} {et~al}\mbox{.}(2007){Zahn}, {Lidz}, {McQuinn}, {Dutta},
  {Hernquist}, {Zaldarriaga}, \& {Furlanetto}}]{Zahn2007}
{Zahn} O., {Lidz} A., {McQuinn} M., {Dutta} S., {Hernquist} L., {Zaldarriaga}
  M., {Furlanetto} S.~R., 2007, \apj, 654, 12

\bibitem[{{Zhou} {et~al}\mbox{.}(2013){Zhou}, {Guo}, {Liu}, {Yue}, {Xu}, \&
  {Chen}}]{Zhou2013}
{Zhou} J., {Guo} Q., {Liu} G.-C., {Yue} B., {Xu} Y.-D., {Chen} X.-L., 2013, Re.
  Astron. Astrophys., 13, 373

\end{thebibliography}


\clearpage
\appendix

\section[]{Mass resolution}
\label{sec:appendix-resolution}

Throughout this work we run \Meraxes\ on the output of the \Tiamat\
collisionless $N$-body simulation.  Here we make use of two complimentary
simulations in the \DRAGONS\ suite to quantify the halo and stellar masses down
to which \Tiamat\ is complete, as well as the total fraction of stellar mass
(and hence ionizing photons) which may be missed due to halo mass resolution.
The names, particle masses and box sizes of each simulation are presented in
Table~\ref{tab:appendix-sims}.

\begin{table}
  \centering
  \caption{
    \label{tab:appendix-sims} The particle masses and box sizes of each of the 
    $N$-body simulations from the \Tiamat\ suite used here.
  }
  \begin{tabular}{l|c|c}
  \hline\hline
  {\bf Simulation} & {\bf Particle mass} & {\bf Box side} \\
  & (${\rm M_{{\sun}}}$) & {\bf length} (Mpc) \\
  \hline
  \Tiamat & $3.9{\times}10^6$ & 100 \\
  \MediTiamat & $1.2{\times}10^6$ & 33.3 \\
  \TinyTiamat & $1.0{\times}10^5$ & 14.8 \\
  \hline\hline
  \end{tabular}
\end{table}

\subsection{Halo mass}
\label{sub:appendix-halomass}

\begin{figure*}
  \begin{minipage}{\textwidth}
    \includegraphics[width=\columnwidth]{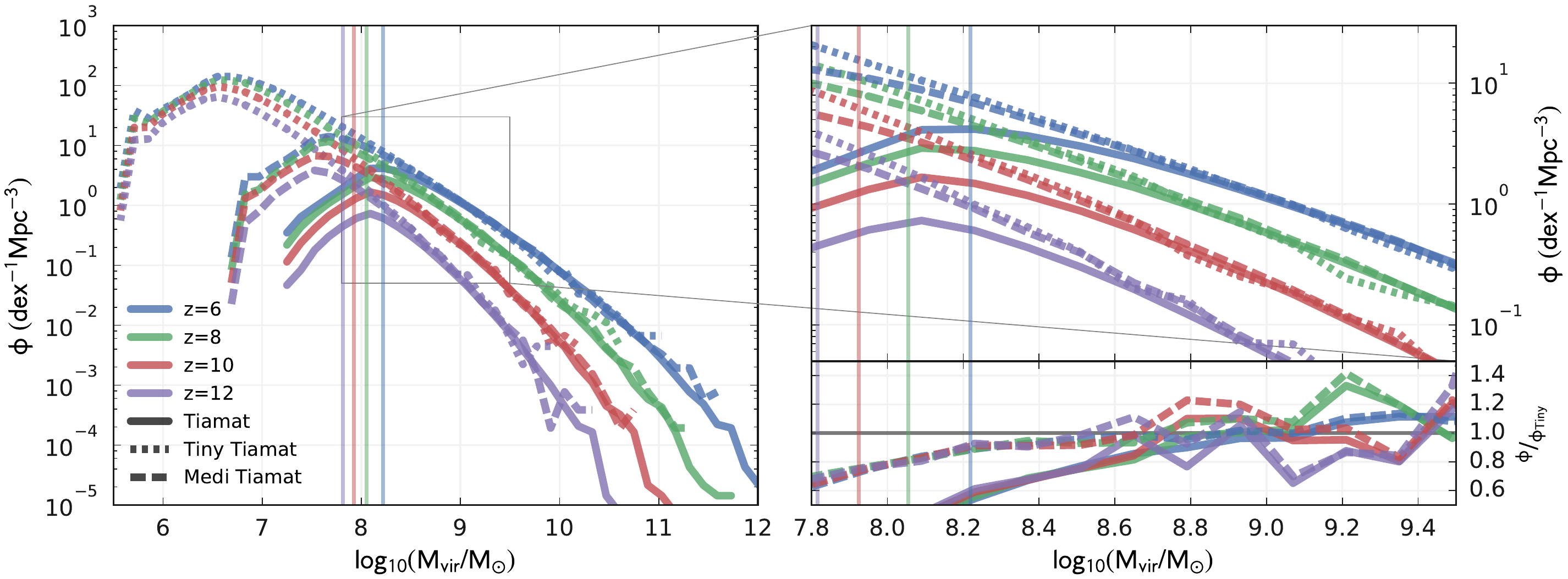}
    \caption{\label{fig:appendix-hmf} The FoF group halo mass functions of the 
      \TinyTiamat\ (dotted), \MediTiamat\ (dashed), and full \Tiamat\ (solid)
      simulations at redshifts 6--12.  Vertical lines indicate the atomic
      cooling mass threshold at each redshift plotted.  The right-hand panel
      shows a zoom in of the mass functions (top), along with the fractional
      difference in number density of the \Tiamat\ and \MediTiamat\ simulations
      with respect to the \TinyTiamat\ results (bottom).}
  \end{minipage}
\end{figure*}

In the left panel of Fig.~\ref{fig:appendix-hmf} we show the FoF group mass
functions for all three simulations at $z$=6, 8, 10 and 12.  The turn overs
at low masses are a direct result of limited mass resolution.  Vertical lines
indicate the atomic cooling mass thresholds ($M_{\rm cool}$) at each of these
redshifts.  In \Meraxes, galaxy formation only occurs in haloes above $M_{\rm
cool}$.  Therefore a fully resolved simulation would ideally identify all haloes
above this value at all redshifts ${\la}15$.  \TinyTiamat\ with its low particle
mass is easily able to achieve this.

The top-right panel of Fig.~\ref{fig:appendix-hmf} shows a zoom-in of the
mass functions near to $M_{\rm cool}$, whilst the bottom-right panel shows
the fractional difference in number density with respect to the \TinyTiamat\
results.  From this we can see that \MediTiamat\ is complete to within $10\%$
down to $M_{\rm vir}{=}M_{\rm cool}$ at $z$=6. \Tiamat\ is complete to
within the same fractional difference down to $M_{\rm vir}{=}10^{8.8}\,{\rm
M_{{\sun}}}$, which is approximately $0.5\, {\rm dex}$ above the atomic cooling
mass threshold at this redshift.  Since the mass resolution of all simulations
remains fixed whilst the value of the atomic cooling mass threshold decreases
with increasing redshift, both \Tiamat\ and \MediTiamat\ effectively miss a
larger fraction of haloes at earlier times.

\subsection{Stellar mass}

\begin{figure*}
  \begin{minipage}{\textwidth}
    \includegraphics[width=\columnwidth]{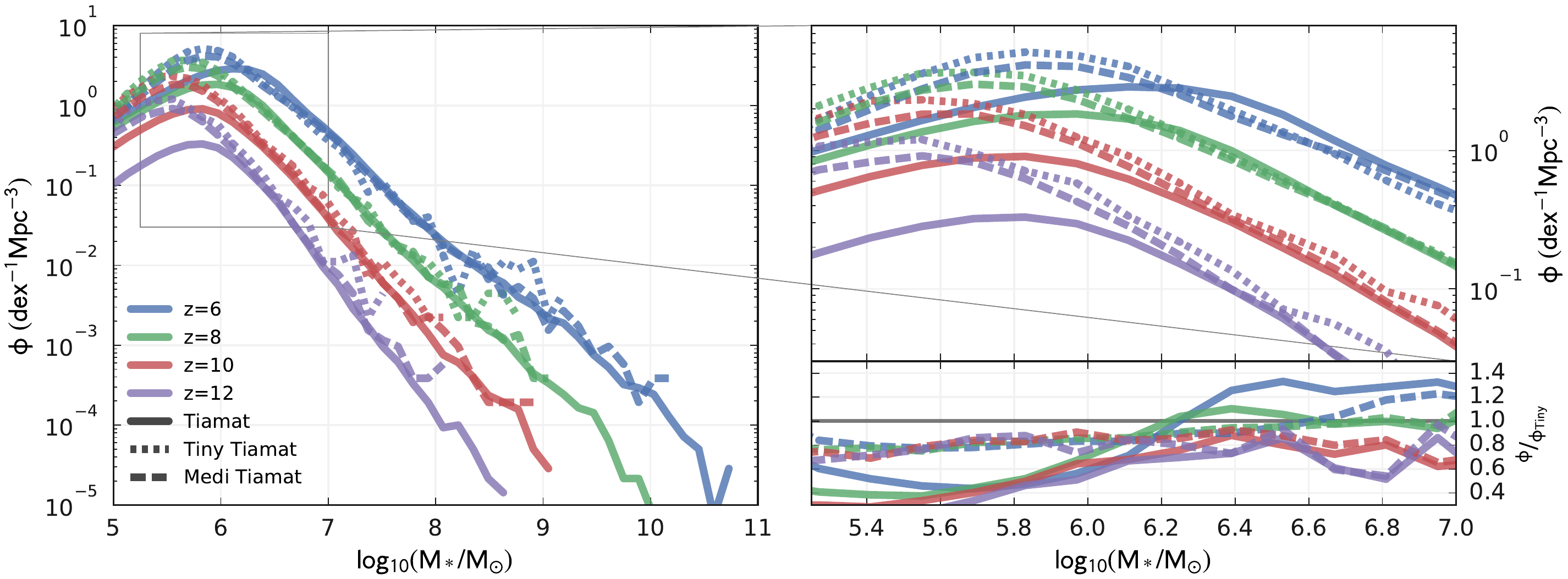}
    \caption{\label{fig:appendix-smf} The galaxy stellar mass function 
      produced by the \noreionfb\ \Meraxes\ model variation run on the merger trees
      extracted from the \TinyTiamat\ (dotted), \MediTiamat\ (dashed), and full
      \Tiamat\ (solid) simulations at redshifts 6--12.  The right-hand panel
      shows a zoom in of the mass functions (top), along with the fractional
      difference in number density of the \Tiamat\ and \MediTiamat\ simulations
      with respect to the \TinyTiamat\ results (bottom).}
  \end{minipage}
\end{figure*}

In Section~\ref{sub:results-relative_eff} we demonstrate that supernova
feedback curtails star formation in low mass haloes.  As a result, although
\Tiamat\ is not complete in halo number density all the way down to the
atomic cooling mass threshold, the fraction of stellar mass (and therefore
ionizing photons) which is unresolved is far less.  In the left hand panel
of Fig.~\ref{fig:appendix-smf} we show the galaxy stellar mass functions
calculated using the \noreionfb\ model run on each $N$-body simulation.  The
free parameters of \Meraxes\ remain fixed between simulations so that any
differences between the results of each are a direct consequence of mass
resolution and volume.

Compared to the FoF group halo mass functions (Fig.~\ref{fig:appendix-hmf}),
the difference in the position of the resolution limit turnovers
between simulations is greatly reduced.  The upper-right panel of
Fig.~\ref{fig:appendix-smf} shows a zoom-in of the mass functions
around the turn over positions.  The lower-right panel again shows the
fractional difference in number density with respect to \TinyTiamat\
which resolves all haloes down to $M_{\rm cool}$ and beyond (see
Section~\ref{sub:appendix-halomass} above).  Here we can see that \MediTiamat\
reaches the turnover in galaxy number density predicted by \TinyTiamat\ to
within $0.1\,{\rm dex}$ at all redshifts shown.  It does, however, predict a
lower normalization in the mass function by around 20\% on average across all
redshifts shown.  This is predominantly a consequence of cosmic variance, driven
by the small box size of both simulations (and in particular \TinyTiamat).  The
turnover in the stellar mass function predictions of \Tiamat\ occurs at masses
approximately 0.2--0.3$\,{\rm dex}$ higher than \TinyTiamat\ at all redshifts.

\begin{figure}
  \includegraphics[width=\columnwidth]{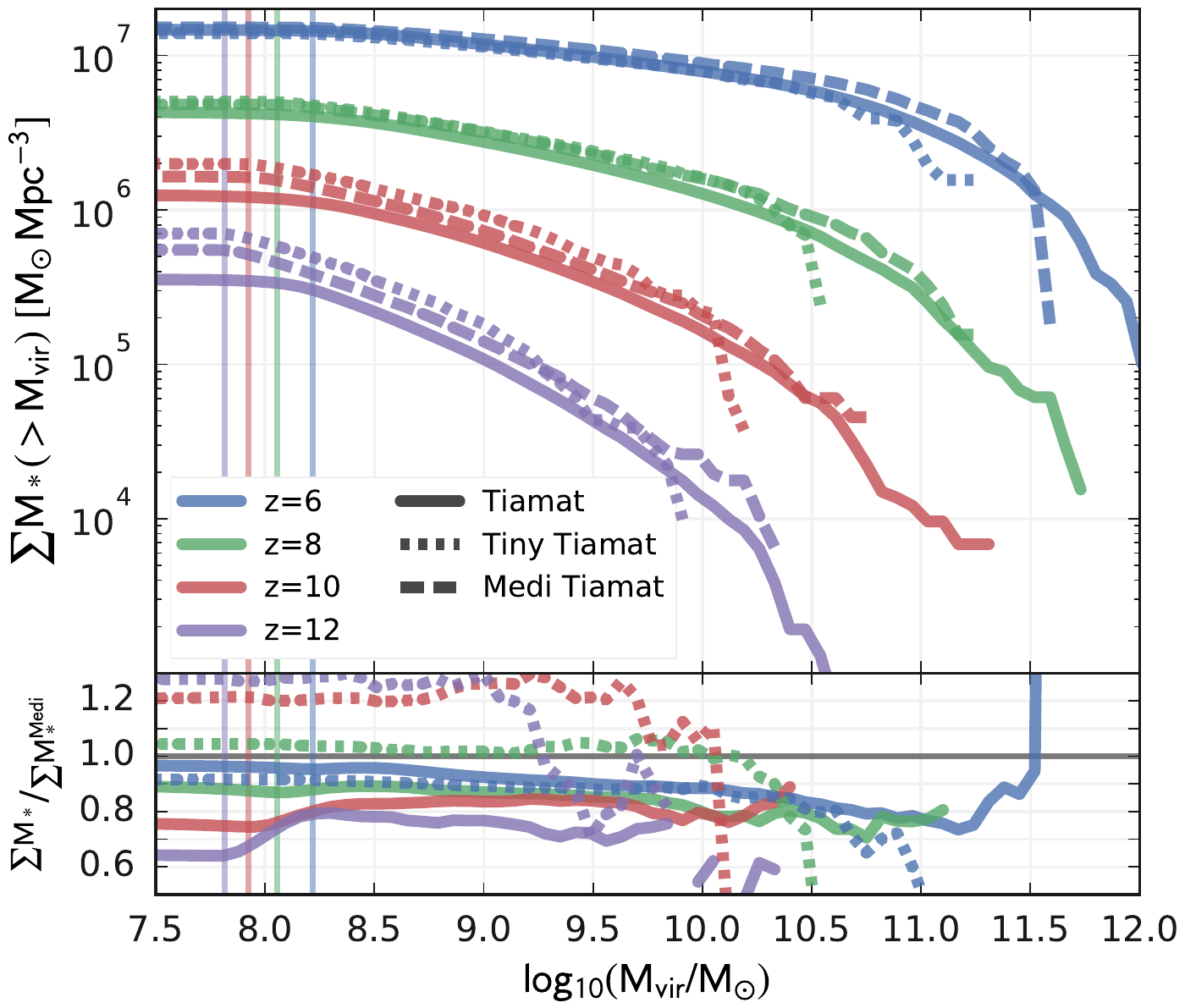}
  \caption{\label{fig:appendix-cumsm} Cumulative stellar mass as a function of
  FoF group mass predicted by the \noreionfb\ \Meraxes\ model variation run
  on the merger trees extracted from the \TinyTiamat\ (dotted), \MediTiamat\
  (dashed), and full \Tiamat\ (solid) simulations at redshifts 6--12.  The
  lower panel shows the fractional difference with respect to the \MediTiamat\
  results.}
\end{figure}

In the top panel of Fig.~\ref{fig:appendix-cumsm} we plot the cumulative
stellar mass of each simulation (again predicted by the \noreionfb\ model) as a
function of FoF group virial mass.  The bottom panel indicates the fractional
difference between the simulations, this time with respect to \MediTiamat.  We
have chosen to utilize \MediTiamat\ as our reference here since it achieves a
good compromise between mass resolution and volume, as demonstrated above.
\TinyTiamat, on the other hand, fails to capture the most massive haloes at any
redshift due to its limited volume.  This biases the cumulative masses predicted
by this simulation downwards.  At $z$=6, we find that \Tiamat\ recovers 97\% of
all stellar mass (and therefore ionizing photons) above the atomic cooling mass
threshold.  The fraction falls to 90\% at $z$=8 and 75\% at $z$=10.

\bsp
\label{lastpage}

\end{document}